\documentclass[epj]{svjour}

\usepackage{color} 
\usepackage{subfigure}

\usepackage{graphicx}
\usepackage{braket}
\usepackage{amsmath}

\usepackage{fancyhdr}
\usepackage{setspace}

\usepackage{pifont}

\newcommand{\etal}{{\it et al. }}
\newcommand{\gray}{$\gamma$-ray}
\newcommand{\grays}{$\gamma$ rays}
\newcommand{\g}{$\gamma$}
\newcommand{\nubar}{$\bar{\nu}$}
\newcommand{\nubarg}{$\bar{N}_\gamma$}
\newcommand{\TKEbar}{$\langle {\rm TKE}\rangle$}

\newcommand{\nf}{($n$,f)}
\newcommand{\nfth}{($n_{\rm th}$,f)}

\newcommand{\CGM}{$\mathtt{CGM}$}
\newcommand{\CGMF}{$\mathtt{CGMF}$}
\newcommand{\FREYA}{$\mathtt{FREYA}$}
\newcommand{\FIFRELIN}{$\mathtt{FIFRELIN}$}
\newcommand{\MCNP}{$\mathtt{MCNP}$}
\newcommand{\mcnp}{$\mathtt{MCNP6.2}$}
\newcommand{\MCNPX}{$\mathtt{MCNPX}$}
\newcommand{\POLIMI}{$\mathtt{MCNPX}$-$\mathtt{PoliMi}$}
\newcommand{\polimi}{$\mathtt{PoliMi}$}
\newcommand{\COH}{$\mathtt{CoH}$}

\newcommand{\checkmark}{\ding{51}} 

\begin{document}

\title{Correlated Prompt Fission Data in Transport Simulations}

\author{P. Talou\inst{1} \and R. Vogt\inst{2,3} \and J. Randrup\inst{4} \and M.E. Rising\inst{5} \and S.A. Pozzi\inst{6} \and J. Verbeke\inst{2} \and M.T. Andrews\inst{5} \and S.D. Clarke\inst{6} \and P. Jaffke\inst{1} \and M. Jandel\inst{7,9} \and T. Kawano\inst{1} \and M.J. Marcath\inst{6} \and K. Meierbachtol\inst{8} \and L. Nakae\inst{2} \and G. Rusev\inst{7} \and A. Sood\inst{5} \and I. Stetcu\inst{1} \and C. Walker\inst{7} }

\institute{
  Nuclear Physics Group, Theoretical Division, Los Alamos National Laboratory,
  Los Alamos, NM 87545, USA
\and 
Nuclear \& Chemical Sciences Division, Lawrence Livermore National Laboratory,
Livermore, CA 94551, USA
\and
Physics Department, University of California at Davis, Davis, CA 95616, USA
\and
Nuclear Science Division, Lawrence Berkeley National Laboratory, Berkeley, CA
94720, USA
\and
Monte Carlo Methods, Codes, and Applications Group, Los Alamos National
Laboratory, Los Alamos, NM 87545, USA
\and
Department of Nuclear Engineering and Radiological Sciences, University of
Michigan, Ann Arbor, MI 48109, USA
\and
Nuclear and Radiochemistry Group, Los Alamos National Laboratory, Los Alamos,
NM 87545, USA
\and
Nuclear Engineering and Nonproliferation, Los Alamos National Laboratory,
Los Alamos, NM 87545, USA
\and
Department of Physics and Applied Physics, University of Massachusetts Lowell, Lowell, MA 01854, USA
}
\date{Draft as of \today}

\abstract{
Detailed information on the fission process can be inferred from the observation, modeling and theoretical understanding of prompt fission neutron and \gray~observables. Beyond simple average quantities, the study of distributions and correlations in prompt data, {\it e.g.}, multiplicity-dependent neutron and \gray~spectra, angular distributions of the emitted particles, $n$-$n$, $n$-\g, and \g-\g~correlations, can place stringent constraints on fission models and parameters that would otherwise be free to be tuned separately to represent individual fission observables. The \FREYA~and \CGMF~codes have been developed to follow the sequential emissions of prompt neutrons and \grays~from the initial excited fission fragments produced right after scission. Both codes implement Monte Carlo techniques to sample initial fission fragment configurations in mass, charge and kinetic energy and sample probabilities of neutron and \g~emission at each stage of the decay. This approach naturally leads to using simple but powerful statistical techniques to infer distributions and correlations among many observables and model parameters. The comparison of model calculations with experimental data provides a rich arena for testing various nuclear physics models such as those related to the nuclear structure and level densities of neutron-rich nuclei, the \gray~strength functions of dipole and quadrupole transitions, the mechanism for dividing the excitation energy between the two nascent fragments near scission, and the mechanisms behind the production of angular momentum in the fragments, etc. Beyond the obvious interest from a fundamental physics point of view, such studies are also important for addressing data needs in various nuclear applications. The inclusion of the \FREYA~and \CGMF~codes into the \mcnp~and \POLIMI~transport codes, for instance, provides a new and powerful tool to simulate correlated fission events in neutron transport calculations important in nonproliferation, safeguards, nuclear energy, and defense programs. This review provides an overview of the topic, starting from theoretical considerations of the fission process, with a focus on correlated signatures. It then explores the status of experimental correlated fission data and current efforts to address some of the known shortcomings. Numerical simulations employing the \FREYA~and \CGMF~codes are compared to experimental data for a wide range of correlated fission quantities. The inclusion of those codes into the \mcnp~and \POLIMI~transport codes is described and discussed in the context of relevant applications. The accuracy of the model predictions and their sensitivity to model assumptions and input parameters are discussed. Finally, a series of important experimental and theoretical questions that remain unanswered are presented, suggesting a renewed effort to address these shortcomings.
}


\maketitle

\newpage

\section{Introduction} \label{sec:introduction}

The nuclear fission process, known for over 75 years now, is at the core of many nuclear technologies and scientific studies in fields such as energy, defense, and astrophysics. Conceptually, it can be seen as a complex collective rearrangement of the nuclear many-body system. From our early qualitative description of this process in terms of the deformation of a charged liquid drop~\cite{Bohr:1939} to today's quantitative calculations based on macroscopic-microscopic~\cite{Moller:2015,Ichikawa:2012} or purely microscopic~\cite{Younes:2011,Regnier:2016,Bulgac:2016} descriptions, an enormous amount of data has been collected and a number of theoretical models proposed to account for the wide variety of fission signatures such as fission cross sections, fission fragment yields, fission half-lives, fission isomers, prompt and $\beta$-delayed neutron and \gray\ emission, ternary fission and fission fragment angular distributions. In addition, a large collection of {\it integral} data pertaining to the use of nuclear technologies and, in particular, to nuclear energy and defense applications, has been collected since the dawn of the atomic age.
The existence of such a wide-ranging set of nuclear fission data
may give the impression that everything in nuclear data is now well known.
However, that is far from true. 
As new regions of the periodic table are explored, whether to understand how, when, and where the elements in the universe were initially formed, to understand exotic nuclear structure configurations away from the valley of stability, or to develop innovative nuclear technologies with distinct fuel and material compositions, more diverse and more accurate nuclear data as well as refined models are required to fill in gaps in data beyond the reach of even modern experimental techniques. In the specific case of nuclear fission, many fundamental questions remain.  At the same time, modern applications require very high accuracy in quantities such as cross sections, angular distributions, and spectra. Studying correlated signatures of the fission process can help shed some light on both domains of interest.

The various characteristics of a fission event are naturally correlated.  However, such correlations are generally absent in the evaluated nuclear databases used by modern transport codes. Those correlations range from fission cross sections with fission fragment angular distributions, fission fragment yields with prompt fission neutrons and \grays\ as well as correlations in the number, energy and angle of emission of neutrons and \grays. In scenarios where average quantities dominate, such as the multiplication factor in a critical assembly, correlations are expected to play only a minor role. In other applications, like neutron multiplicity counting~\cite{Ensslin:1998}, however, the situation is quite different and great care must be taken to describe correlations and distributions adequately, such as the higher moments of the prompt neutron multiplicity distribution $P(\nu)$.

In recent years, several parallel efforts to model the fission process on an event-by-event basis have led to the development of computer codes~\cite{CGMF,FREYA,FIFRELIN,GEF} that can calculate many of these correlations.  Integrating these codes into a transport simulation code like \MCNP\textsuperscript{\textregistered}\footnote{$\mathtt{MCNP}$\textsuperscript{\textregistered} and Monte Carlo N-Particle\textsuperscript{\textregistered} are registered trademarks owned by Los Alamos National Security, LLC, manager and operator of Los Alamos National Laboratory. Any third party use of such registered marks should be properly attributed to Los Alamos National Security, LLC, including the use of the designation as appropriate.  For the purposes of visual clarity, the registered trademark symbol is assumed for all references to \MCNP~within the remainder of this paper.}~\cite{MCNP6} represents a major breakthrough for the accurate simulation of fission events in transport calculations.

In this review, the various correlations that develop naturally in a fission event are described (Section~\ref{sec:fission}) before discussing (Section~\ref{sec:codes}) the \CGMF~and \FREYA~codes that simulate such events in detail. In Section~\ref{sec:transportCodes}, the \MCNP6.2 code~\cite{MCNP6} is then briefly introduced and a more in-depth discussion of the fission models present in \MCNP6.2 is given. The \POLIMI~ code~\cite{Pozzi:2012}, developed at the University of Michigan, is an extension of \MCNPX2.7. While not a standalone code, it has been at the forefront of the modeling of correlated fission data, primarily for detector development, safeguards and nonproliferation applications.  The fission-specific developments made in \POLIMI~ are also reviewed before discussing the integration of the fission event generators \CGMF~and \FREYA~into the new release of \MCNP6.2.

In Section~\ref{sec:simulations}, numerical results on correlated fission observables are compared to available experimental data. Those data span correlations between emitted particles, $n$-$n$, $n$-$\gamma$ and $\gamma$-$\gamma$; correlations between emitted particles and fission fragments; and time correlations in fission chains. The time correlations differ from the rest as they are not related to a single fission event but instead to a suite of fission events characteristic of {\it multiplying} objects. Because of the importance of those correlations in safeguards and nonproliferation applications, they are included here although they are not intrinsically within the scope of the event-by-event codes. 

Section~\ref{sec:status} presents the status of the fission event generators discussed here, lists the fission reactions currently supported, and provides initial estimates of the sensitivity of the results to model input parameters and physics assumptions. A suite of new experimental and theoretical developments that are needed in order to improve the reliability and predictability of the fission event generators are also proposed. Finally, a broad summary is provided in Section~\ref{sec:conclusion}.

\section{The nuclear fission process} \label{sec:fission}

This section introduces some of the basic concepts of fission physics that are most relevant to correlation studies.  It begins with a brief description of fission theory and phenomenology.  It then continues with a discussion of relevant fission observables with an emphasis on multiplicities, spectra and correlations.  Finally, it concludes with an introduction to some of the experiments measuring correlations in fission, emphasizing those being carried out by some of the authors of this work.

\subsection{Theoretical insights}

The fission of a heavy nucleus is generally described as a complex collective rearrangement of nuclear matter in which collective and single-particle effects play important roles. The traditional picture of the liquid-drop model, proposed by Bohr and Wheeler in 1939~\cite{Bohr:1939}, provides a relatively simple basis for a qualitative understanding of many features of the fission process.  However, only a more fundamental quantum description can explain certain well-known fission observables, such as fission isomers, as well as provide more quantitative results.

Fission occurs because the repulsive Coulomb force acting between the protons overwhelms the attractive nuclear force responsible for the nuclear binding. In its simplest representation, describing the minimum potential energy of the nucleus as a function of a single deformation parameter ({\it e.g.}, quadrupole deformation) reveals a ``fission barrier" that the nucleus has to overcome to undergo fission. In reality, the situation is much more complicated: the fission path is not restricted to a one-dimensional landscape.  The fission barrier is often double-humped due to shell corrections leading, in particular, to the existence of fission isomers: the nucleus can emit an alpha particle or even heavier clusters during the fission process. In addition, nuclear dynamics along the fission path can strongly alter this much simpler static picture.

In the simplest description of low-energy fission of actinides, the heavy nucleus breaks apart into two smaller fragments of unequal mass at the scission point.  The 
pre-neutron emission heavy fission fragment yield is strongly peaked near $A_H \sim 140$.  The corresponding peak in the light fragment yield is 
$A_L \sim A_0 - A_H$ where $A_0$ is the mass of the fissioning nucleus.  There is a dip in the yield near symmetry, $\sim A_0/2$.  With increasing excitation energy, the asymmetric fragment mass yield distribution grows more and more symmetric, filling in the dip near symmetry. Fission fragments are characterized by a mass $A$, a charge $Z$, an excitation energy $U$, an angular momentum $J$, and a parity $\pi$. For each pair of complementary fragments produced in a fission event, the total excitation energy available to the fragments is a function of the $Q$ value of the fission reaction and of the total energy carried away as kinetic energy. 

To a very good approximation, the two complementary fragments are emitted back-to-back and have opposite momentum vectors in the center-of-mass frame. However, except in the simplest case, spontaneous fission, the angular distribution of the fission fragments in the laboratory frame is not isotropic. Following Bohr's interpretation~\cite{Bohr:1956} of early experimental fission fragment angular distributions observed in photofission reactions, the nucleus populates only a few well-defined fission transition states on top of the outer saddle barrier, defined by their quantum numbers $(J,K,M)$ where $K$ is the projection of the angular momentum $J$ on the fission axis, and $M$ its projection on the beam axis. Most low-energy nuclear reaction codes such as $\mathtt{CoH}$~\cite{Kawano:2010}, used in modern nuclear data evaluations, rely on this representation of the fission barrier and transition states. Angular distributions of the fission fragments strongly depend on the specific sequence and density of these transition states.

The primary fission fragments are significantly excited and quickly emit neutrons and \grays\ to reach either their ground state or a long-lived isomeric state. Eventually, these secondary fission fragments may further $\beta$ decay, leading to fission {\it products}, which can themselves emit delayed neutrons and photons until they reach a stable configuration. This work is only concerned with {\it prompt} emissions. The definition of prompt is somewhat arbitrary since the time associated with the $\beta$-decay is unique to each secondary fission fragment species. In general, however, any particle emitted within a few hundred nanoseconds of fission, the time scale for late-time $\gamma$ transitions, 
would be categorized as prompt. It is important to note that different experiments record prompt events differently because the detectors have different characteristics. Therefore, any comparison between experiment and theory has to take these into account.

Most modern calculations, including the ones presented here, assume that all prompt neutrons are emitted from the fully-accelerated fragments. However, one cannot rule out the possibility that a small fraction of prompt neutrons is also emitted during the descent from the saddle to the scission point, or even dynamically right at scission, similar to the ternary fission process of $\alpha$-particle emission. The search for such ``scission" neutrons has been ongoing for a long time, but still with rather inconclusive results due to the difficulty of finding a unique signature. 

The average multiplicity of prompt neutrons, \nubar, depends on the average total excitation energy available in the fragments. Prompt \grays\ are mostly emitted after the neutron evaporation cascade ceases.
The average characteristics of prompt neutron emission are relatively well known, at least for a few selected and important spontaneous and neutron-induced fission reactions on key actinides.  However, detailed correlated information is still lacking even for well-studied cases. In addition, few predictive capabilities are available aside from the average prompt fission neutron spectrum (PFNS), 
or more generally, the chi-matrix, relating the incident and outgoing neutron energies,
$\langle \chi(E',E)\rangle$. Even in this case, the accuracy of the predictions strongly depends on the availability of experimental data for neighboring fissile isotopes and/or energies.

For a more comprehensive discussion of the physics and data on the PFNS of actinides, see the recent International Atomic Energy Agency Coordinated Research Project, ``Prompt Fission Neutron Spectra of Actinides''~\cite{Capote:2016}.  Note that little to no discussion of correlated fission data can be found in this report, although some discussion of \FREYA, \CGMF\ and other Monte Carlo codes can be found in Ref.~\cite{Capote:2016}.

\subsection{Distributions and correlations in nuclear fission events}
\label{sec:data}

Correlations arise naturally in the nuclear fission process. For instance, the angular distribution of the fission fragments has long been interpreted~\cite{Bohr:1956} as a signature of the presence of fission transition states at the top of the outer barrier. In turn, those transition states play a key role in the calculation of near-barrier fission cross sections. Specific fragmentations are also related to the particular fission channels or modes that the nucleus goes through on its way to scission~\cite{Brosa:1990,Moller:2001}. Those particular fragmentations and the resulting structure and shape of the initial fission fragments strongly influence the number and energy of the prompt neutrons and \grays\ that are subsequently emitted.

This review is limited to the study of correlations among the {\it prompt} fission neutrons and \grays, as well as their characteristics in specific fragmentations. No further discussion will be made of $\beta$-delayed emission. Even this limited scope already constitutes a vast and rich topic.  

\begin{figure}[ht]
\centering
\includegraphics[width=\columnwidth]{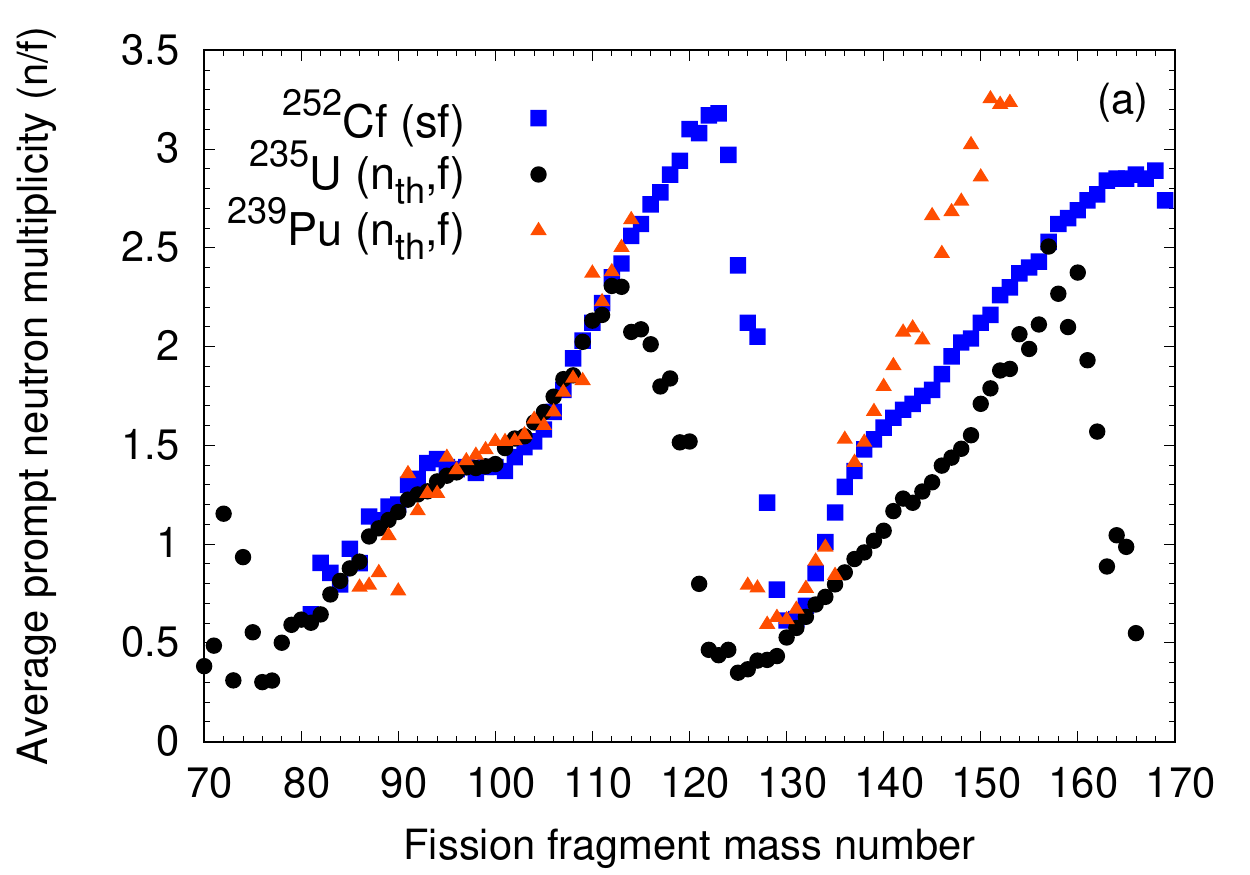} 
\includegraphics[width=\columnwidth]{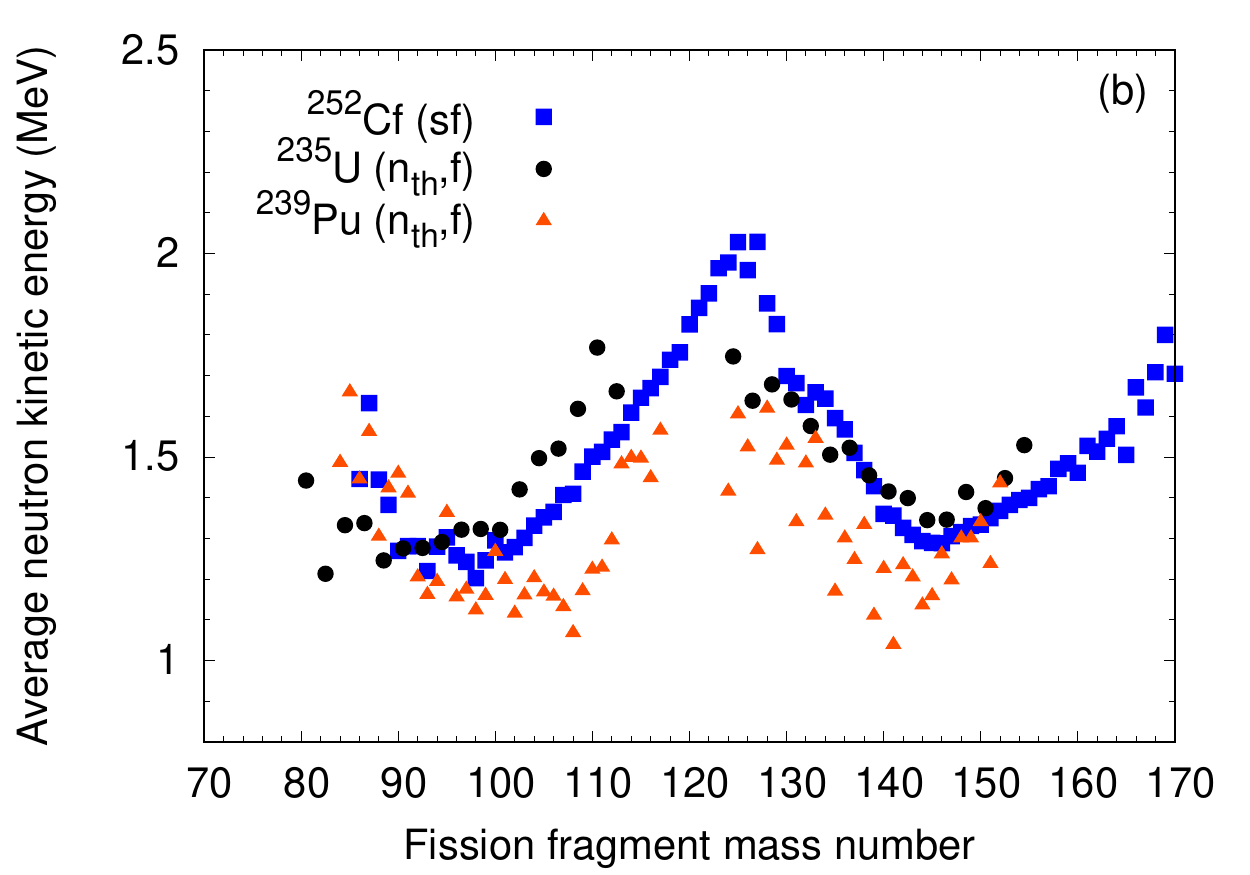}
\caption{\label{fig:Adep}(Color online) (a) Average prompt fission neutron multiplicity data \protect\cite{Dushin:2004,Vorobyev:2010,Nishio:1995} as a function of the fission fragment mass, \nubar$(A)$, for several actinides. (b) Data on the average prompt fission neutron kinetic energy in the center of mass frame as a function of fission fragment mass, $\langle \epsilon_{\rm cm} \rangle (A)$ \protect\cite{Gook:2014,Nishio:1998,Tsuchiya:2000}.}
\end{figure}

\subsubsection{Fragment-dependent characteristics}

The characteristics of prompt neutron and \gray\ emission depends strongly on the parent fission fragments. The measured average multiplicity of emitted neutrons is shown as a function of the pre-neutron emission fission fragment mass number $A$.  It exhibits a well-known ``sawtooth" shape for all actinides 
as shown in Fig.~\ref{fig:Adep} (a). 
The average neutron multiplicity is foremost a reflection of how much excitation energy is present in the fission fragment from which the neutrons are evaporated. The gross structure observed in $\overline{\nu}(A)$ reflects the interplay of the deformation energy and shell structure in the configurations of the fragments near scission. Very compact shapes, as predicted near $A_H \sim 130$, in the region of proton and neutron shell closures, would have little to no deformation and therefore very little extra energy for subsequent neutron emission.  In addition, strong repulsive Coulomb forces will result in high kinetic energies. The complementary light fragments, near mass 122 in the case of $^{252}$Cf(sf), will be strongly deformed further from shell closures.  The difference between the light fragment shapes near scission and their ground-state configurations will result in higher excitation energies and thus higher neutron multiplicities at $A_L \sim 122$.
 
\begin{figure}[ht]
\centering
\includegraphics[width=\columnwidth]{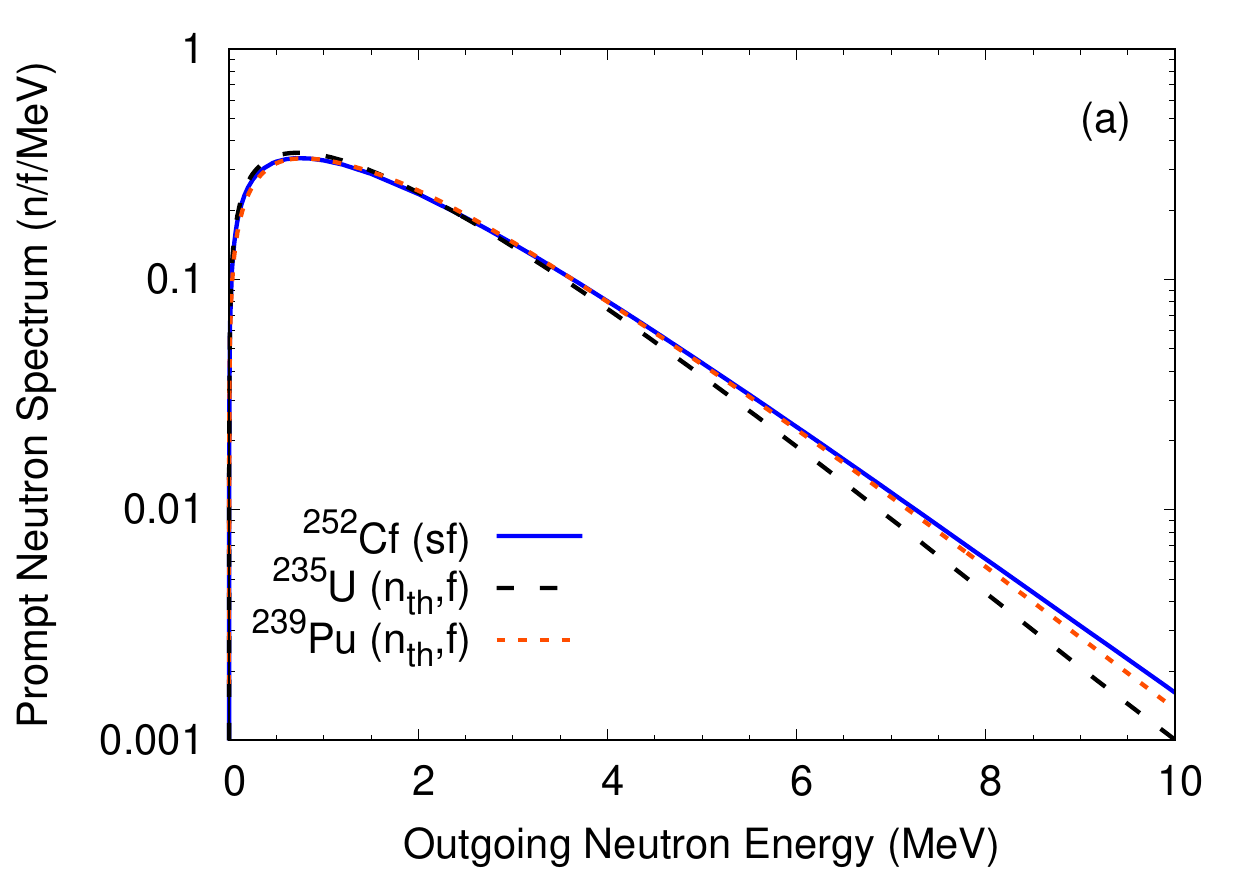} 
\includegraphics[width=\columnwidth]{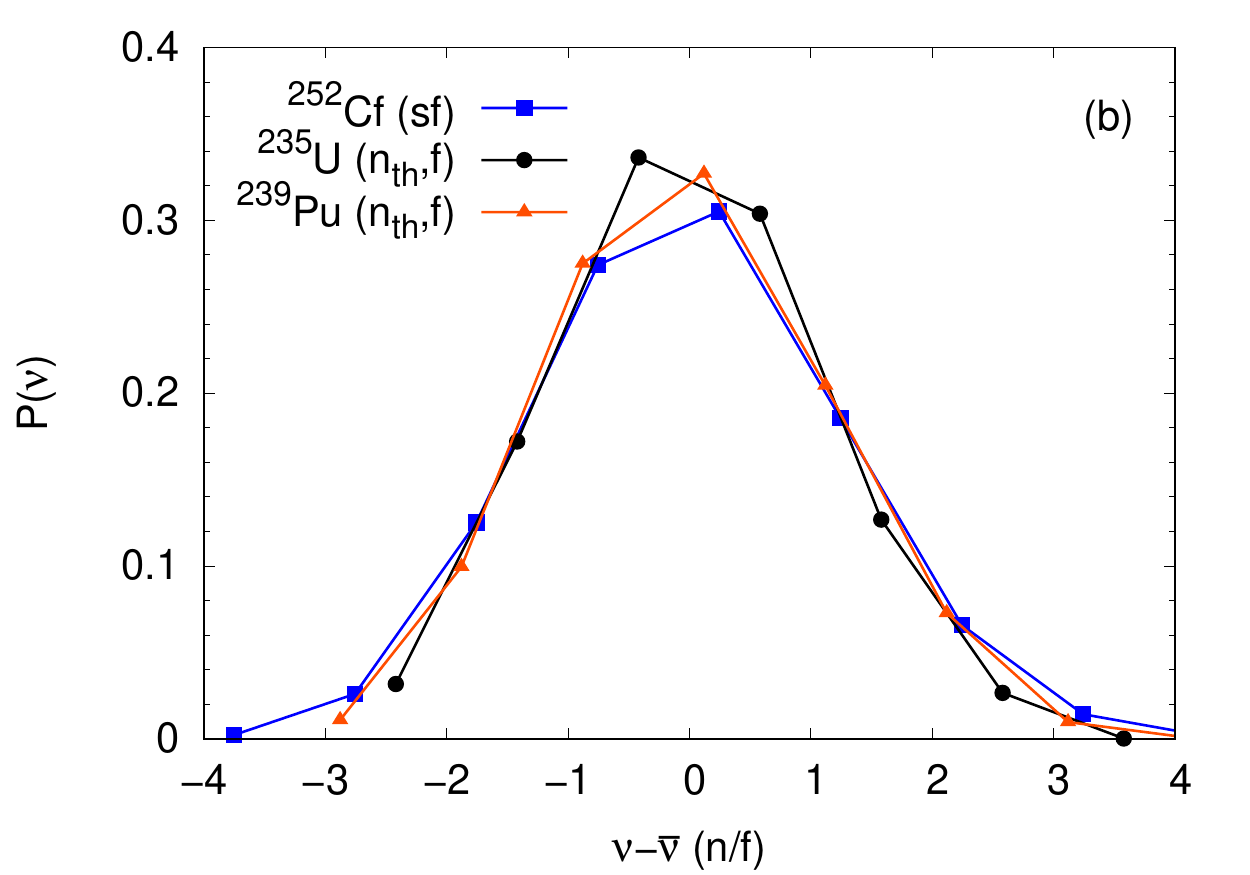}
\caption{\label{fig:Evals}(Color online) (a) Prompt fission neutron spectrum for several actinides, taken from the ENDF/B-VII.1 library \protect\cite{ENDFB71}. (b) Neutron multiplicity distributions for several actinides, both measured and evaluated.  
Each distribution is centered around the average neutron multiplicity. 
The data are from Santi and Miller~\cite{Santi:2008} for $^{252}$Cf(sf), and Holden and Zucker~\cite{Holden:1988} for $^{235}$U and $^{239}$Pu ($n_{\rm th}$,f).}
\end{figure}

The average kinetic energy of the neutrons emitted from each fragment depends on the nuclear structure and on the nuclear level density of the daughter fragment. Figure~\ref{fig:Adep}(b) shows the average prompt fission neutron kinetic energy in the center-of-mass of the fragments as a function of fragment mass. Note that since the neutron kinetic energy is measured in the lab frame, some modeling is required to obtain the result shown in Fig.~\ref{fig:Adep}(b).  
The average neutron kinetic energy, integrated over all fragments, is the first moment of the average prompt fission neutron spectrum. Thus, the hardness of the neutron spectrum depends significantly on the specific fragments that emitted the neutrons.

For most nuclear applications, such details do not matter and only average quantities are relevant. In particular, the average prompt fission neutron spectrum and neutron multiplicity, $\overline \nu$
are two observables that can be evaluated using simplified models~\cite{Madland:1982} which do not require a detailed description of the sequential neutron evaporation process. However, observables such as the PFNS and neutron multiplicity distribution, $P(\nu)$, place important constraints on models that attempt to correctly describe neutron emission. Such constraints are particularly important when using those models in cases where experimental data are missing.  The PFNS and $P(\nu)$, shown as a function of $\nu - \overline \nu$ to give the distributions a common center, are presented in Fig.~\ref{fig:Evals} for the same isotopes as in Fig.~\ref{fig:Adep}. 

\begin{figure}[ht]
\centering
\includegraphics[width=\columnwidth]{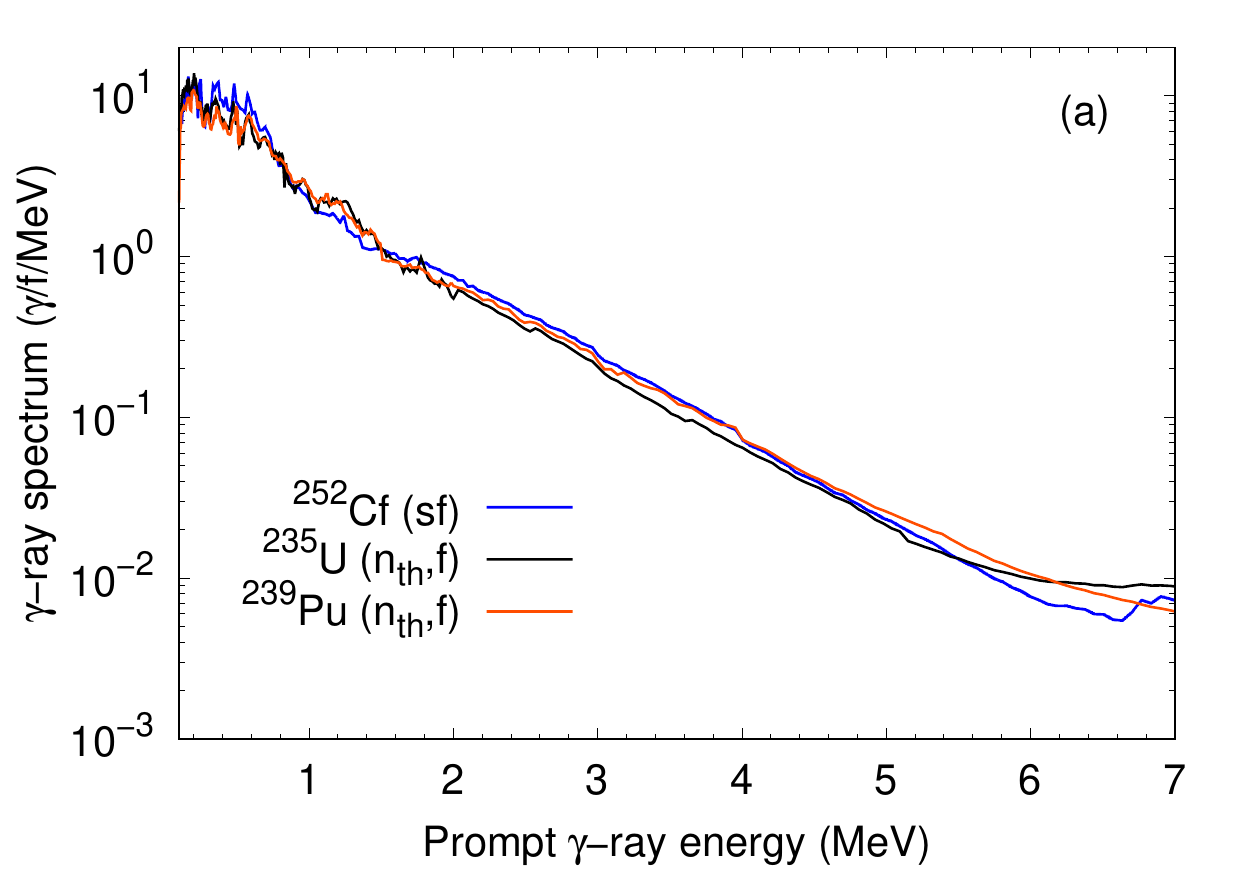}
\includegraphics[width=\columnwidth]{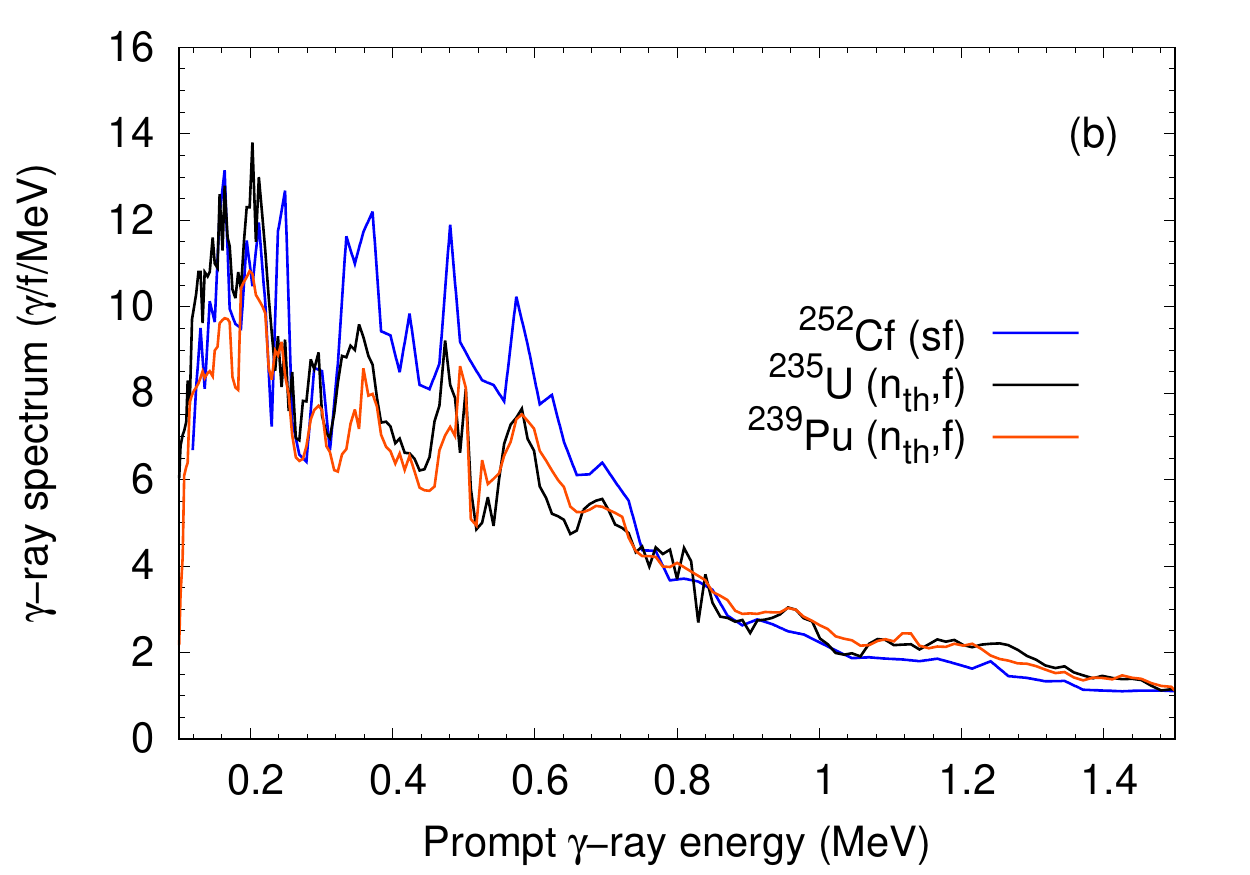}
\caption{\label{fig:pfgs}(Color online) Average prompt fission \gray~spectra for $^{252}$Cf(sf) and thermal neutron-induced fission of $^{239}$Pu and $^{235}$U, as measured by Billnert \etal\cite{Billnert:2013}, Gatera \etal~\cite{Gatera:2017} and Oberstedt \etal\cite{Oberstedt:2013}, respectively.}
\end{figure}

Prompt fission \grays~also depend on the parent fragment. The average prompt fission \gray~spectrum (PFGS) is dominated by statistical \grays~with outgoing energies greater than 1~MeV.  Significant structure appears mostly below 1~MeV~\cite{Verbinski:1973,Billnert:2013}, reflecting specific \g~transitions between low-lying excited states, as seen in Fig.~\ref{fig:pfgs}.  Experimental measurements of the average PFGS for $^{252}$Cf(sf), $^{235}$U($n_{\rm th}$,f), and
$^{239}$Pu($n_{\rm th}$,f) reactions are shown. The presence and intensity of each low-lying \g~line depends mostly on the fission fragment yields resulting from these reactions.

\begin{figure}[ht]
\centering
\includegraphics[width=\columnwidth]{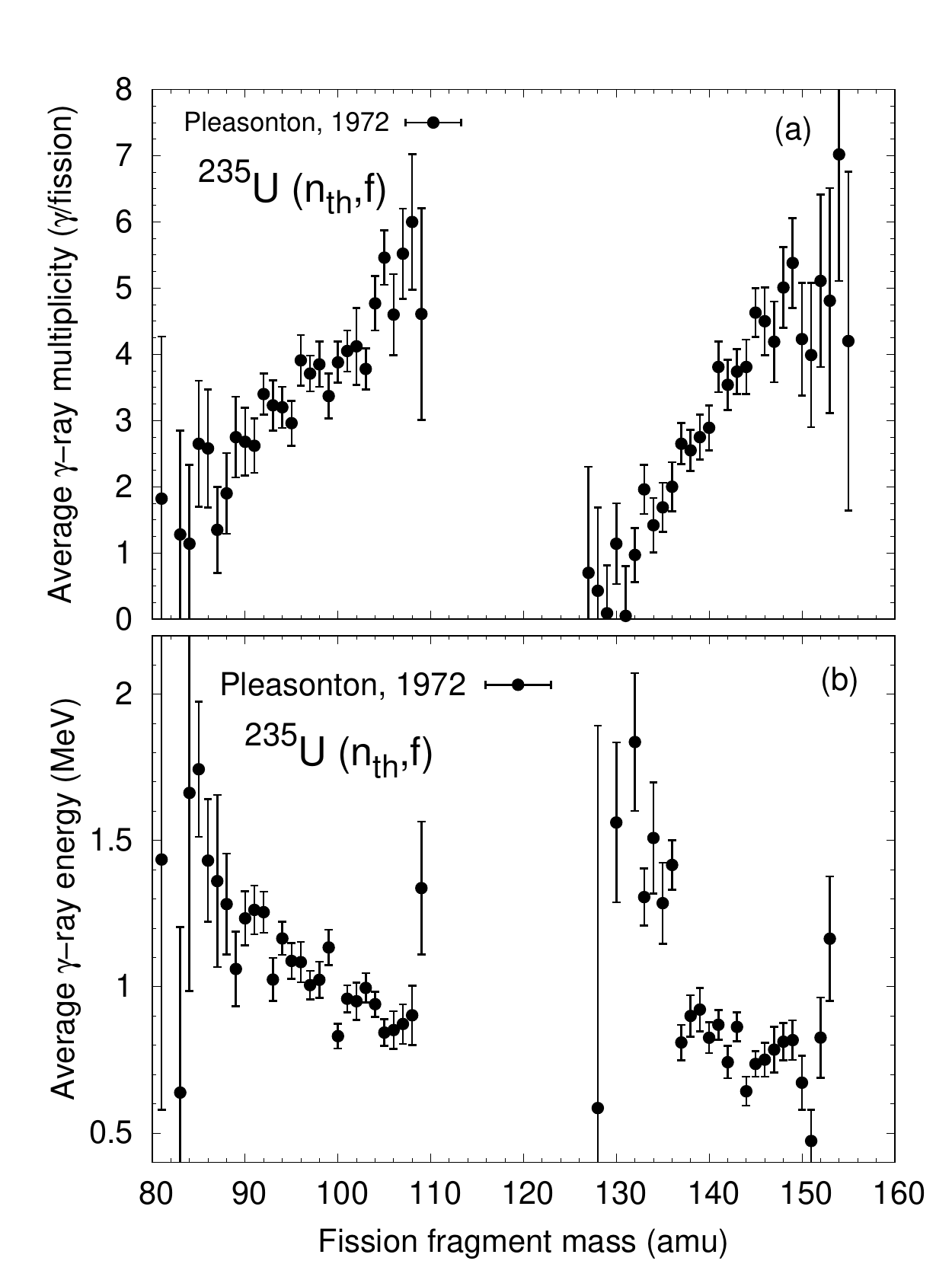}
\caption{\label{fig:GammaA}(Color online) Average prompt fission \gray~multiplicity 
(a) and average energy per emitted \gray\ 
(b) as a function of the fission fragment mass. The experimental data are from Pleasonton \etal~\cite{Pleasonton:1972}.}
\end{figure}

Figure~\ref{fig:GammaA} shows the average \gray~multiplicity, $\overline N_\gamma$, 
Fig.~\ref{fig:GammaA}(a), and energy per $\gamma$, $\langle \epsilon_\gamma \rangle$, 
Fig.~\ref{fig:GammaA}(b), as a function of the fission fragment mass. A similar sawtooth behavior 
can be seen for $\gamma$-ray multiplicity than for the prompt neutron multiplicity (compare Fig.~\ref{fig:Adep}(a) and 
Fig.~\ref{fig:GammaA}(a)). A lower nuclear level density for fragments produced near the double shell closure at $A_H = 132$ can explain the clear increase of $\langle \epsilon_\gamma\rangle(A)$ in this mass region, different from the dependence of the average neutron kinetic energy in Fig.~\ref{fig:Adep}(b) which shows no clear structure for this mass.

\subsubsection{Multiplicity distributions}

\begin{figure}[ht]
\centering
\includegraphics[width=\columnwidth]{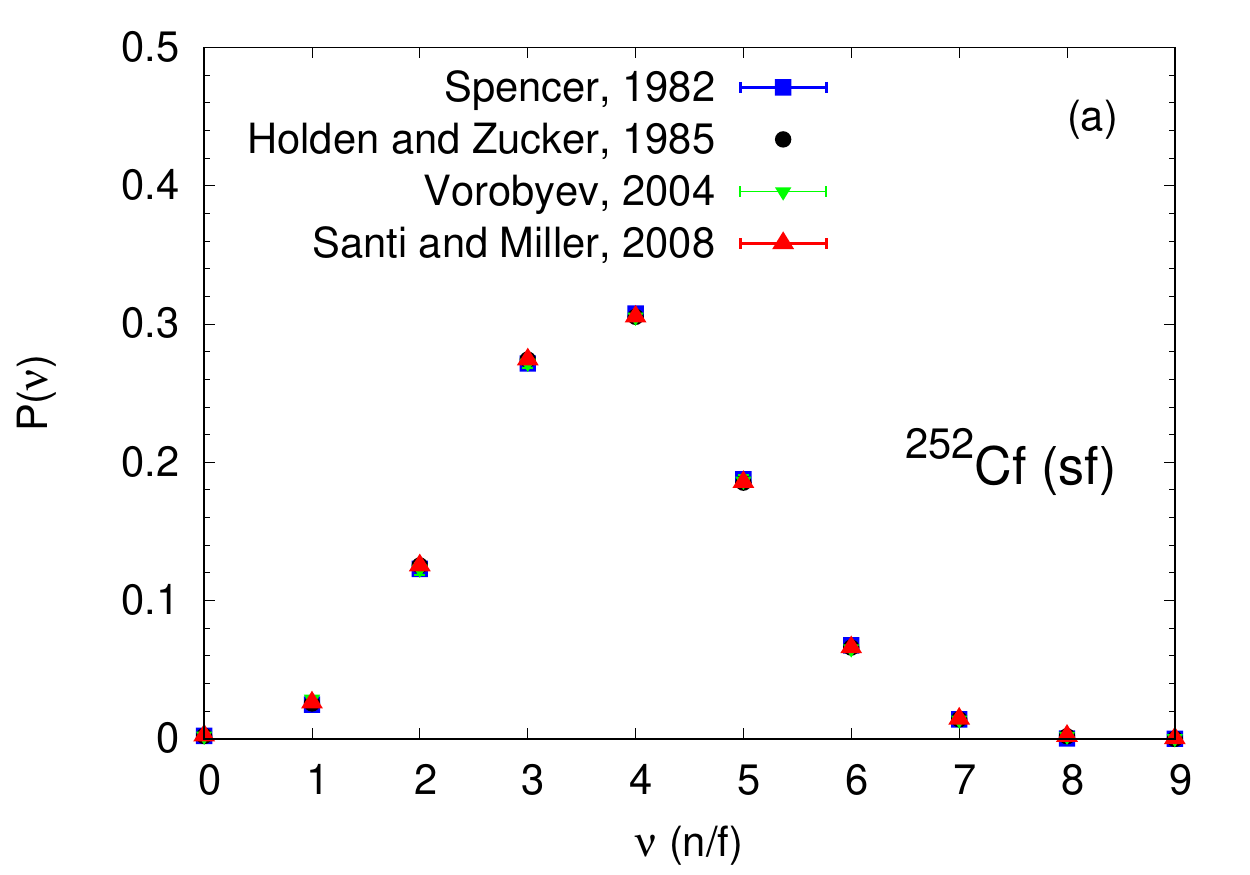}
\includegraphics[width=\columnwidth]{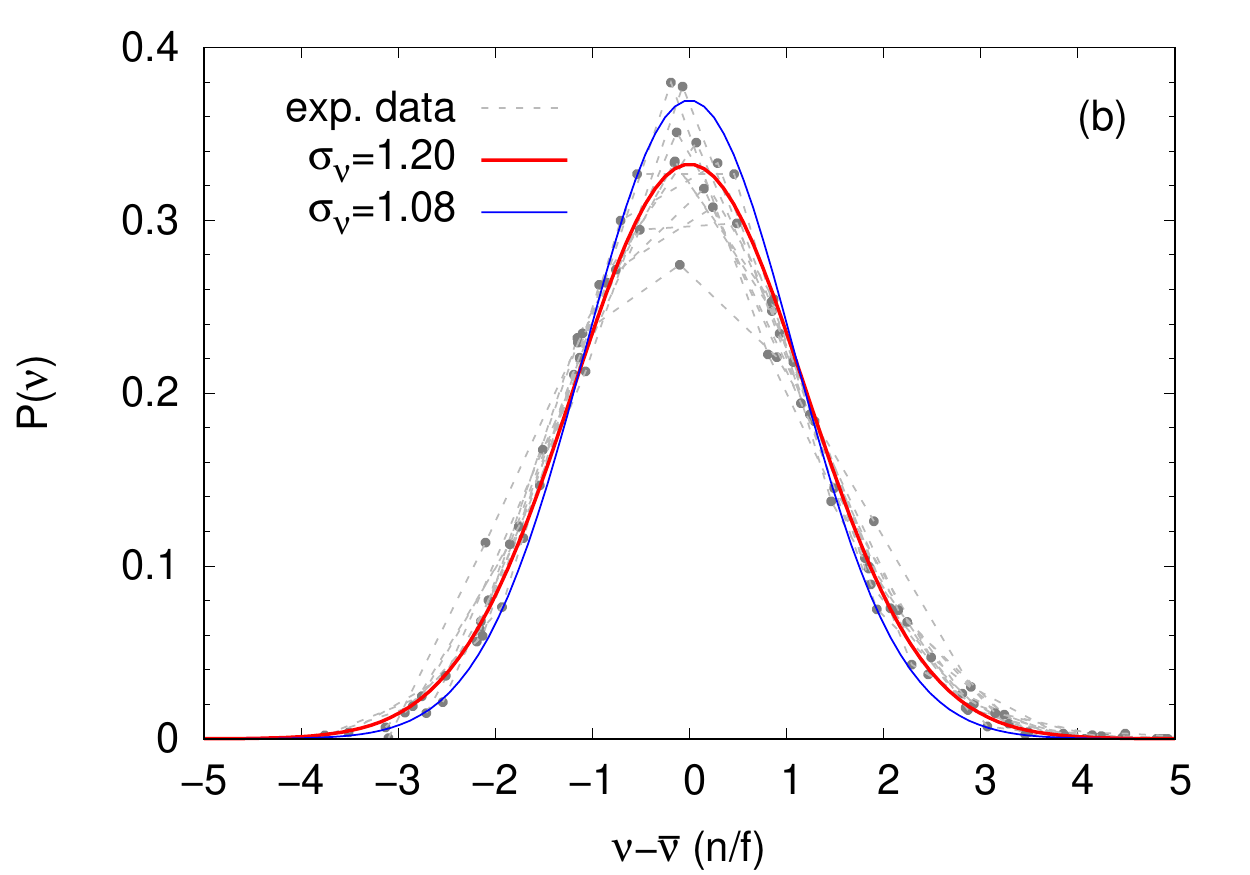}
\caption{\label{fig:Pnu} (Color online) Experimental~\cite{Santi:2008,Spencer:1982,Holden:1985,Vorobyev:2004} and evaluated prompt neutron multiplicity distribution 
(a) in the case of $^{252}$Cf(sf), and experimental neutron multiplicity distributions 
  (b), expressed as a function of $\nu - \overline \nu$ 
for spontaneous fission of $^{236,238,240,242}$Pu, $^{242,244,246,248}$Cm, and $^{246,250,252,254}$Cf (see 
Refs.~\cite{Santi:2008,Verbeke:2010} and references therein).  (Note that $P(\nu)$ supposes integer values of $\nu$.) 
}
\end{figure}

The average prompt neutron multiplicity, \nubar, is a very important quantity for the accurate simulation of many nuclear applications. Evaluated nuclear data libraries rely almost exclusively on experimental data for this quantity. The so-called ``standard" evaluations~\cite{Carlson:2009} rely entirely on experimental data and provide evaluated $\overline \nu$ for selected isotopes with very high accuracy. For instance, the current evaluated uncertainty on \nubar~for $^{252}$Cf(sf) is 0.13\%!  The comparison between a recent evaluation of $P(\nu)$ for $^{252}$Cf(sf) by Santi and Miller \cite{Santi:2008}, the earlier evaluation by Holden and Zucker \cite{Holden:1985} and data \cite{Spencer:1982,Vorobyev:2004} is shown in Fig.~\ref{fig:Pnu}(a). The evaluated distributions for spontaneous fission of many Pu, Cm and Cf isotopes~\cite{Santi:2008} are shown in Fig.~\ref{fig:Pnu}(b) as a function of $\nu - \overline{\nu}$ to facilitate comparison. To a good approximation, those distributions can be represented by a Gaussian of width $\sigma_\nu =1.20$, the red line in Fig.~\ref{fig:Pnu}(b).

Very little is known about the incident energy dependence of $P(\nu)$ for fast neutron-induced fission reactions.  Only one such measurement by Soleilhac \etal\ \cite{Soleilhac:1969} is known. Based on the observation that all measured neutron multiplicity distributions for spontaneous and thermal neutron-induced fission reactions are reasonably Gaussian-like, Terrell inferred a formula for $P(\nu)$~\cite{Terrell:1957}
\begin{eqnarray} \label{eq:Terrell}
\sum_{n=0}^{\nu}{P(n)} = \frac{1}{\sqrt{2\pi}}\int_{-\infty}^{(\nu-\overline{\nu}+1/2+b)/\sigma_\nu}{{\rm exp}(-t^2/2)dt},
\end{eqnarray}
where $t = (E_{\rm inc} - \overline E)/(\sigma_\nu E_0)$, $\overline E$ is the average excitation energy, and $E_0$ is the change in $\overline \nu$ with $E_{\rm inc}$.  Thus Eq.~(\ref{eq:Terrell}) is often used to compute $P(\nu$) as a function of $E_{\rm inc}$.  The parameter $b$ for each value of $E_{\rm inc}$ is determined from the condition that
\begin{eqnarray}
\sum_\nu{\nu P(\nu;E_{\rm inc})} = \overline{\nu}(E_{\rm inc}) \, .
\end{eqnarray}
The value of $b$ was found to be small in all cases.  Terrell used~\cite{Terrell:1957} a Gaussian of width $\sigma_\nu=1.08$ in his analysis (the blue line in Fig.~\ref{fig:Pnu}(b)).

The factorial moments of $P(\nu)$ are defined as
\begin{eqnarray}
  \nu_n = \sum_\nu \frac{\nu!}{(\nu -n)!}\, P(\nu) \, .
\end{eqnarray}
The first three moments are then given by
\begin{eqnarray}
  \nu_1 & = & \overline \nu = \langle \nu \rangle \, , \label{eq:nu1} \\
  \nu_2 & = & \langle \nu (\nu - 1) \rangle \, , \label{eq:nu2}\\
  \nu_3 & = & \langle \nu (\nu - 1) (\nu - 2) \rangle \, . \label{eq:nu3}
\end{eqnarray}
These moments must be known very precisely for applications involving neutron multiplicity counting
and they are very well known for $^{252}$Cf(sf).  
Unfortunately very little is known about the incident-energy dependence of the factorial multiplicity moments
for fast neutron-induced fission reactions.

\begin{figure}[ht]
\centering
\includegraphics[width=\columnwidth]{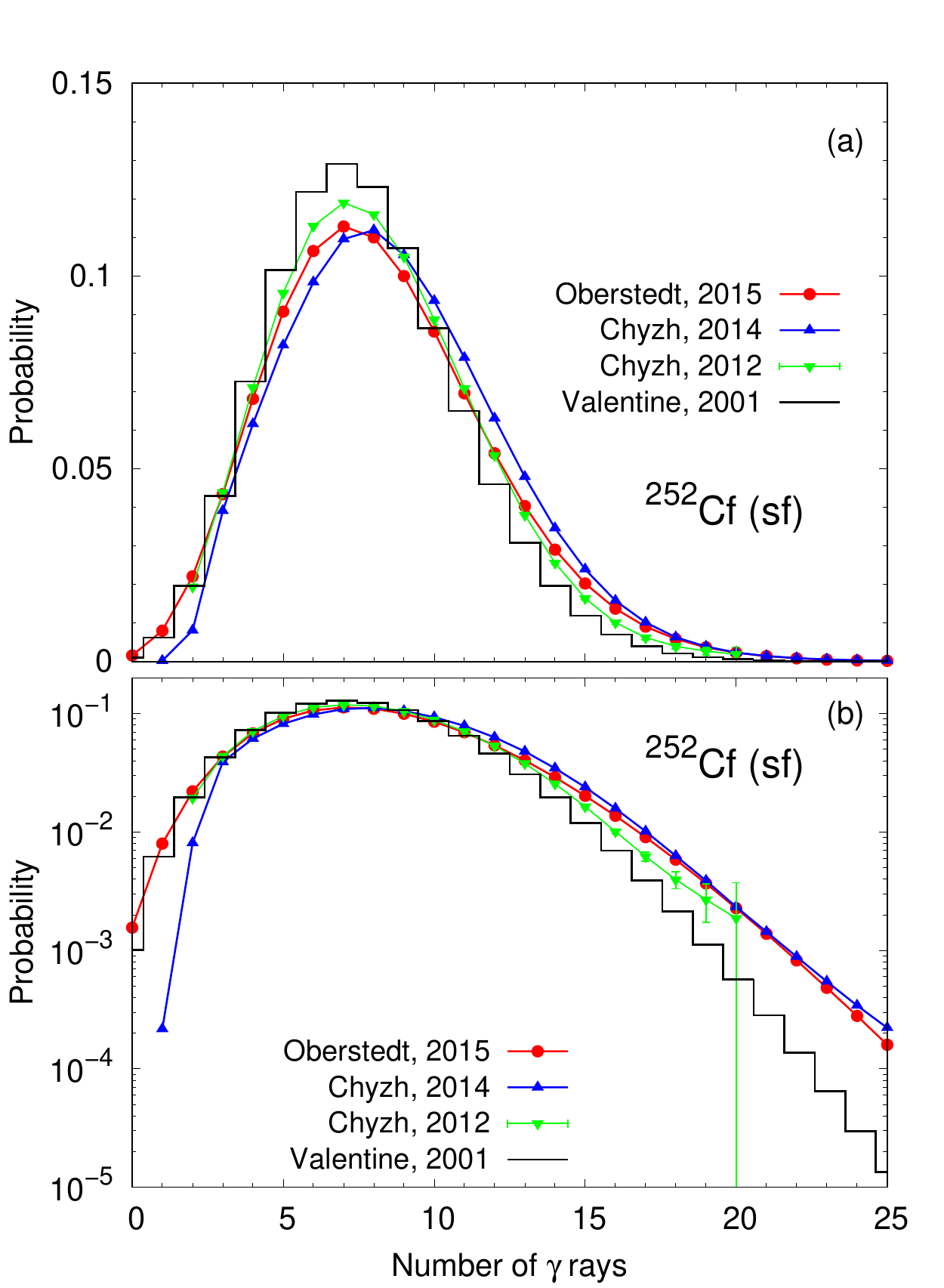}
\caption{(Color online) Prompt fission \gray~multiplicity distribution for $^{252}$Cf (sf), as represented by a negative binomial distribution in Valentine~\protect\cite{Valentine:2001}, and as measured recently by Oberstedt \etal \protect\cite{Oberstedt:2015b} and Chyzh \etal \protect\cite{Chyzh:2014}.  The distributions are shown on a linear scale (a) and a log scale (b) to highlight the mean and tail of the distributions respectively. 
The Oberstedt data were obtained with a 6~ns time coincidence window and 100~keV \g~energy threshold \protect\cite{Oberstedt:2015b}. The 2014 Chyzh data, obtained with the DANCE detector, employed a 10~ns time coincidence window and a 150~keV \g-energy threshold \protect\cite{Chyzh:2014}.  The 2012 Chyzh data~\protect\cite{Chyzh:2012} is also shown here.
}
\label{fig:Cf252sf-PnuGamma}
\end{figure}

The prompt \gray~multiplicity distribution, $P(N_\gamma)$, can also be used in non-destructive assay methods that rely on correlated \gray~fission data. A negative binomial distribution was shown~\cite{Valentine:2001} to agree fairly well with experimental data. However, as can be seen in Fig.~\ref{fig:Cf252sf-PnuGamma}, even for $^{252}$Cf(sf), recent measurements~\cite{Oberstedt:2015b,Chyzh:2014} disagree significantly with past results \cite{Valentine:2001}. Note that the ``experimental''  data by Oberstedt~\cite{Oberstedt:2015} reported here corresponds to the result of a fit using a negative binomial distribution, and cannot be considered raw experimental data.  The comparison of experimental data with model calculations of $P(N_\gamma)$ is complicated by the use of a specific \gray~detector energy threshold below which no \grays~are measured, as well as by the time coincidence window between the emitted \grays~and the fission trigger and have to be considered when comparing to model calculations. Both quantities have a significant impact on the reported experimental distributions since they are in part responsible for the differences between the Oberstedt and Czyzh data in Fig.~\ref{fig:Cf252sf-PnuGamma}.  Other differences in unfolding techniques can likely account for the rest. 

\subsubsection{Neutron-neutron and neutron-\g~correlations}

Strong correlations are expected in prompt neutron and \gray~emission, due in part to the kinematic boost imparted to neutrons emitted from the same or complementary fragments. Two neutrons emitted from the same fragments will be focused in the same direction, 0 degrees, while two neutrons emitted from complementary fragments will be emitted with an angular separation of close to 180 degrees. The initial conditions of the fragments dictate both neutron and \gray~emission, thereby inducing natural correlations.

\begin{figure}[ht]
\centerline{\includegraphics[width=\columnwidth]{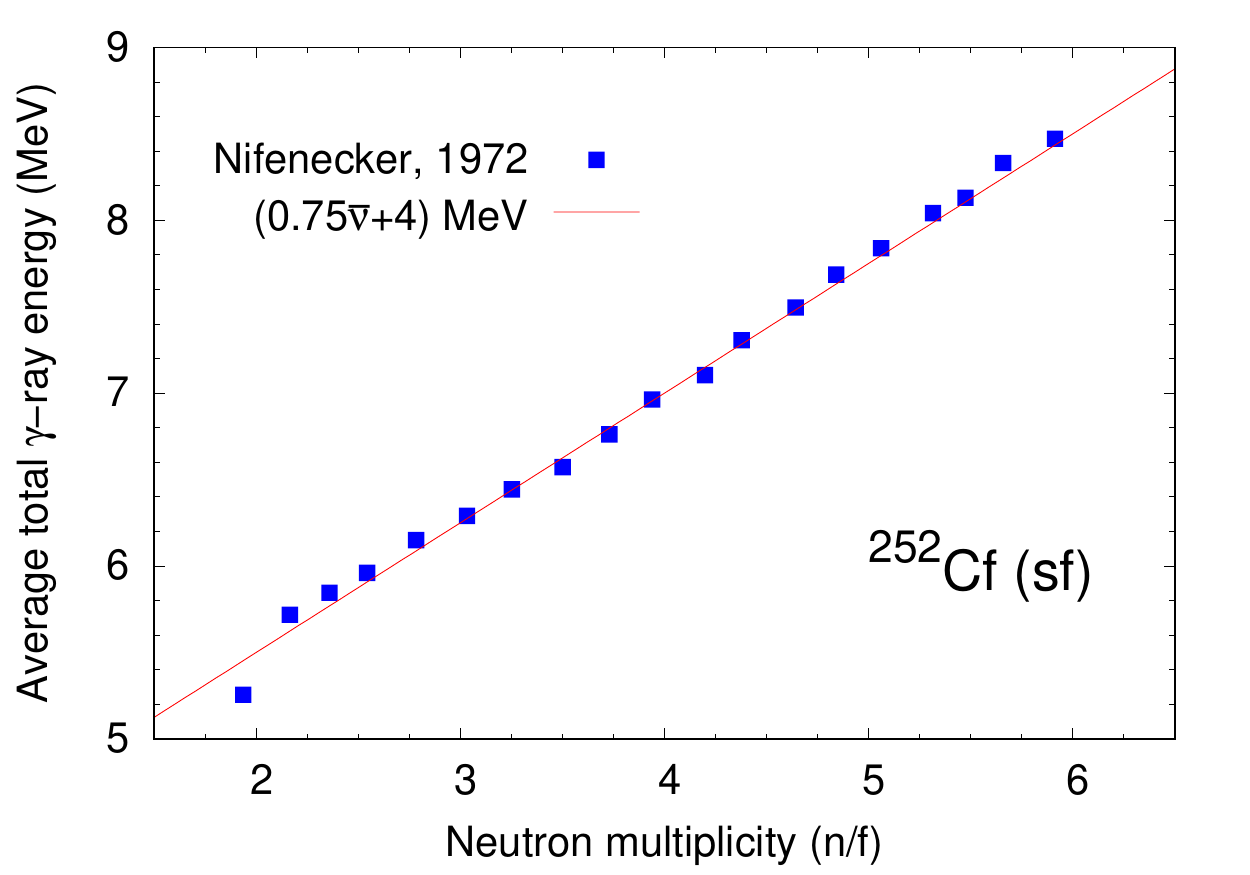}}
\caption{\label{fig:Nifenecker}(Color online) The average total prompt \gray~energy is plotted as a function of the prompt neutron multiplicity measured by Nifenecker \etal \protect\cite{Nifenecker:1972}. 
Experimental points were digitized from Fig. 7 in Ref.~\cite{Nifenecker:1972}.
}
\end{figure}

Experimental data on $^{252}$Cf(sf) from Nifenecker \etal~\cite{Nifenecker:1972}  seem to indicate a positive correlation between the total \gray~energy released and the number of neutrons emitted, as shown in Fig.~\ref{fig:Nifenecker}. Nifenecker inferred the following relation between the neutron multiplicity and \gray~energy for a given fragment:
\begin{eqnarray}\label{eq:Nifenecker}
  \overline{E}_\gamma(A,{\rm KE}) = (0.75 \overline{\nu}(A,{\rm KE}) + 2) \,
           {\rm MeV},
\end{eqnarray}
where $A$ and KE represent the mass and kinetic energy of the fission fragments respectively.  The line in Fig.~\ref{fig:Nifenecker} is the total \gray~energy from a pair of complementary fragments, $\overline E_\gamma^{\rm tot} = (0.75 \overline \nu + 4)$~MeV.

\begin{figure}[ht]
\centerline{\includegraphics[width=\columnwidth]{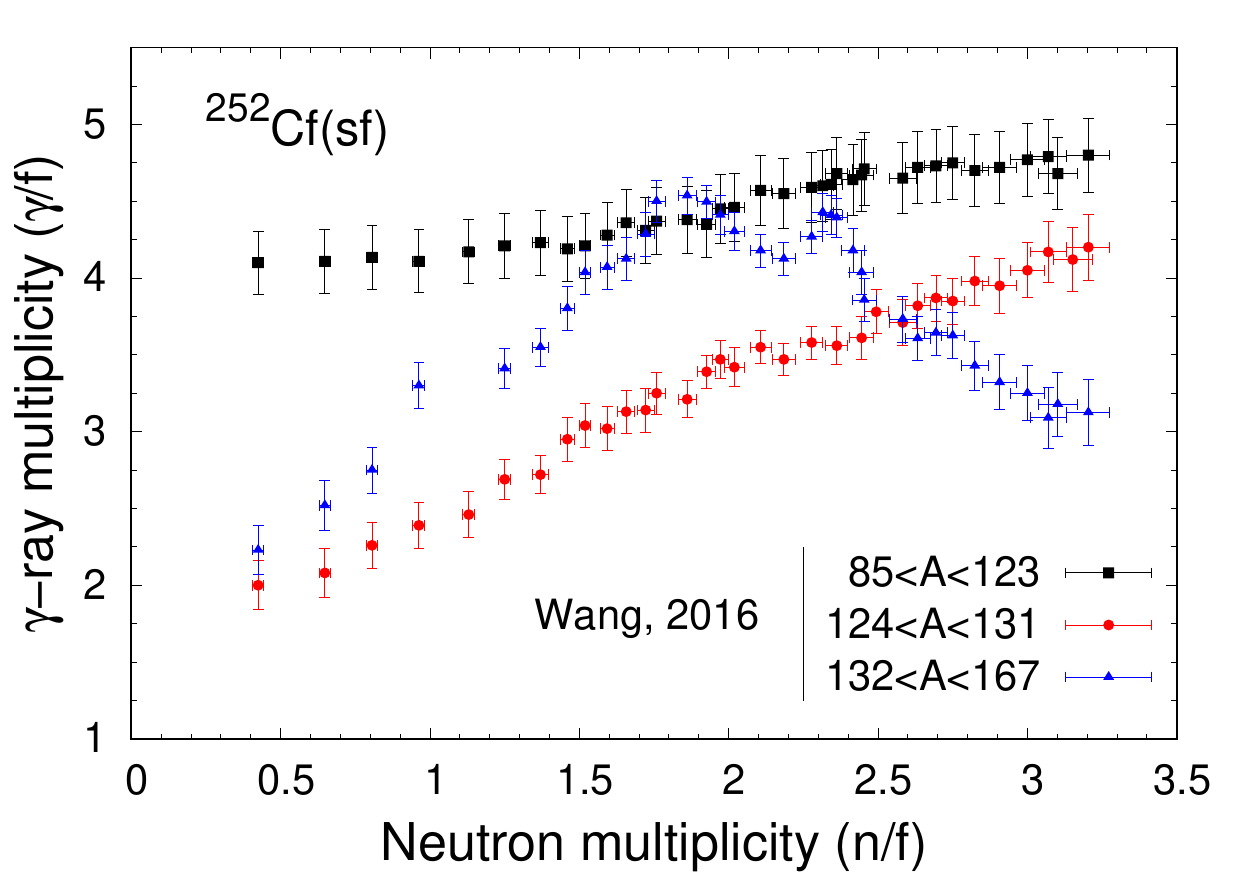}}
\caption{\label{fig:nu-nug}(Color online) The \g~multiplicity is shown as a function of the neutron multiplicity for different fission fragment mass ranges, as measured by Wang \etal \protect\cite{Wang:2016} in $^{252}$Cf(sf).}
\end{figure}

Recently, Wang \etal \cite{Wang:2016} measured correlations between the neutron and \gray~multiplicities, as a function of the mass and total kinetic energy of the fragments, again in $^{252}$Cf(sf). Figure~\ref{fig:nu-nug} shows the strong and complex correlations observed for different fission fragment mass regions, indicating a potentially much more complicated situation than suggested by Eq.~(\ref{eq:Nifenecker}). 

For neutron-induced fission reactions, the ($n,\gamma$f) process, first predicted theoretically in Refs.~\cite{Lynn:1965,Stavinsky:1965}, has been used to interpret variations of the average neutron multiplicity in the $^{235}$U\nf\ and $^{239}$Pu\nf\ reactions below $\sim 100$~eV. This process leads to anti-correlations between \nubar~and \nubarg, since pre-fission $\gamma$ rays increase \nubarg~at the expense of the residual excitation energy available in the fragments for the emission of prompt neutrons. Limited data exist, as reported by Shcherbakov~\cite{Shcherbakov:1990} (and references therein). An alternative explanation for the fluctuations of \nubar~in the presence of resonances has been explored by Hambsch \etal~\cite{Hambsch:1989,Hambsch:2015} as changes in the fission fragment yields in mass and kinetic energy which would influence the number of prompt neutrons emitted. An increase in kinetic energy would result in a smaller number of prompt neutrons but would not impact \gray~emission, making this correlated measurement even more relevant.

\subsubsection{Angular distributions}

\begin{figure}[ht]
\centerline{\includegraphics[width=\columnwidth]{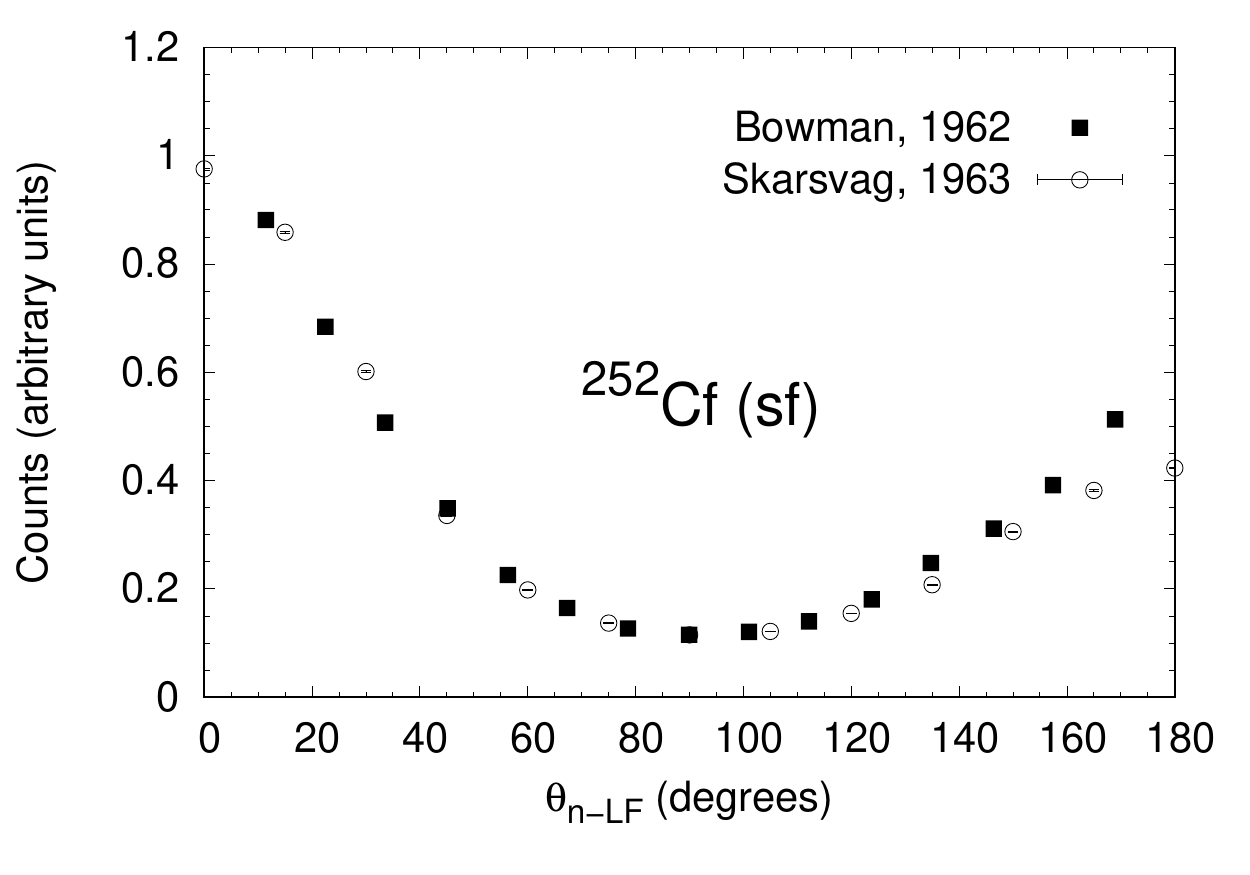}}
\caption{\label{fig:nLF-Cf252sf} (Color online) Measured~\cite{Bowman:1962,Skarsvag:1963} angular distribution of prompt fission neutrons with respect to the light fragment direction, in the case of $^{252}$Cf(sf).  The minimum neutron kinetic energy was 0.5~MeV.}
\end{figure}

Angular correlations between the fission fragments and the emitted neutrons emerge naturally from the kinematics of the reaction. Assuming that neutrons are emitted from fully accelerated fragments, the kinematic boost of the fragments from the center-of-mass to the laboratory frame focuses the neutrons in the direction of the fragments. Therefore, it is expected, and observed, that neutrons are emitted preferentially near 0 and 180 degrees relative to the direction of the light fragment. Figure~\ref{fig:nLF-Cf252sf} illustrates this point in the case of $^{252}$Cf(sf) where experimental data from Bowman~\cite{Bowman:1962} and Skarsvag~\cite{Skarsvag:1963} show increased emissions at 0 and 180 degrees. In addition, the higher peak near 0 degrees indicates that more neutrons are emitted from the light fragment than from the heavy fragment in this particular reaction.

Another potential source of neutron emission anisotropy is due to the rotation
of the fragments. The average spin of the initial fragments is often estimated
to be $\sim 7\hbar -9 \hbar$, and neutrons emitted from the fragments would tend
to align in a plan perpendicular to the direction of the spin.  However, this
effect is small compared to the kinematic focusing just discussed
\cite{Randrup:2014}.

Finally, neutrons emitted during the descent from saddle to scission, at or
near the neck rupture (scission), or during the acceleration of the fragments
could all contribute to an increased anisotropy. It is generally believed that
more than 95\% of the prompt neutrons are emitted from the fully-accelerated
fragments.  Therefore this effect will also be small compared to the kinematic
focusing. However, those contributions should not be neglected when trying to
infer the contribution of scission neutrons by comparing the measured angular
distributions to calculations.

\begin{figure}[ht]
\centerline{\includegraphics[width=\columnwidth]{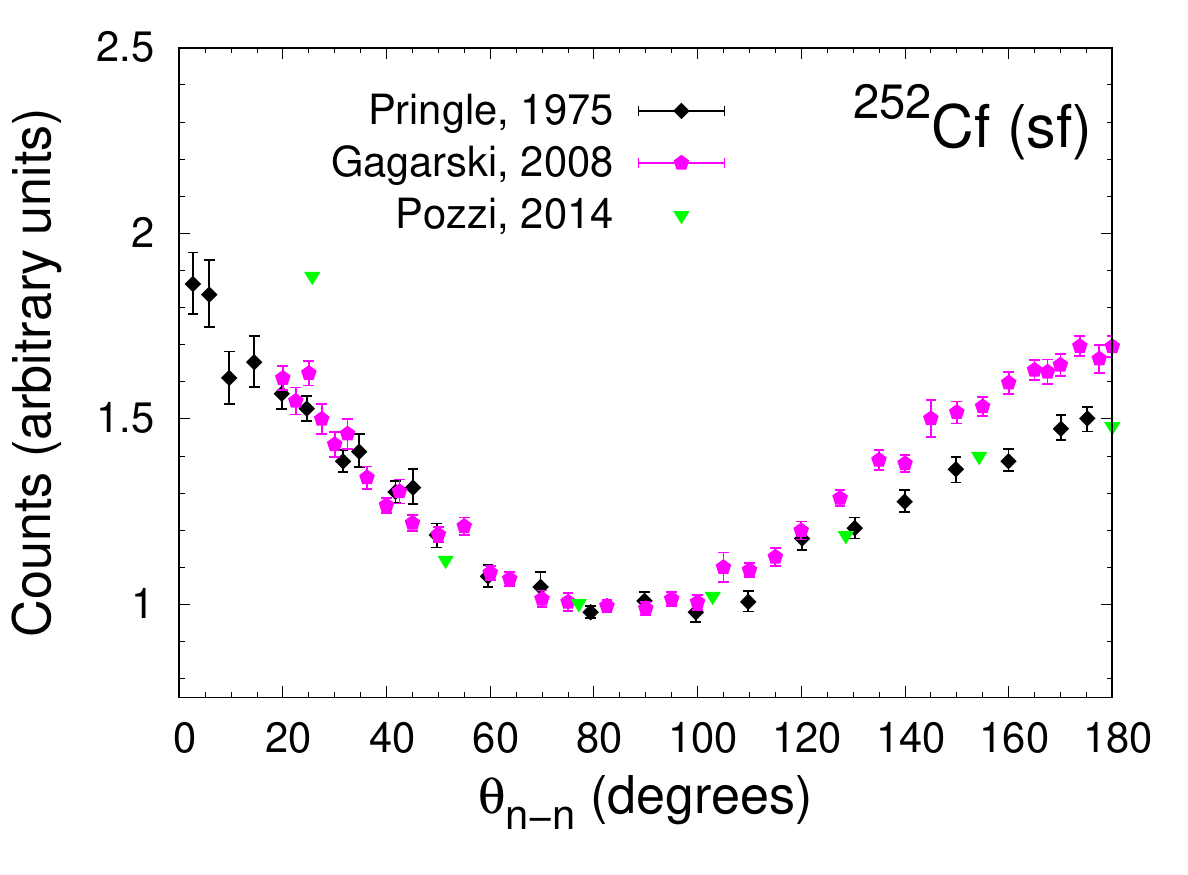}}
\caption{\label{fig:nn-Cf252sf}(Color online) Measured~\cite{Pringle:1975,Gagarski:2008,Pozzi:2014} neutron-neutron separation angle in the case of $^{252}$Cf(sf).  The neutron detection threshold for Gagarski~\cite{Gagarski:2008} and Pozzi~\cite{Pozzi:2014} data is 0.425~MeV, and 0.7~MeV for Pringle~\cite{Pringle:1975}.}
\end{figure}

In the absence of a detector that can track the direction of the fission fragments, $n$-$n$ angular correlations can also represent a signature of the fission process.  Such correlations will not exist in other neutron-induced reactions, such as $(n,2n)$, where the two neutrons would mostly be emitted isotropically. In fission, because of the kinematic focusing discussed above, the neutrons will follow the direction of the fission axis. If the two neutrons are emitted from the same fragment, then their aperture will be very small, $\sim 0$ degrees. On the other hand, if the two neutrons originate from each complementary fragment, then their aperture will be close to 180 degrees. Here again, the distribution of $\theta_{n-n}$ can be expected to peak near 0 and 180 degrees. This can be seen in Fig.~\ref{fig:nn-Cf252sf} for $^{252}$Cf(sf).

A well-known feature of the fission process is the marked anisotropy of the fission fragment angular distribution in the laboratory frame. First observed in the photofission of thorium~\cite{Winhold:1952}, this discovery was quickly interpreted by A. Bohr~\cite{Bohr:1956} as the presence of discrete fission transition states on top of the fission barriers. Most recently, the anisotropy coefficients have been measured for different fission reactions and actinides at CERN~\cite{Leong:2013}, PNPI~\cite{Vorobyev:2015} and LANL~\cite{Kleinrath:2015}. Prompt neutron polarization asymmetries were also measured recently at the HI$\gamma$S facility in the photofission of Th, U, Np, and Pu isotopes~\cite{Mueller:2012}. As expected, the polarizations are strongly impacted by the fission fragment angular distributions. Therefore, for fission reactions other than spontaneous fission or very low-energy fission reactions, the interpretation of any observed $n$-$n$ angular correlations should always be done by folding the prompt neutron anisotropic emissions with the appropriate fission fragment angular distribution.

There is evidence~\cite{Skarsvag:1980} that prompt \grays~also exhibit anisotropic emission from rotating fission fragments. Such data can provide some information about the angular momentum vectors in the fragments and the multipolarity of \g~emission.  It can also help identify rotational and vibrational levels from stretched \grays~for specific fission fragments.

\subsubsection{Time correlations}

Time correlations, in our case, are understood to be correlations between the arrival times of prompt neutrons in a fission chain, {\it i.e.}, neutrons from different fission events, at a detector. Hence, this type of correlations is not intrinsic to a particular fission event but rather a property of multiplying fission objects. Because of its importance in nuclear assay applications, this topic is discussed at greater length in Sec.~\ref{sec:timeCorrelations}.

\subsection{Measuring correlations in fission observables}
\label{sec:measurements}

Experimental studies of the nuclear fission process have been rich and numerous since its discovery. Fission cross sections, average prompt fission neutron spectra, multiplicities and, to some extent, fission yields have been the focus of most efforts, typically driven by applications in nuclear energy and defense programs. However, experimental data on correlations between prompt emission and parent fission fragments are rather limited, and do not provide sufficient constraints on the input parameters of modern fission models and codes such as the ones presented here. In this section, 
our own efforts 
to measure these correlations are discussed.  Other recent and future experiments that can nicely complement these efforts are also mentioned. 


\begin{figure}[ht]
\centering
\includegraphics[width=0.95\columnwidth]{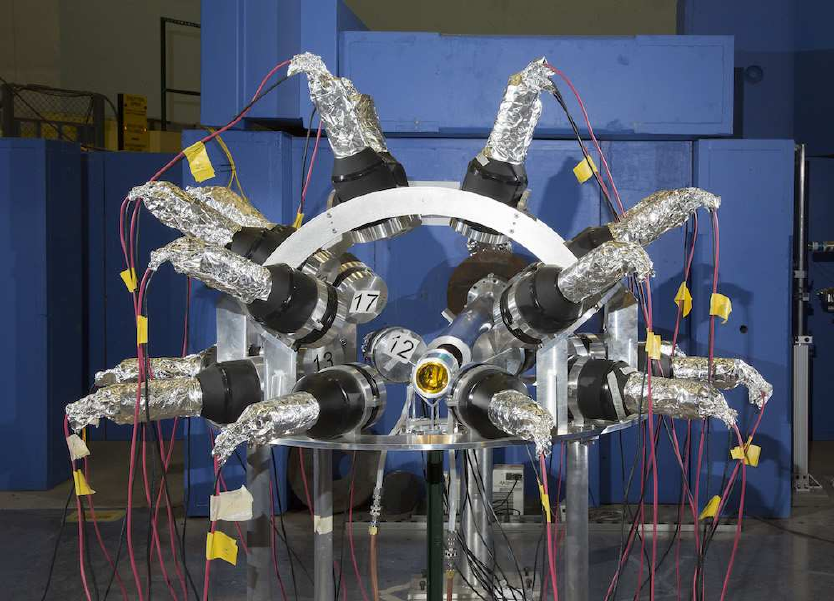}
\caption{(Color online) The low-energy Chi-Nu array consists of 22 $^6$Li glass detectors to measure the PFNS down to $\sim 10$~keV.}
\label{fig:chinu-low}
\end{figure}

\begin{figure}[ht]
\centering
\includegraphics[width=0.95\columnwidth]{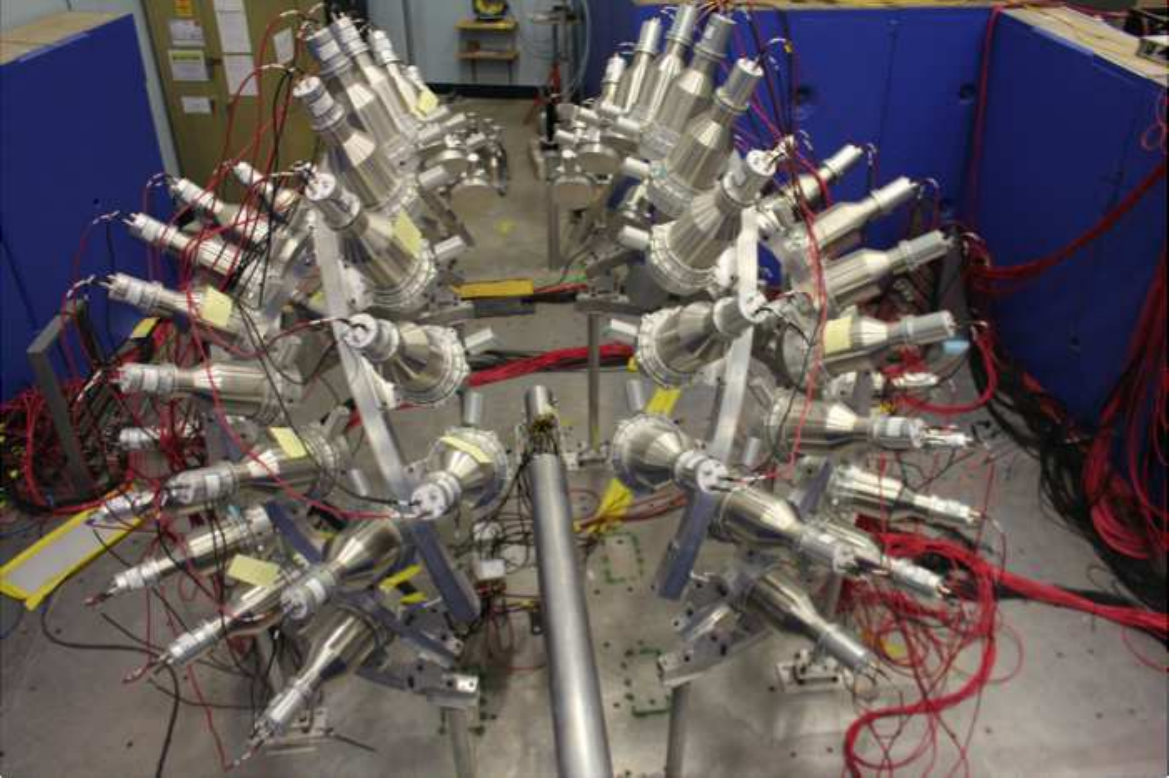}
\caption{(Color online) The high-energy Chi-Nu array consists of 54 EJ-309 liquid scintillators designed to measure the PFNS up to $\sim 15$~MeV with adequate statistics.}
\label{fig:chinu-high}
\end{figure}

The Chi-Nu arrays (Figs.~\ref{fig:chinu-low} and~\ref{fig:chinu-high}) have been developed at Los Alamos National Laboratory (LANL) and are deployed at the Los Alamos Neutron Science Center (LANSCE) to measure the average PFNS of several actinides with great accuracy. In particular, $^{235}$U\nf\ and $^{239}$Pu\nf\ are studied over a broad range of incident neutron energies. Most past PFNS measurements have acquired data in the 700~keV to 7$-$8~MeV range of outgoing neutron energy. A significant number of neutrons are emitted below 700~keV but multiple scattering corrections, neutron background, and low-sensitivity of liquid neutron scintillators have prevented accurate measurements of the PFNS in this region. At the highest energies, statistics become poor and long acquisition times are necessary for an adequate measurement. In addition, most measurements have been performed for spontaneous and thermal neutron-induced fission reactions only. The Chi-Nu project aims to accurately measure the PFNS of neutron-induced fission of $^{235}$U and $^{239}$Pu for incident neutron energies from thermal up to $\sim 200$~MeV and for outgoing neutron energies from 10~keV up to 15~MeV. 

Two Chi-Nu arrays have been built.  The first one, Fig.~\ref{fig:chinu-low}, consists of 22 $^6$Li glass detectors designed to measure the low-energy part of the spectrum, down to $\sim 10$~keV.  The second one, Fig.~\ref{fig:chinu-high}, consists of 54 EJ-309 liquid scintillators that can be used to extend the spectral measurement up to $\sim 15$~MeV with sufficient statistics. Thanks to their segmented nature, the Chi-Nu arrays can also be used to study $n$-$n$ correlations as well as neutron energy and angular correlations. Using pulse-shape discrimination, they can also be used to study prompt fission \grays. The analysis of a large amount of data already collected is now being performed specifically with correlations in mind.

\begin{figure}[ht]
\centering
\includegraphics[width=0.95\columnwidth]{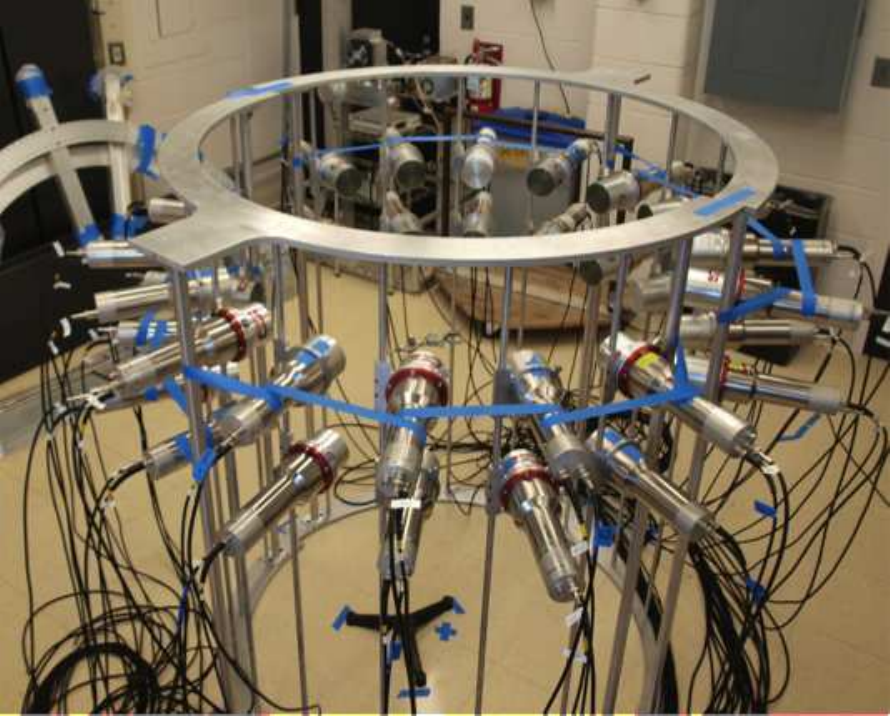}
\caption{(Color online) The array of EJ-309 and NaI scintillators used to measure neutron-neutron correlations in $^{252}$Cf(sf) at the University of Michigan~\cite{Marcath:2016}.}
\label{fig:UMexpsetup}
\end{figure}

While the Chi-Nu arrays were not designed to extract correlated prompt data, the University of Michigan developed specific experiments~\cite{Pozzi:2014} to measure those correlations. One of those experimental setups is shown in Fig.~\ref{fig:UMexpsetup} and consists of 24 EJ-309 and 8 NaI(Tl) scintillators, arranged in two rings surrounding a centrally-located $^{252}$Cf source. A somewhat different setup was used to measure correlations in the spontaneous fission of $^{240}$Pu~\cite{Marcath:2016}. In that experiment, a $\sim 2$~g plutonium sample was placed at the center of the detector assembly and neutron doubles were acquired within a 100 ns time window. To our knowledge, this was the first measurement of neutron-neutron correlations and neutron doubles for this reaction.

\begin{figure}[ht]
\centering
\includegraphics[width=0.9\columnwidth]{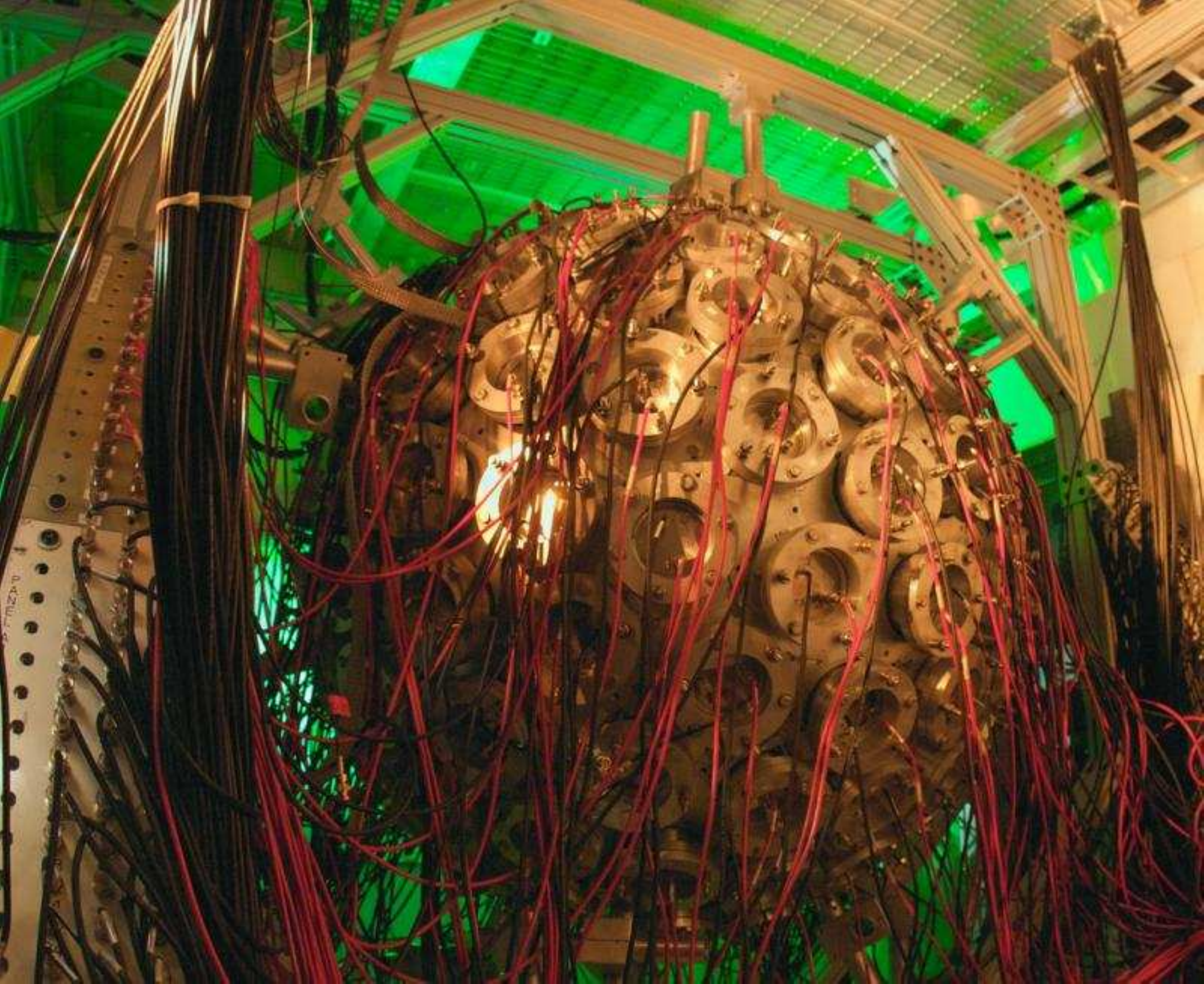}
\caption{(Color online) The DANCE detector array is a 4$\pi$ calorimeter made of 160 BaF$_2$ crystals developed to measure capture cross sections with very small samples or/and very radioactive targets.}
\label{fig:DANCE}
\end{figure}

The DANCE setup (Detector for Advanced Neutron Capture Experiments), installed at the Lujan Center at LANSCE consists of 160 BaF$_2$ crystals arranged in a $4\pi$ geometry, as shown in Fig.~\ref{fig:DANCE}.  Originally designed to measure capture cross sections on very small target samples and/or very radioactive materials, it is a very high efficiency calorimeter that can be used to study the prompt fission \gray~multiplicity and energy spectrum~\cite{Ullmann:2013,Jandel:2014}. 

\begin{figure}[ht]
\centering
\includegraphics[width=0.9\columnwidth]{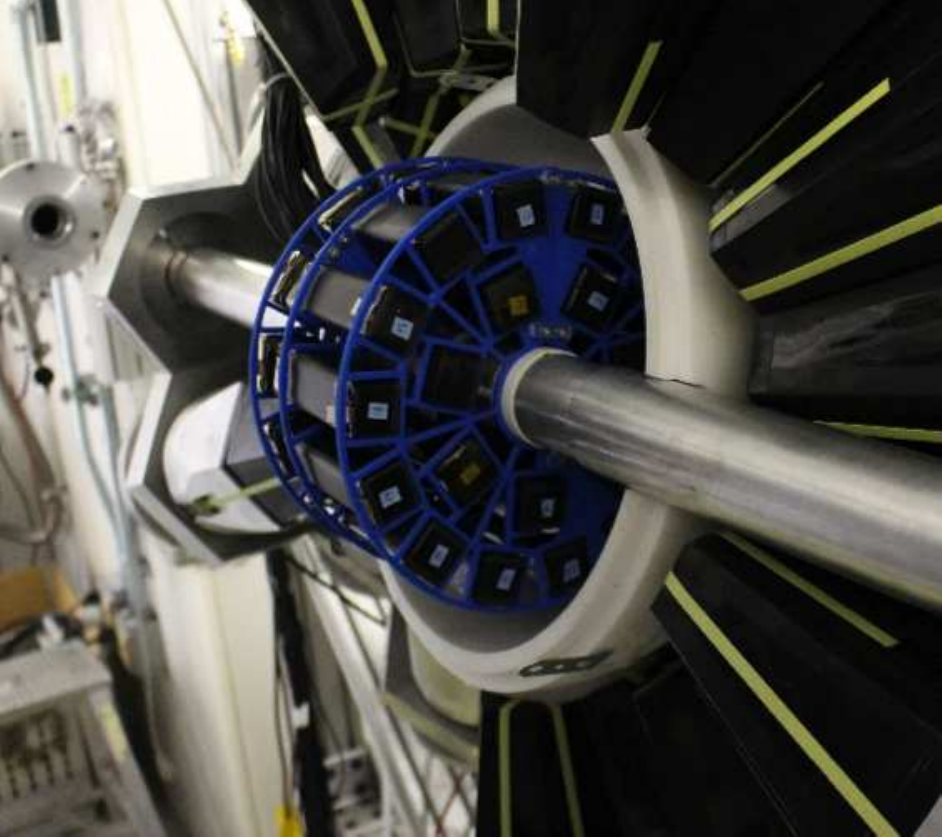}
\caption{(Color online) The NEUANCE detector array 
~\cite{Jandel:2018} consists of 21 stilbene crystals arranged in a compact form to fit around the beam line at the center of the DANCE cavity.}
\label{fig:NEUANCE}
\end{figure}

To enhance the capabilities of DANCE, a new detector array, NEUANCE, was developed to make correlated measurements of prompt fission neutrons and \grays~\cite{Jandel:2018}. In its present configuration, NEUANCE consists of 21 
23~mm $\times$ 23~mm $\times$ 100~mm stilbene crystals arranged cylindrically around the beam line with the target at the center inside the DANCE array. A picture of NEUANCE inside one of the hemispheres of DANCE is shown in Fig.~\ref{fig:NEUANCE}. 

\begin{figure}[ht]
\centering
\includegraphics[width=0.75\columnwidth]{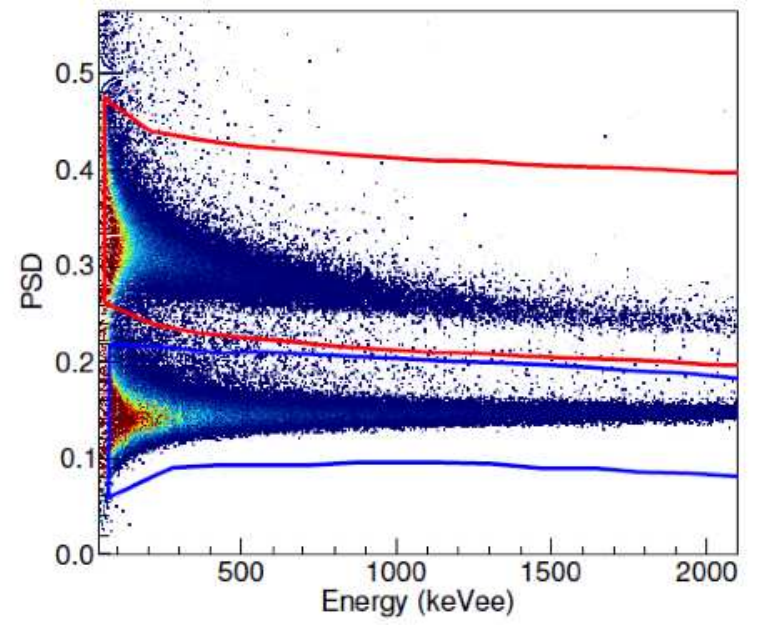}
\caption{(Color online) A pulse-shape discrimination (PSD) plot measured~\cite{Jandel:2018} with a $^{252}$Cf source with one of the NEUANCE stilbene detectors. The upper band  (outlined in red) is a result of detected neutrons while the bottom one (blue) corresponds to \gray~events.}
\label{fig:PSD}
\end{figure}

\begin{figure}[ht]
\centering
\includegraphics[width=\columnwidth]{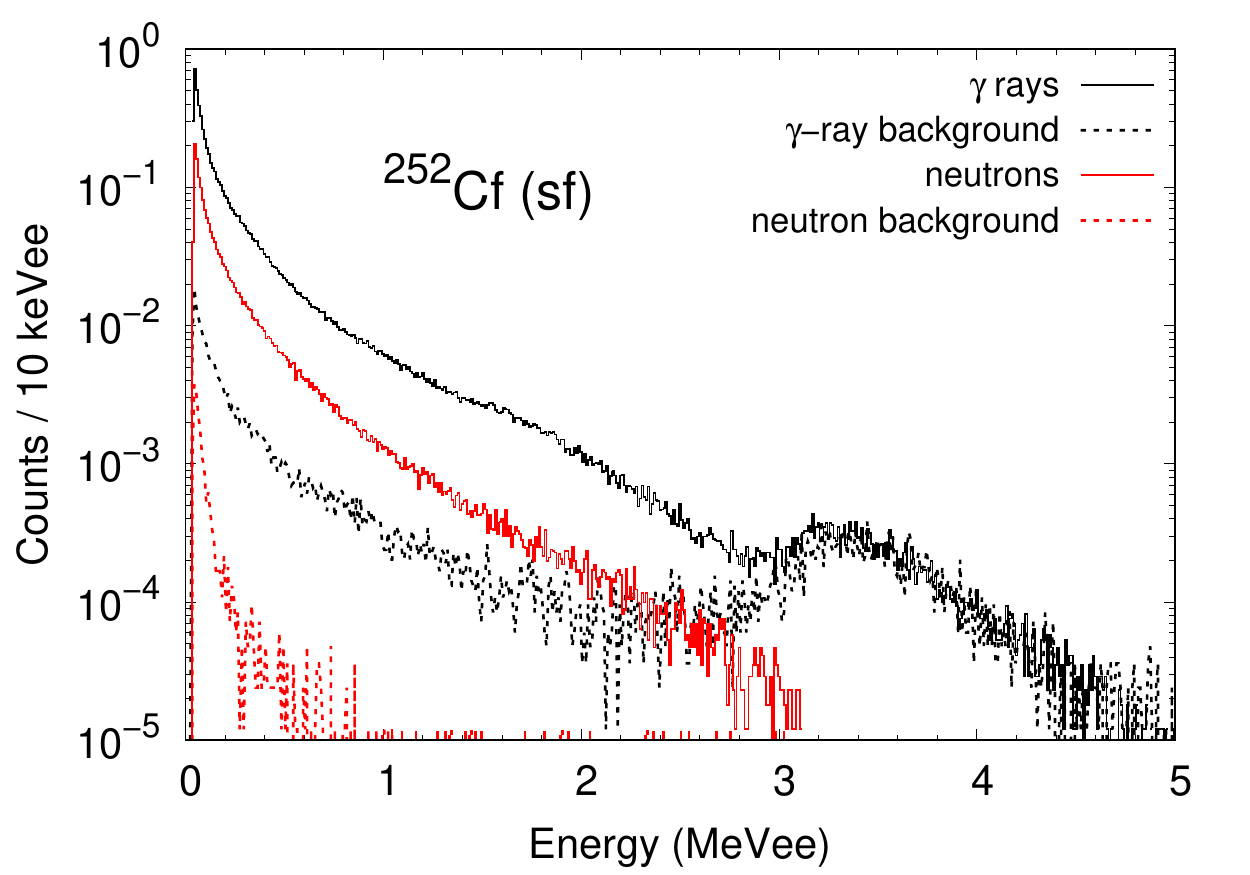} 
\caption{(Color online) Gamma-ray (red lines) and neutron (black lines) spectral intensity observed~\protect\cite{Jandel:2018} in one detector of the 21 stilbene detectors of the NEUANCE array. The thick lines represent the rates observed from the $^{252}$Cf source and the thin lines are the background rates.  The bump at 3.5~MeV in the ambient $\gamma$-ray background results from the saturation of signals from high energy $\gamma$-rays or neutrons (cosmic rays).  See Ref.~\protect\cite{Jandel:2018} for details.}
\label{fig:stilbene-spectra}
\end{figure}

The NEUANCE stilbene detectors have excellent pulse-shape discrimination (PSD) properties, allowing discrimination between neutron and \gray~signals.  A PSD plot for one of the stilbene detectors is shown in Fig.~\ref{fig:PSD}. A $^{252}$Cf source with activity of 739.3 fission events per second was used in a recent measurement. The spectral intensities of \grays~and neutrons are shown in Fig.~\ref{fig:stilbene-spectra} by black and red lines, respectively, for just one of the 21 stilbene crystals of the NEUANCE array. The thick lines represent the rates measured with the NEUANCE detectors for $^{252}$Cf source and thin dotted lines are the background rates.

\begin{figure}[ht]
\centering
\includegraphics[width=0.9\columnwidth]{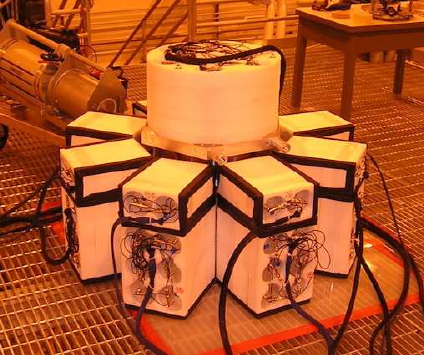}
\caption{(Color online) Photograph of the 77 liquid scintillator array used at LLNL to measure $n$-$n$ correlations in $^{252}$Cf(sf) and $^{240}$Pu(sf) \protect\cite{Nakae:birthdayCake,Verbeke:2017}.}
\label{fig:birthdayCake}
\end{figure}

A detector setup to measure neutron-neutron correlations was also developed at Lawrence Livermore National Laboratory (LLNL), as shown in Fig.~\ref{fig:birthdayCake}. It consists of 77 EJ-301 liquid scintillators, each read out by a single photomultiplier tube. Each tower of 8 scintillators is symmetrically arranged into octants with an array inner diameter of~60 cm. Thirteen identical scintillators compose the top of the detection system. The detector was designed for fast multiplicity counting and assaying of fissile material. The fast scintillator decay time of a few ns allows faster count rates than $^3$He well counters. The relatively tightly-packed system has an overall geometric efficiency of 50\% (2$\pi$). Measurements have been carried out with spontaneous fission sources of $^{252}$Cf and $^{240}$Pu placed at the center of the detection system. Some results are discussed in Sec.~\ref{sec:simulations}.

As a step toward a more portable neutron-gamma detector setup, a small (six liquid scintillators) and flexible experimental setup has been built at 
LANL. Relative detector angles and distances from the fission source are adjustable. Data acquisition software provides list-mode data collection. The flexibility of this setup is important to validate transport simulations in a wide range of configurations to study $n$-$n$ angle, multiplicity and energy correlations. Of particular interest is the measurement and characterization, via accurate simulations, of cross talk between adjacent detectors and scatter from surrounding objects. By definition, cross talk occurs when a particle recorded in one of the detectors scatters and triggers an adjacent detector. Simulations of the detector response for neutrons are in progress, while \gray~capabilities will be added in the near future.

\begin{figure}[ht]
\centering
\includegraphics[width=0.9\columnwidth]{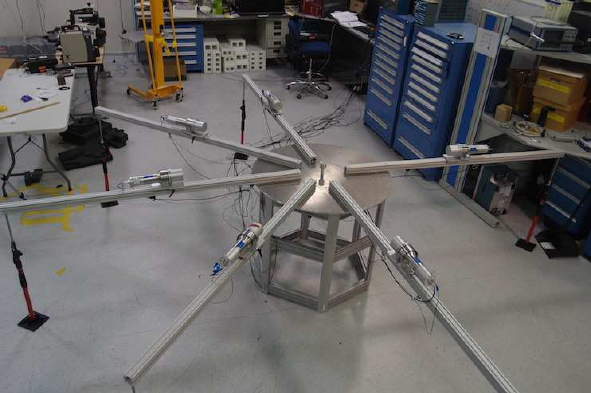}
\caption{(Color online) The prototype for a versatile, flexible, and portable neutron detector array 
~\cite{Andrews:2017} has been developed and is being tested at LANL to study neutron-neutron correlations in various geometrical configurations.}
\label{fig:NISC}
\end{figure}


Other experimental setups have been devised to measure various correlated fission data. At the JRC in Geel, Belgium, the SCINTIA array~\cite{Hambsch:2015} has been developed to measure neutron energy and multiplicity in coincidence with the fission fragment mass. When complemented by \gray~detectors, it can provide very useful information on $n$-\g~energy and multiplicity correlations as a function of resonances in the $n+^{235}$U cross section, and help infer the respective, possibly complementary roles of the ($n,\gamma$f) process and fission fragment $Y(A$,TKE) yield fluctuations.

The SOFIA (Study on Fission with Aladin) experimental program~\cite{Martin:2014,Pellereau:2017}, carried out at GSI, measures fission yields in inverse kinematics for a broad range of fissioning nuclei with very accurate information on the mass and charge of the fragments.  The average neutron multiplicity can be inferred, albeit not for monoenergetic reactions.

Measurements of the prompt neutron multiplicity distribution typically involves the use of a Gd-loaded scintillator tank~\cite{Vorobyev:2005} in order to capture and thermalize all prompt neutrons with an almost 100\% efficiency. Many experimental data sets are available for spontaneous fission and, to a lesser extent, for thermal neutron-induced fission reactions. The measurement of $P(\nu)$ 
for $E_{\rm inc}$ larger than thermal is rendered much more difficult due to the important background from neutron scattering in the surrounding material.  Only one such measurement has been reported~ for $^{235,238}$U\nf\ and $^{239}$Pu\nf\ up to $E_{\rm inc} = 10$~MeV \cite{Soleilhac:1969}. 

As shown in Fig.~\ref{fig:Pnu}(b), the prompt neutron multiplicity distribution for actinides can typically be represented by a Gaussian distribution.  The average multiplicity is 2-4 neutrons, with non-negligible contributions from up to 6-7 neutrons, see Fig.~\ref{fig:Pnu}(a). In a few cases, the average neutron spectrum is known with great accuracy. However, very little is known about the dependence of the neutron spectrum on the neutron multiplicity. Is the average spectrum for 6 neutrons emitted the same as when only 2 neutrons are emitted? A similar question can be asked for the prompt \grays, whose multiplicity distribution spans an even larger range, with $N_\gamma$ up to $\sim 20$ (see Fig.~\ref{fig:Cf252sf-PnuGamma}).

Measuring the angular distribution of prompt neutrons and \grays~emitted in a particular fission reaction also brings useful information.  However, it can be somewhat more difficult to interpret as it represents a convolution of the angular distribution of the fission fragments with the angular distribution of the emitted particles with respect to the direction of emission of the fragments. 

Finally, interesting quasi-differential measurements~\cite{Daskalakis:2014} of neutron scattering off fissile material provide useful benchmarks for prompt fission neutron emission angular distributions, although only as part of other contributing reaction channels such as elastic and inelastic scattering.

\section{Modeling prompt fission emission}

This section introduces the two complete event fission models, \CGMF\ and \FREYA, that have been incorporated into the \MCNP6 transport code.  First the physics encapsulated in these two codes is described, along with some discussion of their similarities and differences.  Next, a general introduction to radiation transport codes is presented, highlighting the concept of incorporating complete event models and how this can enhance the simulation of fission in such codes.  The section ends with a brief demonstration that incorporation into \MCNP6 does not affect the \CGMF\ and \FREYA\ results.

\subsection{Complete event fission models} \label{sec:codes}

Although various physics models and codes have been developed and used to describe different aspects of prompt fission neutron and \gray~emission in specific limited studies, the models implemented in transport simulations and used to evaluate nuclear data libraries have been mostly limited to average multiplicity and spectra. For instance, the Los Alamos model~\cite{Madland:1982} has been and is still used~\cite{Neudecker:2015} for nuclear data evaluations of the $\chi(E',E)$ matrix of the PFNS as a function of incident neutron energy for most evaluated libraries including the U.S. ENDF/B-VII.1 library 
~\cite{ENDFB71}. This model provides an average spectrum with few adjustable parameters that can be tuned to match existing PFNS data. The accuracy of this approach is strongly limited by the availability of experimental PFNS data for neighboring nuclei and energies. This model makes use of several important physical assumptions, some of which have been lifted in modern extensions of the original model and averages over only a few mass yields. No additional detailed information, such as the average neutron multiplicity as a function of fragment mass $\overline{\nu}(A)$ or the neutron multiplicity distribution $P(\nu)$, can be extracted.  (See Ref.~\cite{Capote:2016} for more details about evaluations with the Los Alamos model and its extensions.)

In recent years, several computer codes have been developed to simulate the sequence of prompt neutron and \gray~emission in detail. Event-by-event simulators have been implemented in fast numerical codes that can be integrated into transport simulations of fissioning systems. Here two such codes, \CGMF~and \FREYA, are presented in some detail.  Note that there are also several similar codes developed independently: $\mathtt{FIFRELIN}$~\cite{FIFRELIN} developed at the CEA in France, $\mathtt{GEF}$~\cite{GEF} developed at GSI in Germany and CENBG in France, $\mathtt{FINE}$ developed by Kornilov~\cite{Kornilov:2015}, and more recently $\mathtt{EVITA}$
~\cite{Morillon:2017}, based on the $\mathtt{TALYS}$ deterministic code, 
developed by CEA in France. Some limited code comparisons can be found in Ref.~\cite{Capote:2016}.  A separate model code developed by Lestone at LANL \cite{Lestone:2016} has been used successfully to simulate the neutron-neutron and neutron-fragment correlations for $^{252}$Cf(sf) and neutron-induced fission of $^{235}$U and $^{239}$Pu. Although this code uses more available experimental data as input, and therefore is more limited in scope, it represents a very viable, fast and complementary alternative to the efforts discussed in the present paper.  There can be significant differences in the physics implemented in those different codes, hence one can expect differences in the calculated results, especially for more differential observables.

Here only the broad outlines of the \CGMF~and \FREYA~codes are presented. For more detail, see the publicly available user manuals in Refs.~\cite{CGMF} and \cite{FREYA} respectively and references therein.

\subsubsection{The \CGMF~code} \label{sec:CGMF}

\begin{figure}[ht]
\centering
\includegraphics[width=\columnwidth]{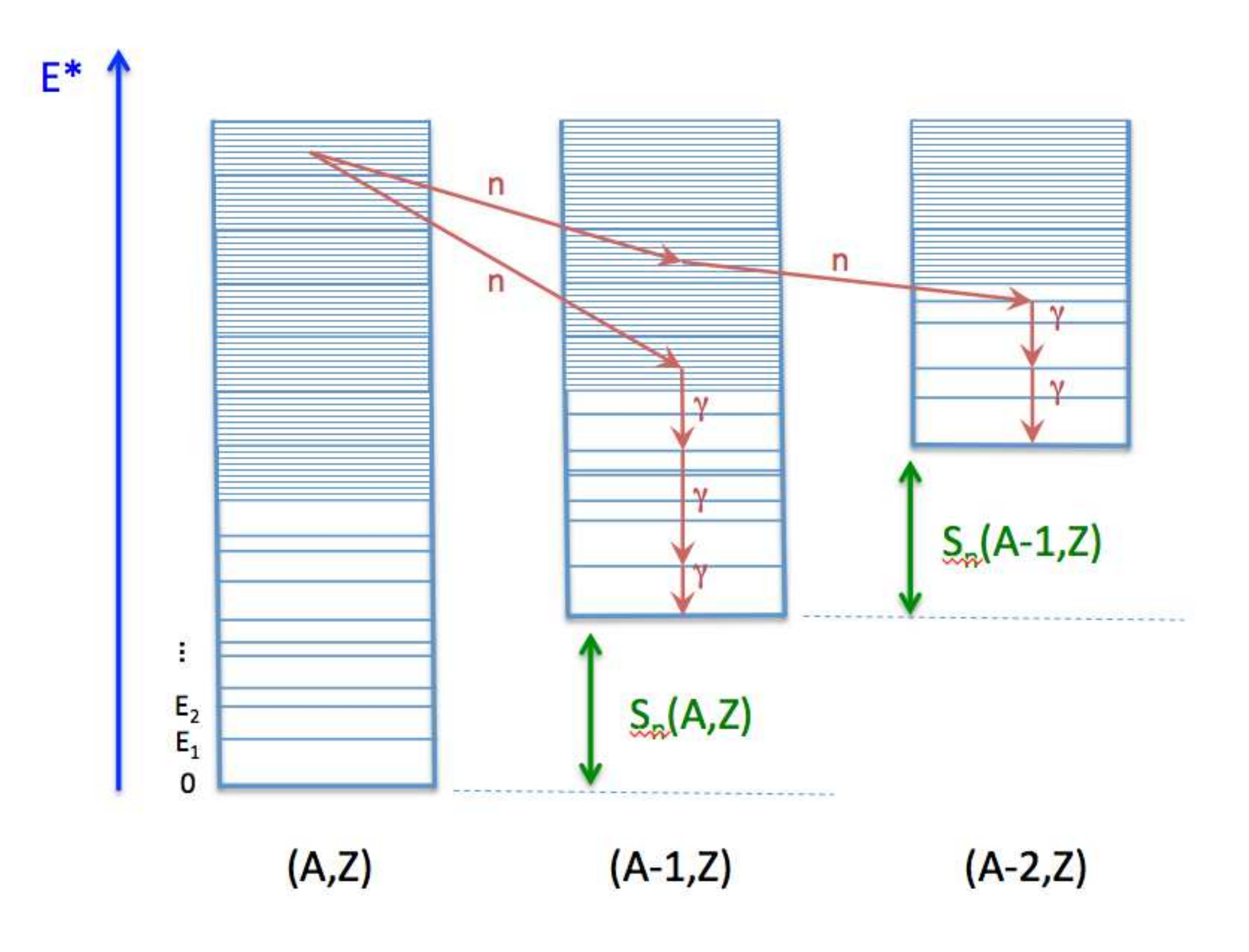}
\caption{(Color online) Schematic representation~\cite{CGMF} of the decay of the fission fragments by successive evaporation of prompt neutrons and \grays.}
\label{fig:decay}
\end{figure}

The \CGMF~code, developed at Los Alamos National Laboratory, is a Monte Carlo implementation of the statistical Hauser-Feshbach nuclear reaction theory~\cite{Hauser:1952} applied to the de-excitation of the primary fission fragments. At every stage of the decay (see Fig.~\ref{fig:decay}), the code samples probability distributions for the emission of neutrons and \grays. Each fission fragment is described as a compound nucleus with an initial excitation energy $E^*_i$, spin $J_i$ and parity $\pi_i$. Neutrons are emitted, removing their kinetic energy from the fragment intrinsic excitation energy, while doing little to change the angular momentum $J$. 
On the other hand, \gray~emissions, generally after all neutrons are evaporated, tend to decrease $J$. Several nuclear models as well as nuclear structure information are needed in order to perform these calculations, as discussed below. Typical results of a \CGMF~run can be collected as a (long) series of data strings that represent each fission event. The initial characteristics of the fission fragment in mass, charge, kinetic energy, excitation energy, spin, parity, and their momentum vectors in the laboratory frame, as well as the kinematic information on all emitted neutrons and photons in the laboratory frame are recorded. The statistical analysis of those recorded events provides the needed output that can be compared to experimental data. All types of distributions and correlations in multiplicity, energy and angular distribution can be inferred from such history files in a rather straightforward manner.

For a particular fission reaction, such as $^{239}$Pu\nf\ with $E_{\rm inc} =2$~MeV, \CGMF\ requires the fission fragment yields in mass, charge and total kinetic energy, 
$Y(Z,A,$TKE) produced in this reaction as input. Those yields are sampled using Monte Carlo techniques to obtain the initial fission fragments from which the sequence of neutron and \gray~evaporations can start. Experimental information on the fission fragment yields is rather scarce at this time although important recent theoretical and experimental developments, such as discussed in Sec.~\ref{sec:measurements}, 
show great promise. Depending on the reaction studied, different prescriptions for the reconstruction of the full 3D distribution are used. When available, experimental information, even partial, has been preferred~\cite{Talou:2011} to less accurate phenomenological models. For instance, experimental data on $Y(A)$, $\langle {\rm TKE}\rangle(A)$, $\sigma_{\rm TKE}(A)$, or even  $Y(A,$TKE) are available for some limited isotopes and energies. In the case where no experimental data exist, which is particularly true for higher incident neutron energies, then the simple five-Gaussian prescription used in \FREYA, discussed in the following section, is also used in \CGMF. The mass-dependent charge distributions $Y(Z|A)$ are taken from Wahl's systematics~\cite{Wahl:1988}.

Once a light fission fragment 
$(Z_L,A_L)$ is chosen randomly, its complementary heavy partner is obtained by mass and charge conservation such that $Z_H=Z_0-Z_L$ and $A_H=A_0-A_L$, where $(Z_0,A_0)$ are the charge and mass of the fissioning parent nucleus. At this point, \CGMF\ treats 
binary fission only, no ternary or more exotic fission events are considered. 
The total excitation energy (TXE) available for these fragments is given 
by the $Q$ value for that particular split ($Q_{LH}$)
minus the total kinetic energy carried away by these fragments (TKE),
${\rm TXE} \, = \, Q_{LH} - \, {\rm TKE}$, where
\begin{eqnarray}
  Q_{LH} & = & E_0^* + M_n(Z_0,A_0)c^2 \label{eq:Qfiss} \\
 & & -M_n(Z_L,A_L)c^2 -M_n(Z_H,A_H)c^2  \nonumber
\end{eqnarray}
and $M_n(Z,A)$ is the nuclear mass. 
The excitation of the fissioning nucleus depends on how the fission was initiated: it vanishes for spontaneous fission, $E_0^*({\rm sf})=0$;
it is given by $E_0^*(\gamma,{\rm f})=E_\gamma$ for photofission
and for (low-energy) neutron-induced fission it is equal to $E_0(n,{\rm f})=E_{\rm inc}+S_n$ 
where $S_n$ is the neutron separation energy.
All nuclear masses and binding energies are taken from the AME2012 Atomic Mass Evaluation~\cite{Audi:2012}, complemented by FRDM-2012 calculations~\cite{Moller:2015} when no experimental data exist.

The total excitation energy, TXE, available for neutron and \gray~emission is then shared among the two complementary fragments. Several prescriptions exist for sharing this energy. In its current version, \CGMF~uses a mass-dependent parameter $R_T(A)$ in order to best reproduce the experimental mass-dependent neutron multiplicity, $\overline{\nu}(A)$. A second input parameter, $\alpha$, is used to modify the initial spin distribution, given as
\begin{eqnarray}
\rho(J,\pi)=\frac{1}{2}(2J+1)\exp \left[ -\frac{J(J+1)}{2B^2(Z,A,T)}\right],
\end{eqnarray}
where $B^2$ is defined in terms of the fragment temperature,
\begin{eqnarray}
B^2(A,Z,T) = \alpha\frac{\mathcal{I}_0(Z,A)}{\hbar^2},
\end{eqnarray}
and $\mathcal{I}_0(Z,A)$ is the ground-state moment of inertia for the fragment $(Z,A)$. The adjustable input parameter $\alpha$ can then be used to tune this initial spin distribution to reproduce the average prompt fission \g~multiplicity.

Once the initial conditions $(U_i,J_i,\pi_i)$ in energy, spin and parity of each fragment are set, the Hauser-Feshbach statistical theory of de-excitation of a compound nucleus can be applied. Only neutrons and \grays~have a reasonable probability of being emitted for the fragments and energies considered. Charged particles are hindered by the Coulomb barrier. Also, the dynamical emission of particles or clusters, such as ternary $\alpha$ particles, is not considered.

The probability for a fragment to emit a neutron of energy $\epsilon_n$ is given by
\begin{eqnarray}
P_n(\epsilon_n)d\epsilon_n \propto T_n(\epsilon_n)\rho(Z,A-1,U-\epsilon_n-S_n)d\epsilon_n,
\end{eqnarray}
where $\rho$ describes the nuclear level density in the residual nucleus $(Z,A-1)$ at the residual excitation energy
$U-\epsilon_n-S_n$, and $S_n$ denotes the neutron separation energy. The neutron transmission coefficients $T_n$ are obtained through optical model calculations. Because of the large number of fragments produced, only a global potential can be used. Until now, the global optical potential of Koning and Delaroche~\cite{Koning:2003} has been used for most \CGMF~calculations.  The transmission coefficients for photon emission are obtained from the \gray~strength function $f_\gamma(\epsilon_\gamma)$, assuming the Brink hypothesis, {\it i.e.}, the equivalence between the $(n,\gamma)$ and $(\gamma,n)$ reaction channels, and using the Kopecky-Uhl formalism~\cite{Kopecky:1990}
\begin{eqnarray}
T_\gamma(\epsilon_\gamma) = 2\pi\epsilon_\gamma^{2l+1}f_\gamma(\epsilon_\gamma),
\end{eqnarray}
where $l$ is the multipolarity of the electromagnetic transition. In \CGMF, only E1, M1 and E2 transitions are considered. The strength function parameterizations of the RIPL-3 library~\cite{RIPL3} are used.

In \CGMF, as in most other Hauser-Feshbach codes, the fragment, {\it i.e.}, the compound nucleus, is represented by a discrete level region at low excitation energies, completed by a continuum region at higher energies where discrete excitations cannot be resolved any longer. Known discrete levels are read in from the RIPL-3 library~\cite{RIPL3}, which itself derives them 
from the ENSDF nuclear structure data library~\cite{Tuli:1996}. At higher excitation energies, the level density is calculated using the Gilbert-Cameron mixed model of a constant temperature followed by a Fermi gas region to represent the continuum. \CGMF~also implements Ignatyuk's prescription~\cite{Ignatyuk:1975} that dampens the shell corrections with increasing energy.

At each stage of the decay of the fragments (see Fig.~\ref{fig:decay}), the neutron and \gray~emission probabilities are sampled and a particular transition picked, leading to a new configuration characterized by a new set of $(Z,A,E^*,J,\pi)$. In addition, the kinematics of the neutrons, \grays~and fragments are followed exactly in the classical limit. Very small relativistic corrections are ignored. In the current version of the code, neutrons and \grays~are evaporated isotropically in the center-of-mass of the parent fragment. The (small) recoil of the fragments due to the emission of the particles is taken into account. The boost of the center-of-mass to laboratory frames is responsible for the strong focusing of the particles along the fission axis.

Neutrons have a much higher probability of being emitted at high excitation energy, while \grays~compete mostly at lower energies. However, high spins can lead to larger decay width ratios,  $\Gamma_\gamma/\Gamma_n$, allowing for the emission of \grays~for nuclear excitation energies higher than the neutron separation energy. An additional complication is the presence of long-lived isomers in the fission fragments. By default, \CGMF~calculates the prompt fission \g~spectrum for a time coincidence window of 10~ns, which is typical of recent experimental setups used to measure this spectrum~\cite{Oberstedt:2013,Chyzh:2014}. However, the exact coincidence window for a specific experiment should be used to compare theory and experiment. 

In neutron-induced fission reactions with increasing incident neutron energy, neutrons can be emitted from the parent nucleus before it fissions, leading to multi-chance fission reactions, labeled as 
($n,n'$f), ($n,2n$f), ($n,3n$f), $\cdots$ where $n'$ indicates that the emitted neutron is not the same as the incident one. Above about 10~MeV, pre-equilibrium (PE) emissions can also occur, leading to a pre-fission neutron spectrum different than a compound nucleus evaporation spectrum. In this case, the emitted neutron is the incident neutron.  The two-component exciton model~\cite{Gadioli:1992} is used to calculate the probability of PE emission at a given incident neutron energy, as well as the PE neutron spectrum. Probabilities for first-, second-, third- and up to fourth-chance fission are calculated separately using the \COH~nuclear reaction code~\cite{Kawano:2010}.

\subsubsection{The \FREYA~code}\label{sec:FREYA}

The computational model \FREYA, developed at Lawrence Berkeley and Lawrence Livermore National Laboratories, generates complete fission events, {\em i.e.}\ it provides the full kinematic information on the two product nuclei as well as all the emitted neutrons and photons. In its development, an emphasis had been put on speed, so large event samples can be generated fast.  \FREYA\ therefore relies on experimental data supplemented by simple physics-based modeling.

In its standard version, to treat a given fission case, \FREYA\ needs the fission fragment mass distribution $Y(A)$ and the average total kinetic energy $\langle {\rm TKE}\rangle (A)$ for the particular excitation energy considered. $Y(A)$ is taken either directly from the measured yields or from a five-gaussian fit to the data, see Ref.~\cite{Vogt:2012} for details, which makes it possible to parameterize its energy dependence of the yields.

In order to generate an event, \FREYA\ first selects the mass split based on the provided $Y(A)$. The fragment charges are then sampled from the normal distributions suggested by experiment \cite{Vogt:2012}. The linear and angular momenta of the two fragments and their internal excitations are subsequently sampled as described below. After their formation, the fully accelerated fragments de-excite first by neutron evaporation and then by photon emission. In addition to spontaneous fission, \FREYA\ treats neutron-induced fission up to $E_{\rm inc}=20$~MeV.  The possibility of pre-fission evaporations up to fourth-chance fission is considered as well as pre-equilibrium neutron emission, as described in 
Ref.~\cite{Vogt:2012}. 

\FREYA\ contains a number of adjustable parameters that control various physics aspects.  They are listed here as a group but are also mentioned in the text where they were used.
\begin{description}
\item[$d{\rm TKE}$,] an overall shift of $\overline{\rm TKE}$ relative to the input TKE($A$), used to adjust the average neutron multiplicity $\overline{\nu}$; 
\item[$e_0$,] the overall scale of the Fermi-gas level density parameters; 
\item[$x$,] the advantage in excitation energy given to the light fragment. It is currently single valued and energy independent but could be made mass dependent, like the $R_T(A)$ distribution in \CGMF\ and \FIFRELIN~\cite{FIFRELIN} for cases with sufficient data;
\item[$c_S$,] the ratio of the 
``spin temperature" to the ``scission temperature".
\item[$c_T$,] the relative statistical fluctuation in the fragment excitations.
\end{description}
So far, none of the \FREYA\ parameters are assumed to depend on fragment mass.  The $d$TKE is adjusted as a function of $E_{\rm inc}$  to match $\overline \nu(E_{\rm inc})$.  As described shortly, the prescription for the calculation of the level density parameter is energy dependent even though $e_0$ itself is not.  There is currently insufficient information available to assume any energy dependence of $x$, $c_S$ or $c_T$.

The emission of \grays\ in \FREYA\ is governed by $c_S$.  There are two additional settings in \FREYA\ that influence the $\gamma$ results:
\begin{description}
\item[$g_{\rm min}$,] the minimum \gray~energy measurable by a given detector;
\item[$t_{\rm max}$,] the maximum half-life of a level during the photon decay process (which stops when it reaches a level having a half-life exceeding $t_{\rm max}$).
\end{description}
The quantities $g_{\rm min}$ and $t_{\rm max}$ are detector dependent with $t_{\rm max}$ corresponding to the time coincidence window for \CGMF\ mentioned in the previous section.

In the remainder of this section, the physics modeling in \FREYA~is described.

For a given split of compound nucleus $A_0$ into light and heavy fragments, $A_L$ and $A_H$ respectively, the fission $Q$ value for \FREYA\ is defined the same way as for \CGMF, see the discussion around Eq.~(\ref{eq:Qfiss}).
For a given total fragment kinetic energy TKE, the energy available for rotational and statistical excitation of the two fragments is $E^*_{\rm sc}=Q-\overline{\rm TKE}$ for the sampled value of TKE.  The corresponding 
``scission temperature", $T_{\rm sc}$, which is scaled by the parameter $c_S$, is obtained from $E^*_{\rm sc}=(A_0/e_0)T_{\rm sc}^2$.

\FREYA\ explicitly conserves angular momentum. The overall rigid rotation of the dinuclear configuration prior to scission, caused by the absorption of the incoming neutron and the recoil(s) from any evaporated neutron(s), dictates certain mean angular momenta in the two fragments. In addition, due to the statistical excitation of the scission complex, the fragments also acquire fluctuations around those mean values. \FREYA\ includes fluctuations in the wriggling and bending modes (consisting of rotations in the same or opposite sense around an axis perpendicular to the dinuclear axis) but ignores tilting and twisting (in which the fragments rotate around the dinuclear axis). These dinuclear rotational modes are assumed to become statistically excited 
during scission. They are therefore described by Boltzmann distributions,
\begin{eqnarray}
P_\pm(\mbox{\boldmath$s$}_\pm)\,ds_\pm^xds_\pm^y\ \sim\ {\rm e}^{-s_\pm^2/2{\cal I}_\pm T_S}ds_\pm^xds_\pm^y\, ,
\end{eqnarray}
where $\mbox{\boldmath$s$}_\pm=(s_\pm^x,s_\pm^y,0)$ is the spin of the normal modes with plus referring to the wriggling modes (having parallel rotations) and minus referring to the bending modes (having opposite rotations). The corresponding moments of inertia are denoted ${\cal I}_\pm$ \cite{Randrup:2014,Vogt:2013}. The degree of fluctuation is governed by the `spin temperature' $T_S=c_ST_{\rm sc}$ which can be adjusted by changing the parameter $c_S$. The fluctuations vanish for $c_S=0$ and the fragments would then emerge with the angular momenta dictated by the overall rigid rotation of the scission configuration (usually very small for induced fission - and entirely absent for spontaneous fission). The default value, $c_S = 1$, leads to $\overline S_L \sim 6.2\hbar$ and  $\overline S_H \sim 7.6\hbar$ for $^{252}$Cf(sf) and gives reasonable agreement with the average energy of \grays\ emitted in fission (see Ref.~\cite{Randrup:2014} for details).

After accounting for the total rotational energy of the two fragments, $E_{\rm rot}$, there is a total of $E_{\rm stat}=E^*_{\rm sc}-E_{\rm rot}$ remaining for statistical fragment excitation. It is distributed between the two fragments as follows.  First, a preliminary partition, $E_{\rm stat}=\acute{E}^*_L+\acute{E}^*_H$, is made according to the heat capacities of the two fragments which are assumed to be proportional to the corresponding Fermi-gas level density parameters, {\em i.e.}\ $\acute{E}^*_L/\acute{E}^*_H=a_L/a_H$, where
\begin{eqnarray}
a_i({\acute{E}}_i^*) = {A_i\over e_0}
\bigg[ 1 + \frac{\delta W_i}{U_i} \left(1 - e^{-\gamma U_i}\right) \bigg]\ ,
\label{aleveldef}
\end{eqnarray}
with $U_i=\acute{E}_i^*-\Delta_i$ and $\gamma=0.05/$MeV \cite{Lemaire:2005}.  The pairing energy of the fragment, $\Delta_i$, and its shell correction, $\delta W_i$, are tabulated based on the mass formula of Koura {\em et al.}~\cite{Koura:2000}. The overall scale $e_0$ is taken as a model parameter but it should be noted that if the shell corrections are negligible, $\delta W_i\approx0$, or the available energy, $U_i$, is large, then $a_i\approx A_i/e_0$, {\em i.e.}\ $a_i$ is simply proportional to the fragment mass number $A_i$, and the energy-dependent renormalization is immaterial. The value determined in Ref.~\cite{Vogt:2012}, $e_0 \sim 10/$~MeV, is used in the present studies.  The level density treatment in \FREYA\ is consistent with that of \CGMF\ and close to the empirical results of Budtz-J\o rgensen and Knitter in Ref.~\cite{Budtz:1988a} 

If the two fragments are in mutual thermal equilibrium, $T_L\!=\!T_H$, the total excitation energy will, on average, be partitioned as above. But because the observed neutron multiplicities suggest that the light fragments tend to be disproportionately excited, the average excitations are modified in favor of the light fragment,
\begin{eqnarray}
\overline{E}^*_L = x \acute{E}^*_L\ ,\ 
\overline{E}^*_H = E_{\rm stat}-\overline{E}^*_L\ ,
\label{eeshift}
\end{eqnarray}
where the adjustable parameter $x$ is expected to be larger than unity. It was found that $x \sim 1.3$ leads to reasonable agreement with $\overline{\nu}(A)$ for $^{252}$Cf(sf), while $x = 1.2$ is suitable for $^{235}$U\nf\ \cite{Vogt:2011}.  

After the mean fragment excitation energies have been assigned as described above, \FREYA\ considers the effect of thermal fluctuations. In Weisskopf's statistical model of the nucleus, which describes the excited nucleus as a degenerate Fermi gas, the mean excitation of a fragment is related to its temperature  $T_i$ by $\overline{E}_i^*=\tilde{a}_iT_i^2$ \cite{Madland:1982,Terrell:1959,Weisskopf:1937} and the associated variance in the excitation is $\sigma_{E_i}^2=-\partial^2\ln\rho_i(E_i)/\partial E_i^2=2\overline{E}_i^*T_i$. Therefore, for each of the two fragments, an energy fluctuation $\delta E_i^*$ is sampled from a normal distribution of variance $2c_T\overline{E}_i^*T_i$ and the fragment excitations is adjusted accordingly, arriving at
\begin{eqnarray}
E_i^*\ =\ \overline{E}_i^*+\delta E_i^*\ ,\ i=L, H .
\end{eqnarray}
The factor $c_T$ multiplying the variance was introduced to explore the effect of the truncation of the normal distribution at the maximum available excitation. Its value affects the neutron multiplicity distribution $P(\nu)$. Previous work used the default value, $c_T = 1.0$; $c_T\geq 1.0$ is expected.

Energy conservation is accounted for by making a compensating opposite fluctuation in the total kinetic energy,
\begin{eqnarray}\label{tkefinal}
{\rm TKE}\ =\ \overline{\rm TKE} - \delta E^*_L - \delta E^*_H\ .
\end{eqnarray}
The average TKE, $\overline{\rm TKE}$, has been adjusted by $d$TKE to reproduce the average neutron multiplicity, $\overline \nu$. 

The evaporated neutrons are assumed to be isotropic in the frame of the emitting nucleus, apart from a very slight flattening due to the nuclear rotation. Their energy is sampled from a black-body spectrum, $dN_{n}/dE_{n}\sim E_{n}\exp(-E_{n}/T_{\rm max})$, where $T_{\rm max}$ is the maximum possible temperature in the daughter nucleus, corresponding to emission of a very soft neutron \cite{Randrup:2009}.  

\FREYA\ generally assumes that neutron evaporation continues until the nuclear excitation energy is below the threshold $S_n + Q_{\rm min}$, where $S_{n}$ is the neutron separation energy and $Q_{\rm min}$ the energy above the neutron separation threshold where photon emission takes over from neutron emission. The value of $Q_{\rm min}$ is fixed at 0.01~MeV so that neutron evaporation continues as long as energetically possible, independent of angular momentum \cite{Vogt:2017}.  Neutron emission is treated relativistically in \FREYA.

After neutron evaporation has ceased, the excited product nucleus will emit photons sequentially.  This emission is treated in several stages. The most recent version of \FREYA\ uses the RIPL-3 data library \cite{RIPL3} for the discrete decays towards the end of the decay chain.

The first stage is statistical radiation. These photons are emitted isotropically with an energy sampled from a black-body spectrum modulated by a giant dipole resonance, GDR, form factor,
\begin{eqnarray}
{dN_\gamma \over dE_\gamma}\ \sim\ {\Gamma_{\rm GDR}^2 E_\gamma^2 
\over (E_\gamma^2-E_{\rm GDR}^2)^2 + \Gamma_{\rm GDR}^2E_\gamma^2 }\,
	      E_\gamma^2\,{\rm e}^{-E_\gamma/T_{\rm }}\ .
\end{eqnarray}
The position of the resonance is $E_{\rm GDR}/{\rm MeV}=31.2/A^{1/3}+20.6/A^{1/6}$ \cite{Berman:1975}, while its width is $\Gamma_{\rm GDR}=5$~MeV. It is assumed that each emission reduces the magnitude of the angular momentum by $dS$, the standard value being $dS=1\,\hbar$.

The RIPL-3 library \cite{RIPL3} tabulates a large number of discrete electromagnetic transitions for nuclei throughout the nuclear chart, but complete information is available for only relatively few of them. However, by invoking certain assumptions, see Ref.~\cite{Vogt:2017}, it is possible to construct, for each product species, a table of the possible decays from the lowest discrete levels, {\em i.e.}\ the level energies $\{\varepsilon_\ell\}$, their half-lives $\{t_\ell\}$, and the branching ratios of their various decays. Then whenever the decay process described above leads to an excitation below any of the tabulated levels, \FREYA\ switches to a discrete cascade based on the RIPL-3 data. The discrete cascade is continued until the half-life $t_\ell$ exceeds the specified value of $t_{\rm max}$. When comparing with experimental data, $t_{\rm max}$ should be adjusted to reflect the response time of the detection system.  If the RIPL-3 tables do not include any transitions, in this case \FREYA\ allows statistical excitation until near the yrast line and the remaining de-excitation occurs by emission of 
``collective'' \grays\ that each reduce the angular momentum by $2\,\hbar$.  When the \gray\ energy is below $g_{\rm min}$, that \g~ray is not registered in \FREYA\ and does not count toward the total multiplicity.

\subsection{Transport codes} \label{sec:transportCodes}

\subsubsection{Overview}

The ultimate goal of advanced computational physics is to use physical data and numerical algorithms to simulate and predict the behavior of nature.  More specifically, radiation transport codes intend to model the detailed behavior of radiated particles as they interact with various materials.  In order for radiation transport codes to predict natural phenomena as accurately as possible, many details need to be realized: physical assumptions and numerical simplifications need to be minimized; nuclear data-like cross sections need to be well understood; and relevant experimental data are needed for proper validation in various application areas. While this list is not exhaustive, it does include the most basic components of an accurate and robust radiation transport code.

To address the first point, the Monte Carlo method is widely considered to be the radiation transport method with the fewest physical and numerical approximations.  Because the approximations related to the spatial, angular and energy variables associated with the state of a particle are essentially negligible in continuous-energy Monte Carlo codes like \MCNP, comparisons to a variety of experimentally measured quantities are feasible.

Nuclear data and validation are addressed in this paper. Historically, \MCNP~has been used for many applications including, but not limited to, radiation shielding and protection, reactor physics and design, and nuclear criticality safety.  A report on \MCNP~verification and validation is issued annually by the developers of \MCNP~\cite{Brown:2017}, and provides details on the extensive verification and validation work necessary to ensure the trust in \MCNP~results in these applications. A similar goal exists for the new event-by-event nuclear fission models (\FREYA\ and \CGMF) introduced in \MCNP~for use in nonproliferation and safeguards applications.

\subsubsection{Nuclear fission physics models in \MCNP6}

The latest release of \MCNP6.2 not only includes the two new correlated fission multiplicity models, \CGMF~and \FREYA, but it also includes fission multiplicity options dating back to many of the previous releases, including \MCNP6.1.1~\cite{MCNP611}.  

In \MCNP, the bounded integer sampling scheme is employed by default to simulate secondary neutrons emitted from fission reactions.  In this scheme, given the average number of neutrons emitted in fission, \nubar, when a fission event occurs the number of neutrons emitted is either the integer number $n=\lfloor \bar{\nu} \rfloor$ or $n+1$.  The probabilities for selecting $n$ and $n+1$ are chosen to preserve the expected value of \nubar.  In the case of thermal neutron-induced fission of $^{235}$U for instance, \nubar=2.42, and only $\nu$ values of 2 and 3 are sampled.

Similarly, the production of photons from neutron interactions is done using the ratio of the photon production cross section, taking into account photon production interaction probabilities and photon yields, to the total interaction cross section.  This ratio is the expected value of the number of photons produced per interaction at a given incident neutron energy.  In general, if this ratio is small, then \MCNP~uses Russian roulette to determine if a single photon is produced.  If this ratio is large, \MCNP\ produces a few photons (less than ten) and gives a higher weight to each of these photons to preserve the overall expected photon production rate~\cite{MCNP5}.

By enforcing the expected number and/or weight of the fission neutrons and photons produced, the expected values of quantities such as the flux, reaction rates, and criticality, $k_{\rm eff}$, are also maintained.  However, if the objective is to analyze the event-by-event nature of these reactions, such as simulating the behavior of neutron multiplicity counters, the detailed microscopic behavior of the particles (neutrons and photons) emitted during fission is needed.

With the release of \MCNP6, the capabilities in both \MCNP5~\cite{MCNP5} and \MCNPX~\cite{MCNPX} were merged so that users have the ability to select various fission multiplicity treatments. Several sampling algorithms are available to sample from a Gaussian distribution for a given isotope based on data~\cite{Ensslin:1998,Terrell:1957,Santi:2005,Lestone:2005}. While the \MCNP~input options do offer flexibility in simulating spontaneous and neutron-induced prompt fission neutron multiplicities, it is limited in its use in applications due to a variety of assumptions.  First, the direction of travel of each neutron emitted in fission is independently sampled from an isotropic distribution in the laboratory frame.  Next, the energy of neutrons emitted in each fission event is sampled independently from the same average PFNS.  And finally, these specific features do not allow simulation of \gray~multiplicities in spontaneous or neutron-induced fission.

These limitations were somewhat lifted with the implementation of the LLNL Fission Library version 1.8~\cite{Verbeke:2010} in \MCNPX\ and included in the recent releases of \MCNP6. Before the Library was included, all photons produced from all neutron reaction channels in \MCNP\ were sampled prior to the selection of the neutron reaction itself.  While this does not bias the calculation of integral quantities such as flux and $k_{\rm eff}$, it is still impossible to simulate fission event-by-event.

While the LLNL Fission Library \cite{Verbeke:2010} addressed a few of the limitations of the standard \MCNP~multiplicity treatments, the issues with missing event-by-event energy, angular and particle correlations remained.  With the explicit Monte Carlo modeling of the fission process done in both \FREYA~and \CGMF, these last concerns are finally addressed.

\subsubsection{The \POLIMI~code}

The \POLIMI~extension to \MCNPX\ was developed to better simulate coincidence measurements and subsequent time analyses~\cite{Pozzi:2003}. The \polimi~code includes built-in correlations for key spontaneously fissioning isotopes ($^{252}$Cf, $^{238}$U, $^{238,240,242}$Pu, $^{242,244}$Cm) and for \MCNPX-supported induced fission, event-by-event tracking, and conservation of energy and momentum on an event-by-event basis.

\POLIMI~has the option to track and record event information collision-by-collision in specified detector regions. For each collision, key information is recorded: history number, particle number, particle type, collision type, target nucleus, collision cell, and collision time. This recorded collision information can be used to accurately model nonlinear detector responses event-by-event. 

Built-in spontaneous fission sources have prompt neutron multiplicity distributions and multiplicity-dependent neutron energy spectra. As the emitted neutron multiplicity increases, the neutron energy spectrum softens~\cite{Pozzi:2012}. The \POLIMI~algorithm uses the average light and heavy fragment masses of each fissioning isotope to impart momentum from the fragments to the emitted neutrons.

These built-in spontaneous fission sources also have prompt \gray~multiplicity distributions. The $^{252}$Cf photon energy is sampled from the spectrum evaluated by Valentine~\cite{Valentine:1996}. All other isotopes are sampled from a $^{235}$U evaluation. Photons are emitted isotropically.

Both neutrons and photons are generated from independent but full multiplicity and evaluated energy distributions for induced fission.

\subsubsection{Implementation of event-by-event models in \mcnp}

In the \MCNP 6.2 release, 
both the \FREYA~and \CGMF~fission event generators are included.  In the most recent prior version of \MCNP, $\mathtt{MCNP}$6.1.1~\cite{MCNP6,MCNP611}, two low-energy neutron-photon multiplicity packages were released: the LLNL Fission Library~\cite{Verbeke:2010} and the Cascading Gamma-ray Multiplicity (\CGM) code from LANL~\cite{Kawano:2010}.  Version 1.8 of the LLNL Fission Library \cite{Verbeke:2010} included neutron and photon multiplicity distributions but did not include any correlations between emitted particles by default.  Likewise, the released version of the \CGM~code handles a variety of reactions, but does not include particle emission from fission reactions.  The newest versions of these event generators, to be included in the \MCNP6.2 release, are significantly improved over their predecessors by addressing some of these immediate deficiencies, as described below.

The main \MCNP6.2 source code remains separate from the event-by-event source codes.  A clean interface was developed to call the necessary routines and pass the information to and from \MCNP. In the worst case scenario, the interface caused an overhead of less than 1\% on the total computation time. In any realistic transport calculation using these fission event generator models through the new interface, the added computational cost of the interface will be negligible due to the usual amount of computational time used in Monte Carlo codes tracking particles and looking up cross sections.

\begin{figure}[ht]
\centering
\includegraphics[width=\columnwidth]{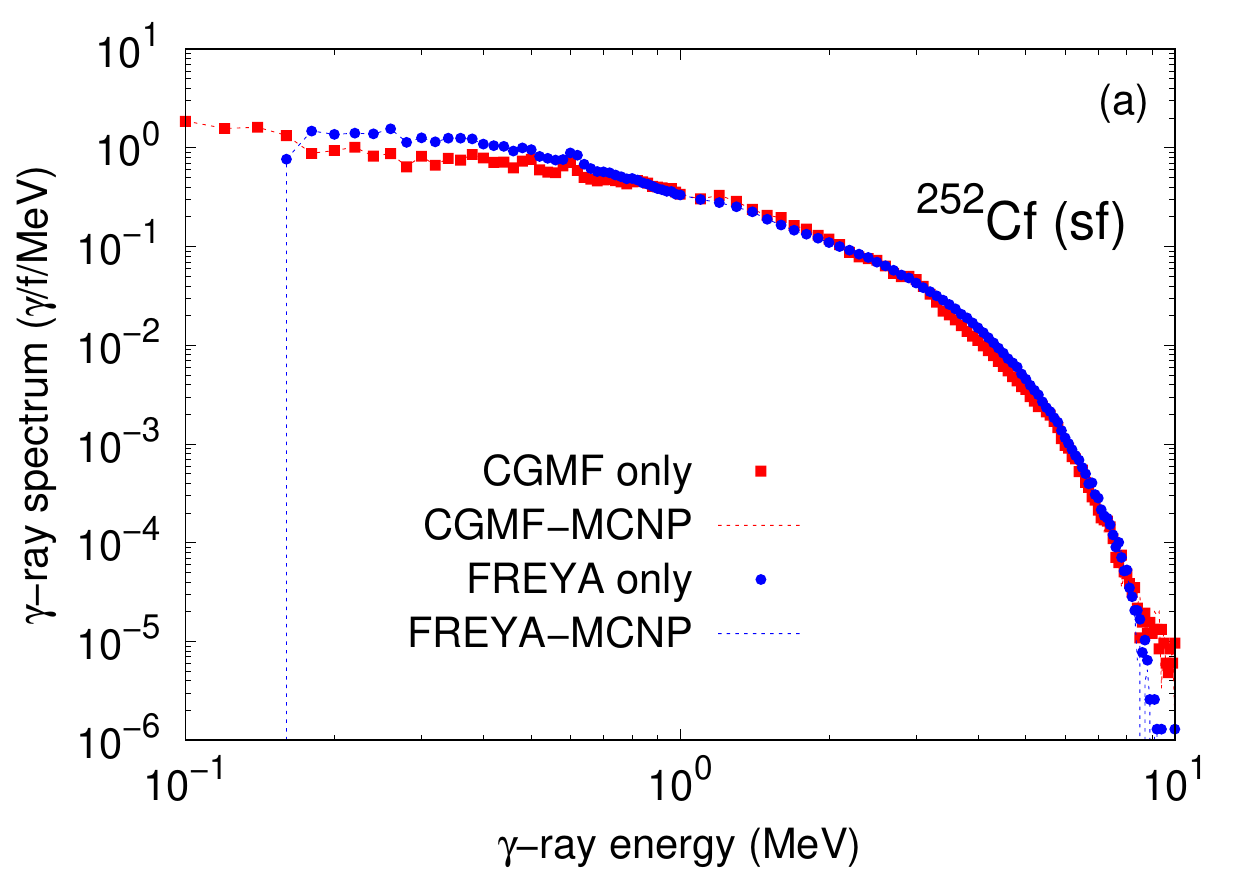} 
\includegraphics[width=\columnwidth]{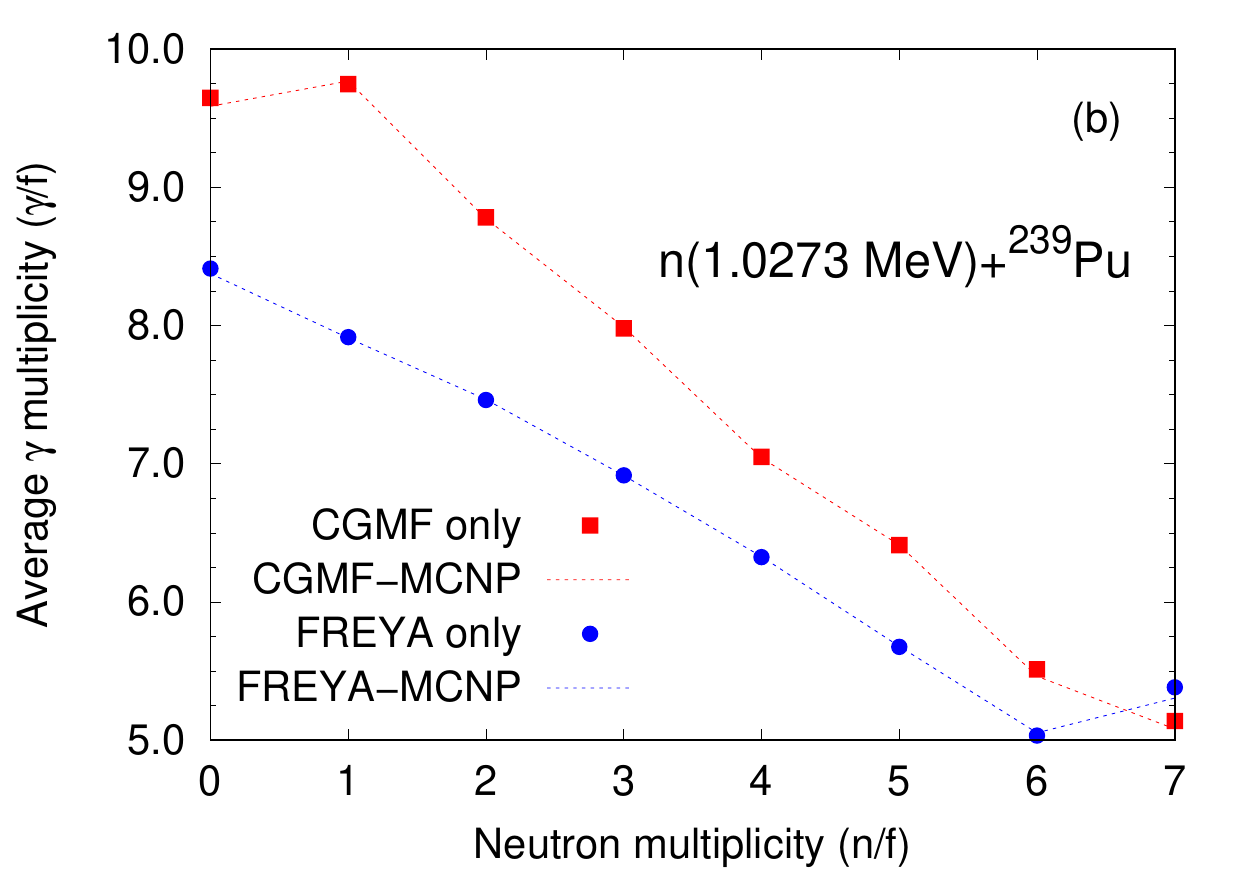} 
\caption{(Color online) The \gray~spectra of the $^{252}$Cf spontaneous fission reaction (a) calculated using \MCNP6, \FREYA~and \CGMF. The average photon multiplicity, $N_\gamma$, as a function of the neutron multiplicity, $\nu$, using \MCNP6, \FREYA\ and \CGMF\ (b).}
\label{fig:standalone}
\end{figure}

As part of the routine \MCNP~code-integration strategy, several tests were performed to check that the integrated and standalone versions of the \FREYA~and \CGMF~codes provide equivalent results to an appropriate numerical accuracy. The following quantities were checked: average multiplicities, \nubar~and \nubarg, and energies, $\bar{\chi}_n$ and $\bar{\chi}_\gamma$; multiplicity distributions, $P(\nu)$ and $P(N_\gamma)$, and correlations in $n$-$\gamma$ multiplicities, $P(\nu,N_\gamma)$, and $n$-$n$ emission angles, $\vec{\Omega}_n\cdot\vec{\Omega}_{n'}$. Each simulation included approximately $10^6$ fission events. While there are sometimes significant differences between the calculated results of \CGMF~and \FREYA, in particular for \gray~average energies and spectra, there is very good agreement between the results from the standalone codes and their \MCNP-integrated counterparts, as shown in Fig.~\ref{fig:standalone} for the average \gray~spectrum of $^{252}$Cf(sf) (a) and $n$-$\gamma$ multiplicity correlations in the $n$(1.0273~MeV)+$^{239}$Pu fission reaction (b). The latter quantity represents an interesting correlation predicted by these models. Increasing the neutron multiplicity results in a decrease in the average photon multiplicity, as shown in Fig.~\ref{fig:standalone}(b).  There is extremely good agreement between standalone and integrated codes.  This shows that \MCNP~does generate the negative correlation between the neutron and photon multiplicities produced by both \FREYA~and \CGMF.  

\begin{figure}[ht]
\centering
\includegraphics[width=\columnwidth]{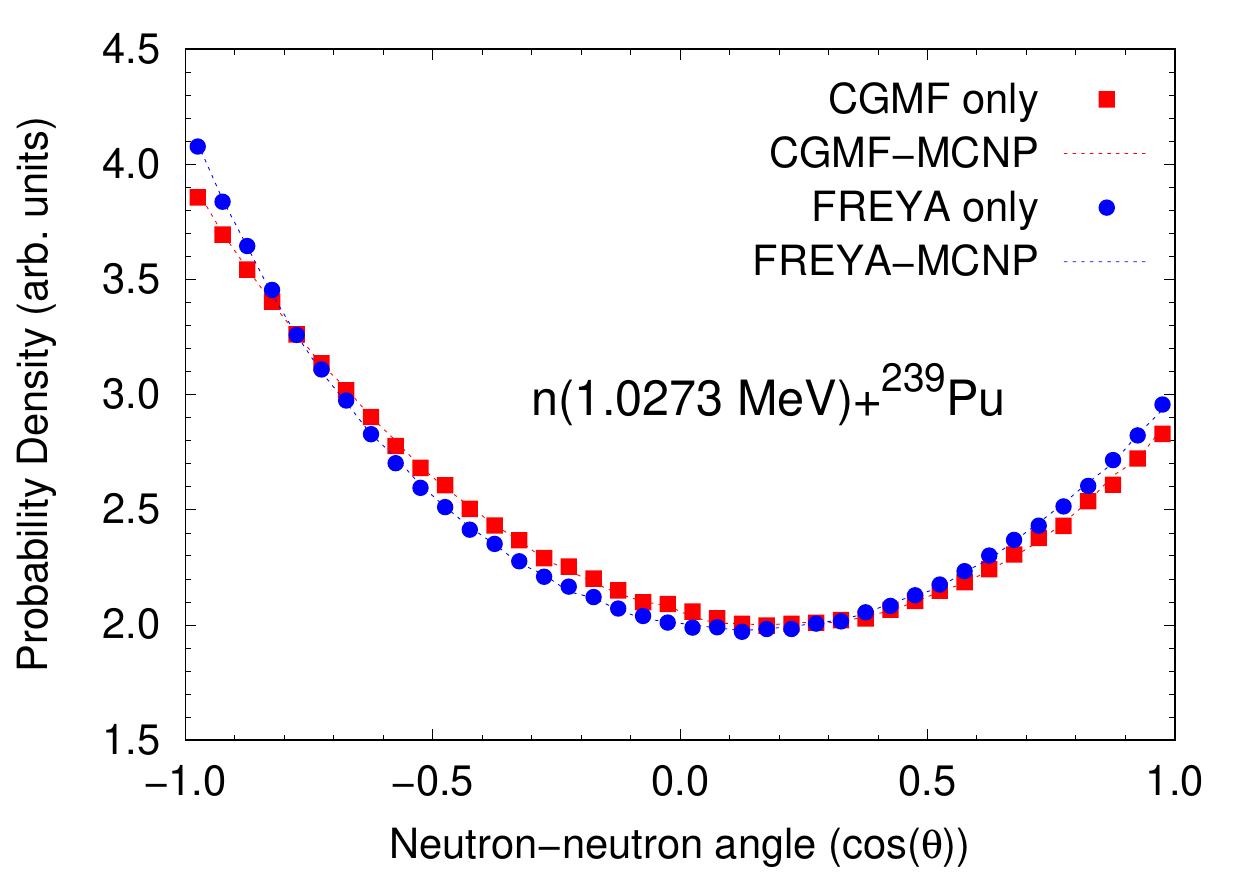} 
\caption{(Color online) The neutron-neutron angular correlations for the $n$(1.0273~MeV)+$^{239}$Pu fission reaction calculated using \MCNP6, \FREYA~and \CGMF.}
\label{fig:PuAng}
\end{figure}

As a final verification test, the neutron-neutron angular correlations observed in these fission event generator models are shown in Fig.~\ref{fig:PuAng} for $n$(1.0273~MeV)+$^{239}$Pu. Note that these quantities are not readily available from \MCNP6 in any standard output or tallies, but can only be computed by analyzing the list-mode data instead.

\section{Simulation results} \label{sec:simulations}

Here results of simulations with \CGMF\ and \FREYA\ are shown for some fission observables and, where possible, compared to the data presented in Sec.~\ref{sec:data}.  First the code results are compared for observables that depend on fragment mass and kinetic energy.  Next, multiplicity-dependent spectral results are shown, followed by calculations of multiplicity distributions, for both neutron and $\gamma$ emission.  The results presented here from \CGMF\ and \FREYA\ use the inputs employed in the public versions of the codes unless otherwise noted.

The remainder of the section is devoted to correlations. First, $\gamma$-$n$ multiplicity correlations are discussed.  The following two subsections are devoted to neutron-light fragment and neutron-neutron angular correlations.  The last parts deal with time dependence of the results, first the dependence of $\gamma$ emission on the time coincidence window of the detector, followed by a discussion of time-chain correlations in multiplicity counting.  This last topic, while not pertaining only to a single event, is included here because of its importance to nuclear assay applications.

\subsection{Dependence on fission fragment mass and kinetic energy}

\begin{figure}[ht]
\centerline{\includegraphics[width=\columnwidth]{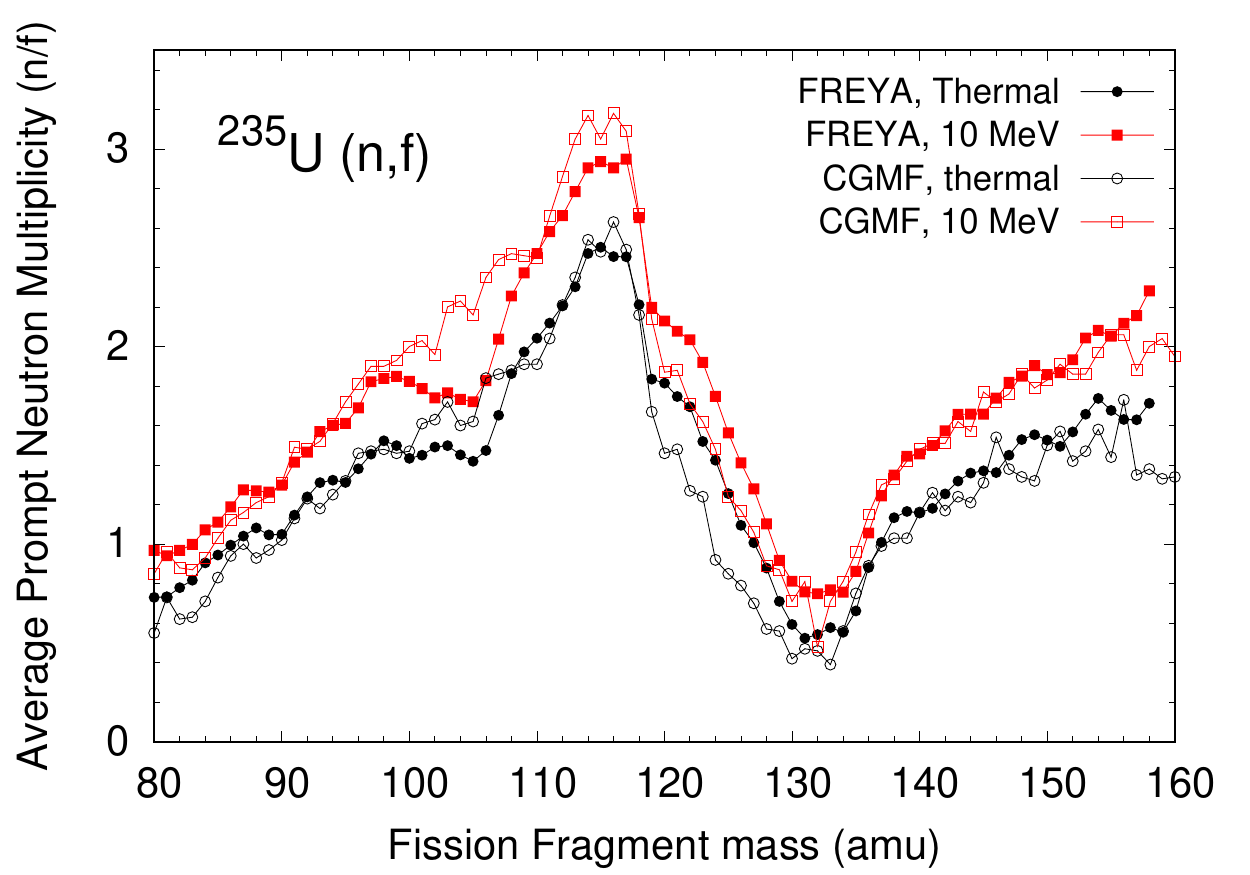}}
\caption{\label{fig:nubarA-U235-En}(Color online) Average prompt neutron multiplicity as a function of fragment mass for the neutron-induced fission reaction on $^{235}$U for thermal and 10~MeV neutrons, calculated by \FREYA\ and \CGMF.}
\end{figure}

The average neutron multiplicity as a function of fragment mass, \nubar$(A)$, calculated by \CGMF~and \FREYA~for several incident neutron energies in the $^{235}$U\nf\ reaction are shown in Fig.~\ref{fig:nubarA-U235-En}. Near mass $A= 132$, characteristics of both neutron ($N=80$) and proton ($Z=52$) spherical shell closures, the average number of emitted neutrons reaches its minimum. There, the expected extra collective energy due to the deformation of the fragments near scission compared to their ground-state configuration is expected to be very small. On the contrary, the complementary fragment, near mass 104, is very elongated. The extra deformation energy will transform into an additional intrinsic excitation energy in the fragments after scission, eventually leading to the release of more prompt neutrons.

The overall agreement between \CGMF\ and \FREYA\ is rather good.  There is some discrepancy between the two results for $A$ between 100 and 110, where \FREYA\ emits fewer neutrons, and in the region between symmetry and $A = 132$ where \FREYA\ emits slightly more neutrons.

The evolution of this dependence as a function of excitation energy has been the focus of several studies in recent years~\cite{Schmidt:2010,Morariu:2012}, although only limited and indirect experimental data exist. Those limited data sets, however, indicate that as the total excitation energy in the fragments increases, the average neutron multiplicity increases almost solely in the heavy fragments. At even higher energies, the situation becomes somewhat more complex since more and more neutrons are evaporated from the compound nucleus prior to fission, in the multi-chance fission process.  The pre-fission neutrons cannot be attributed to either one of the fragments since they are not associated with the fragments.  For example, at 10~MeV, $\sim 0.6$ neutrons on average are emitted from the $^{236}$U compound nucleus prior to fission in \FREYA. 
Upcoming versions of the \CGMF\ and \FREYA\ codes will address this important question more thoroughly in the near future.

\begin{figure}[ht]
\centerline{\includegraphics[width=\columnwidth]{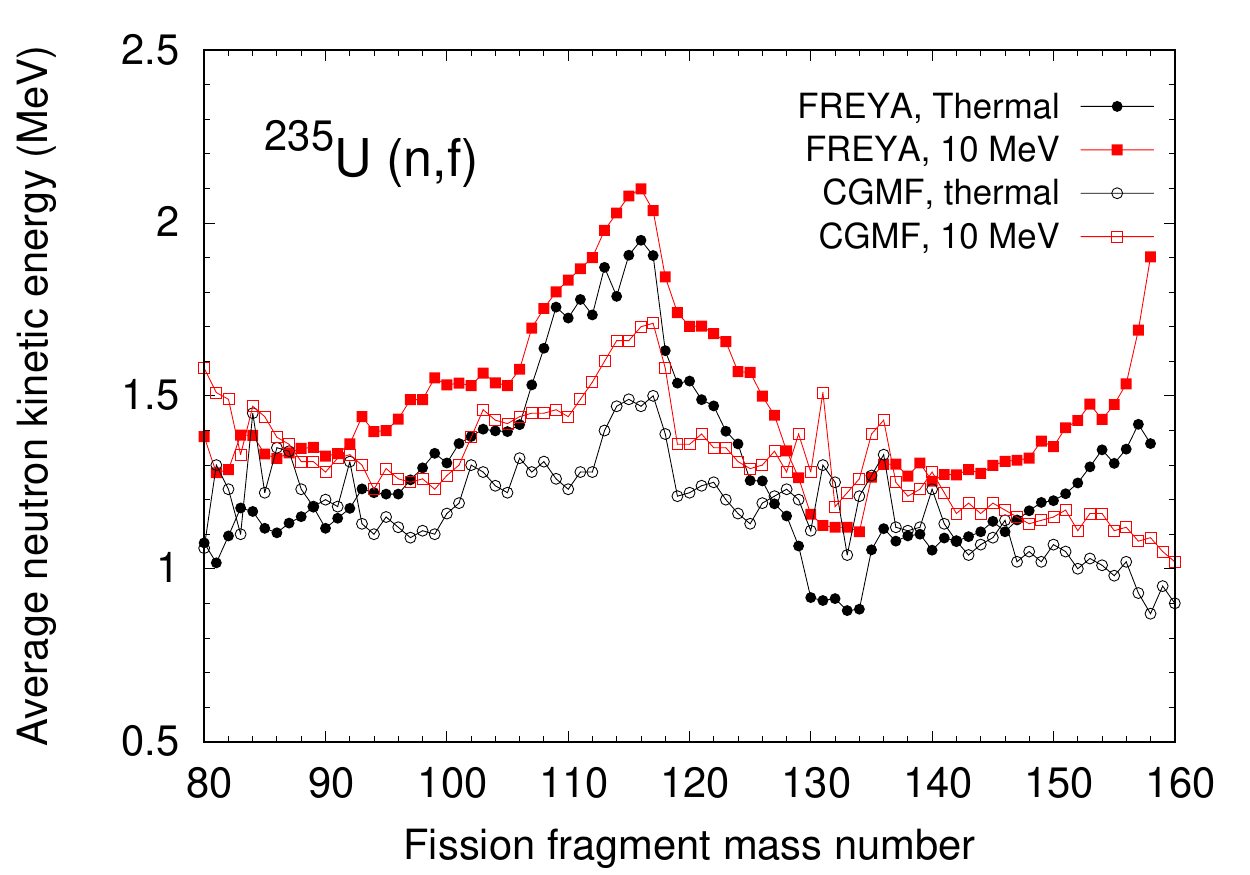}}
\caption{\label{fig:EcmbarA-U235-En}(Color online) The average center-of-mass neutron energy as a function of fragment mass for $^{235}$U\nf\ with thermal and 10~MeV neutrons, calculated by \FREYA\ and \CGMF.}
\end{figure}

The average neutron kinetic energy in the center-of-mass neutron energy as a function of the fragment mass has been measured for several fission reactions, as shown in Fig.~\ref{fig:Adep}(b). The \CGMF~and \FREYA\ results predicted for $^{235}$U\nf\ are shown in Fig.~\ref{fig:EcmbarA-U235-En}.
The dependence of the neutron kinetic energies on $E_{\rm inc}$ is similar to that for $\nu(A)$.  The $A$ dependence shows a similar trend but the overall average energy is somewhat higher for \FREYA.

\begin{figure}[ht]
\centerline{\includegraphics[width=\columnwidth]{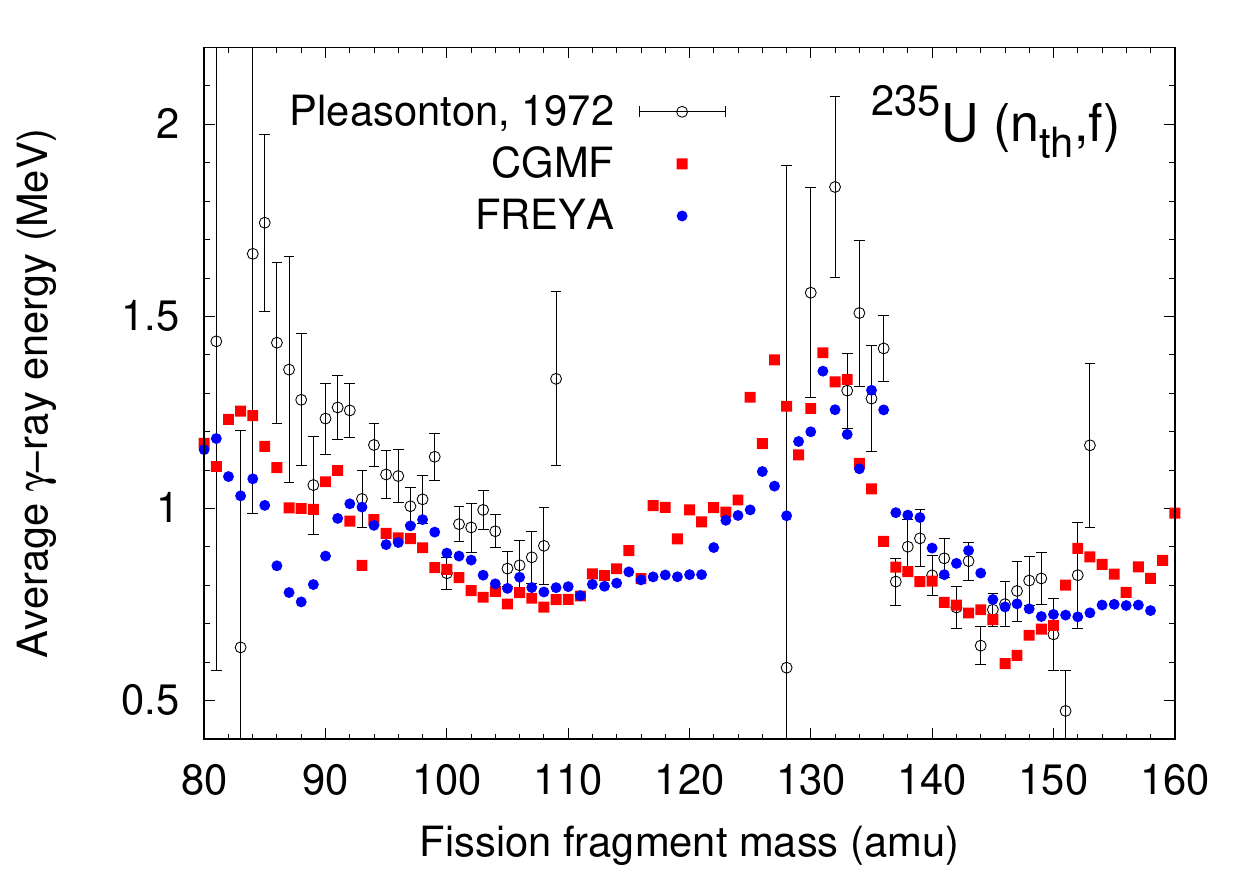}}
\caption{\label{fig:EgAcalc}(Color online) The average prompt fission \gray~energy as a function of the fission fragment mass in the thermal neutron-induced fission reaction on $^{235}$U. Experimental data are from Pleasonton \etal~\cite{Pleasonton:1972}.}
\end{figure}

Prompt \gray~characteristics as a function of the fragment mass are also very interesting to study as they provide complementary information and constraints on the physics models of the fission event generators. In particular, the $\gamma$-$n$ competition is governed by the distribution of angular momentum in the fragments. The average prompt \gray~energy per photon, $\langle \epsilon_\gamma \rangle$, as a function of fragment mass is shown in Fig.~\ref{fig:EgAcalc} for $^{235}$U\nfth. The significant increase of $\langle \epsilon_\gamma\rangle$ for masses near 132 can be explained by the lower density of levels in these near-spherical fragments, thereby increasing the average energy of the \g~transitions between excited levels. This result should be approximately independent of the fissioning nucleus since it depends on the characteristics of the fragments themselves. However, different fission fragment mass yields as observed in different fission reactions will lead to different hardness of the \g~spectrum, from the convolution of $Y(A)$ with $\overline{\epsilon}_\gamma(A)$ shown in Fig.~\ref{fig:EgAcalc}.

As noted earlier, the \gray~multiplicity is very sensitive to the energy threshold, $g_{\rm min}$ in \FREYA, and fission time coincidence window of the detector, $t_{\rm max}$ in \FREYA.  The calculations in Fig.~\ref{fig:EgAcalc} were made with the values for these quantities given in Ref.~\cite{Pleasonton:1972}, $g_{\rm min} = 0.09$~MeV and $t_{\rm max} = 5$~ns.  Given the sensitivity of the \gray~multiplicity to these quantities, the average $\gamma$-ray energy shown in Fig~\ref{fig:EgAcalc} also has some sensitivity to $g_{\rm min}$ even though the total \gray~energy is almost insensitive to the cutoff of most experiments.

\begin{figure}[ht]
\centerline{\includegraphics[width=\columnwidth]{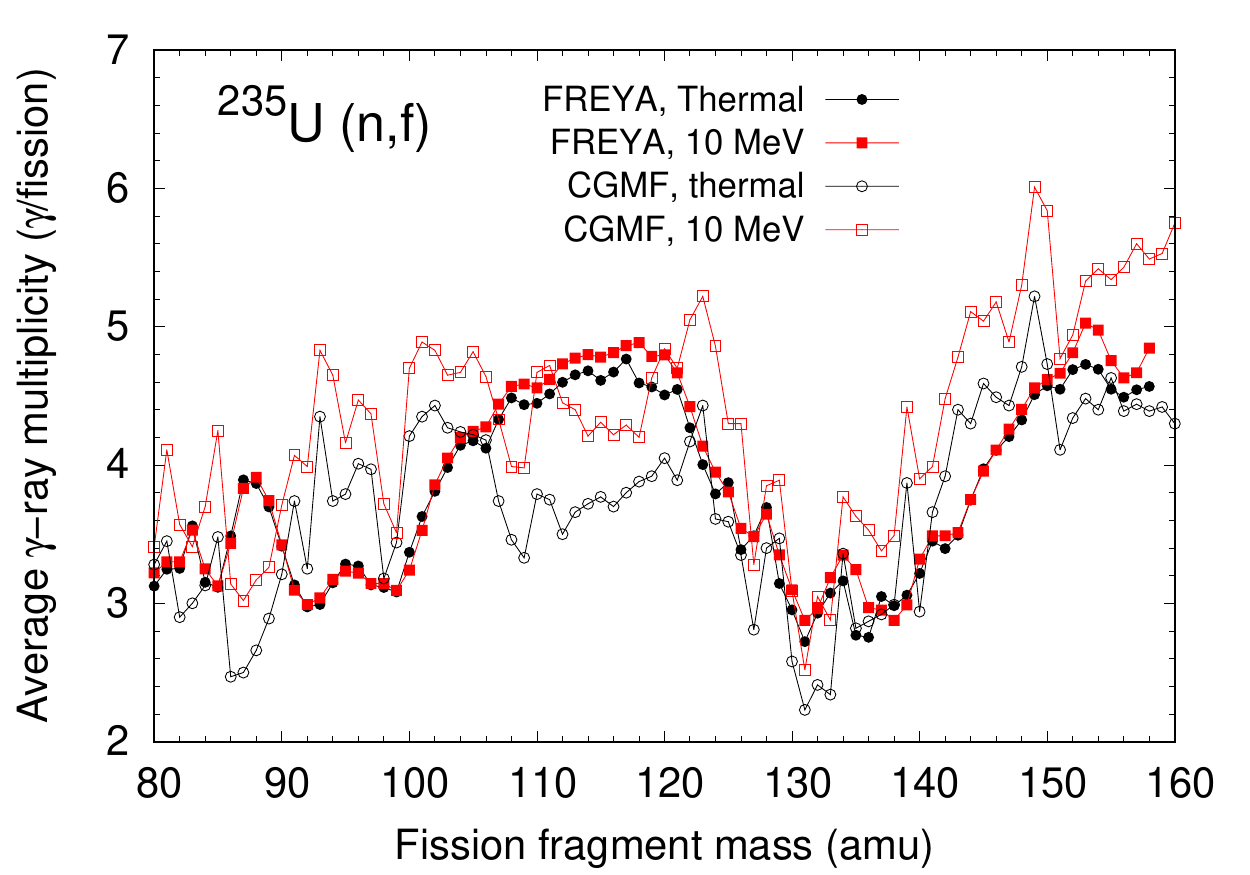}}
\caption{\label{fig:NgAcalc}(Color online) The average prompt fission \gray~multiplicity as a function of the fission fragment mass, for thermal and 10~MeV incident neutrons in the $^{235}$U\nf\ reaction, as calculated with \FREYA\ and \CGMF.}
\end{figure}

The situation with the average prompt fission \gray~multiplicity as a function of fragment mass, is less clear however, as comparable experiments provide somewhat inconsistent results due to the different energy thresholds and time windows for different detectors.  The \FREYA\ calculations shown in Fig.~\ref{fig:NgAcalc} use $g_{\rm min} = 0.1$~MeV and $t_{\rm max} = 10$~ns.  The results are essentially independent of $E_{\rm inc}$ because, in \FREYA, neutron emission continues until the neutron separation energy, $S_n$, is reached.  Thus nearly the same residual excitation energy is left for $\gamma$ emission in \FREYA, regardless of $E_{\rm inc}$. A very similar conclusion is reached with \CGMF, which treats the $n$-$\gamma$ competition slightly differently.  The two results are rather similar as a function of $A$ except for $A<105$ where the \CGMF\ multiplicity is higher.

\begin{figure}[ht]
\centerline{\includegraphics[width=\columnwidth]{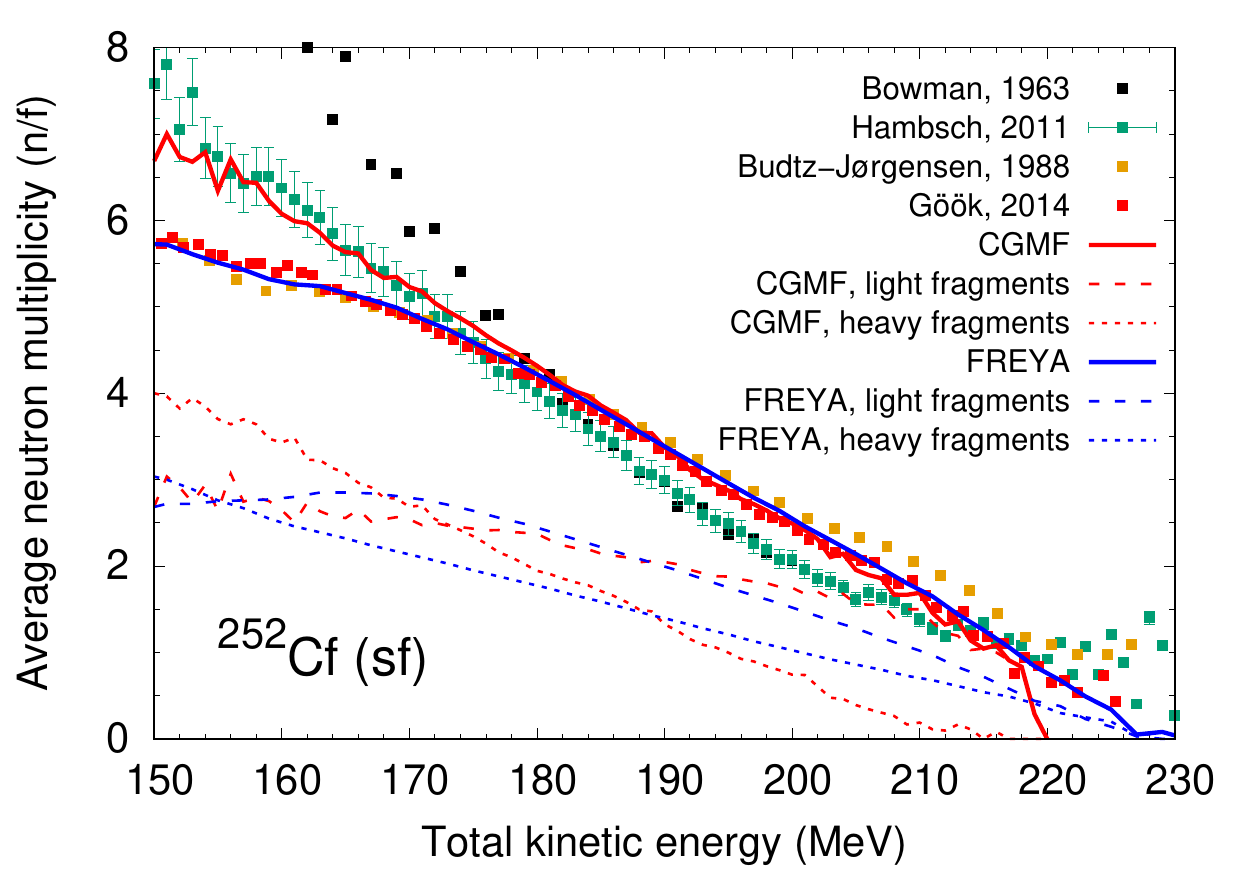}}
\caption{\label{fig:nuTKE}
(Color online) The average prompt fission neutron multiplicity as a function of the TKE of the fission fragments for $^{252}$Cf(sf). The experimental data are from Bowman \etal~\cite{Bowman:1963}, Hambsch\etal~\cite{Hambsch:2011}, Budtz-J\o rgensen and Knitter~\cite{Budtz:1988a}, and G\``o\``ok \etal~\cite{Gook:2014}.}
\end{figure}

Measurements of the evolution of prompt neutron and \gray~emission data as a function of the total kinetic energy of the fission fragments have been reported in a few fission reactions. The slope $d$\nubar$/d$TKE is an indicator of how much energy is required to emit a neutron. Figure~\ref{fig:nuTKE} shows \nubar~as a function of TKE for several $^{252}$Cf(sf) measurements. They all exhibit a rather linear behavior in the range $180 \,\, {\rm MeV} \, < {\rm TKE} < 220$~MeV. At energies ${\rm TKE} \, >220$~MeV, statistics are low since very little excitation energy is left for neutron emission.  At ${\rm TKE} < 180$~MeV, the measurements diverge.  It has been suggested that the flatter low TKE behavior exhibited by some of the experiments is due to a nonlinear dependence of the average fission $Q$ value with TKE.

\begin{figure}[ht]
\centerline{\includegraphics[width=\columnwidth]{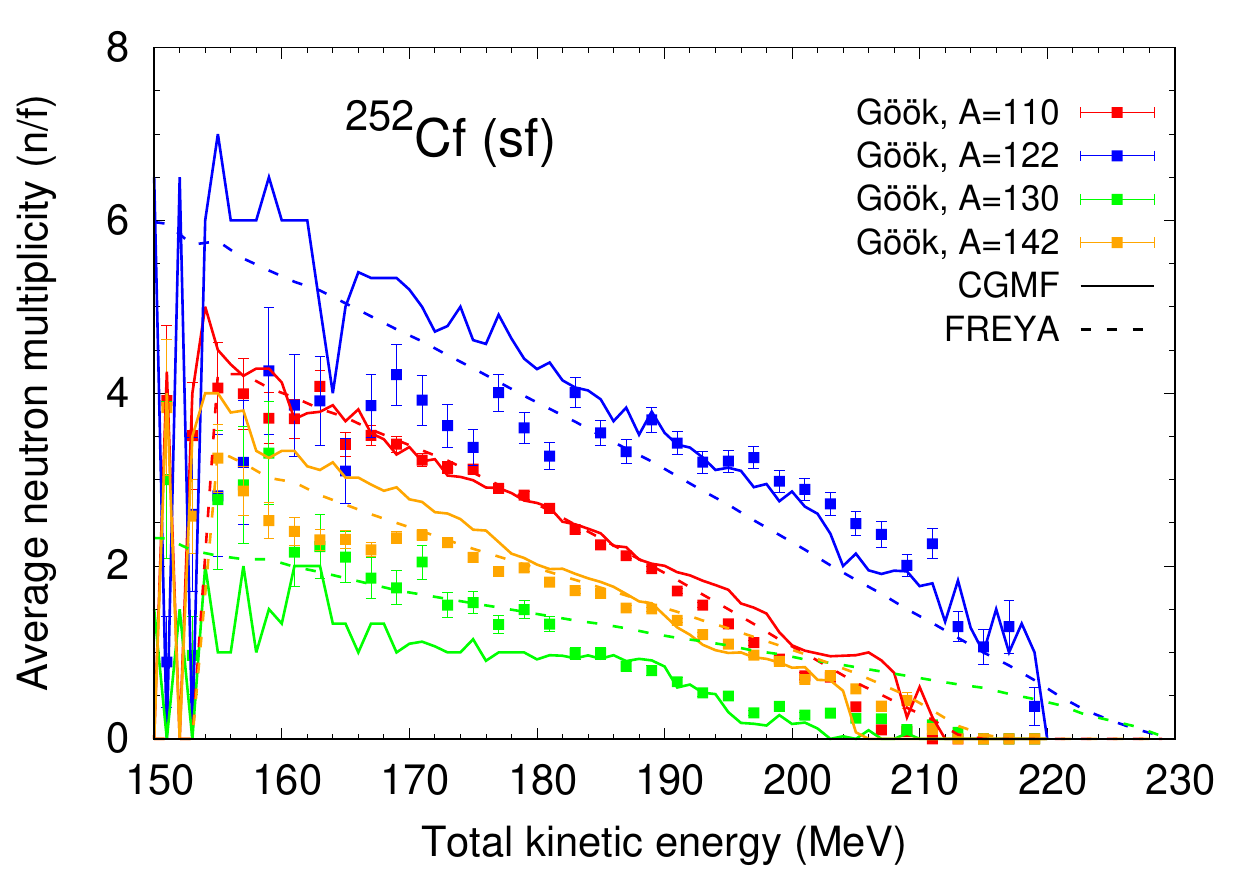}}
\caption{\label{fig:nuTKE-selectA}(Color online) Same as Fig.~\ref{fig:nuTKE} but for selected fragment masses, compared to experimental data by G\"o\"ok \etal~\cite{Gook:2014}.}
\end{figure}

The neutron multiplicity as a function of TKE has also been measured recently~\cite{Gook:2014} as a function of fragment mass, as shown in Fig.~\ref{fig:nuTKE-selectA}.  The pattern corresponds to the sawtooth shape of $\overline \nu(A)$ with the largest \nubar(TKE) for $A = 122$, near the top of the 
sawtooth for $^{252}$Cf(sf), as seen in Fig.~\ref{fig:Adep} where as $A = 110$ and $A = 142$ are masses where the 
sawtooth is rising.  Finally, $A = 130$ is near the doubly-closed shell, near the minimum of \nubar$(A)$, giving the lowest \nubar(TKE).  This behavior gives some insight into how much energy is needed to emit a neutron for a given fragment mass and deformation. 

The \FREYA\ results for $A = 110$ and 142 are in good agreement with the data since, here, the agreement between \FREYA\ and data on $\overline \nu(A)$ is also very good.  At $A = 122$ and 130, however, \FREYA\ over- and underestimates $\overline \nu(A)$ respectively with the single-valued parameter $x$ governing the excitation energy sharing.  The \CGMF\ agreement is closer overall since it uses a mass dependent parameter, 
$R_T(A)$, to match $\overline \nu(A)$.

\begin{figure}[ht]
\centering
\includegraphics[width=\columnwidth]{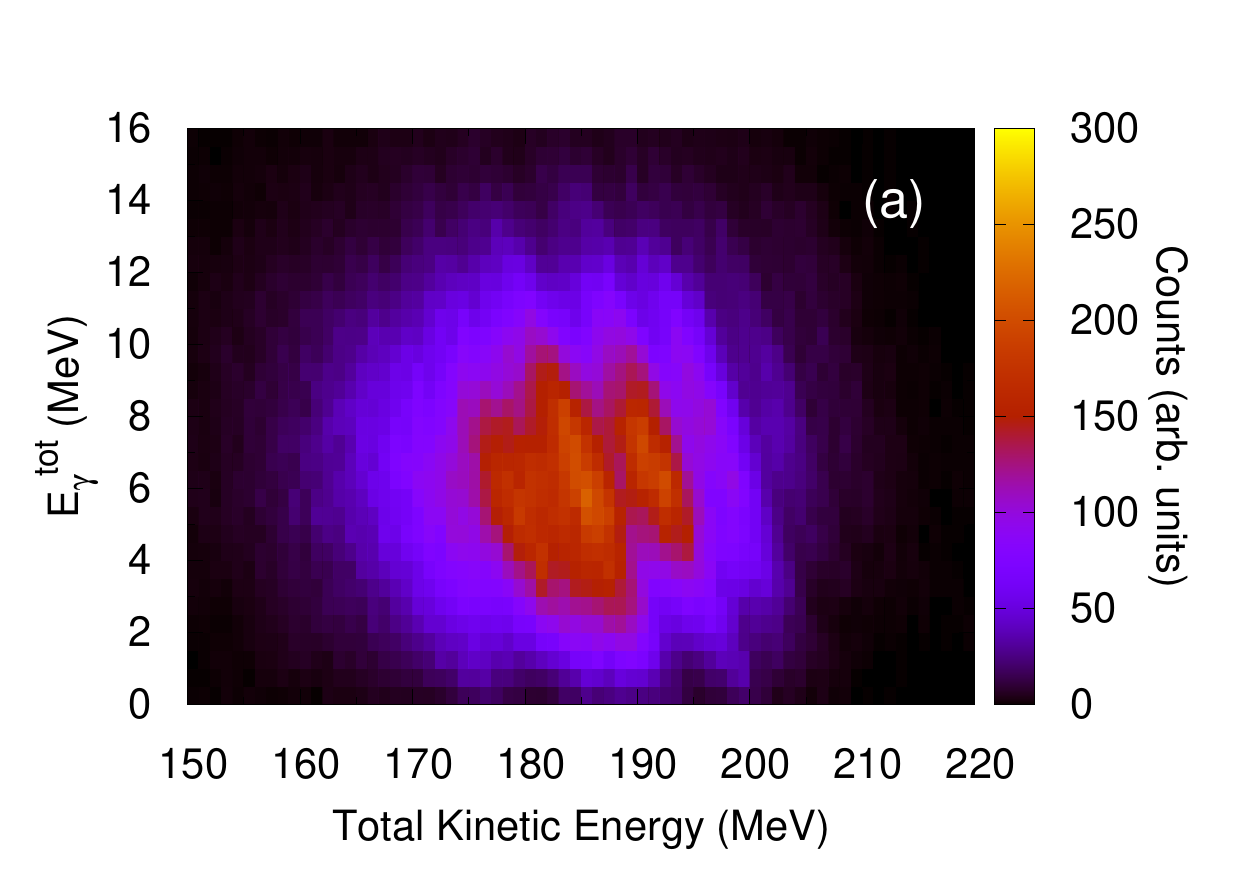}
\includegraphics[width=\columnwidth]{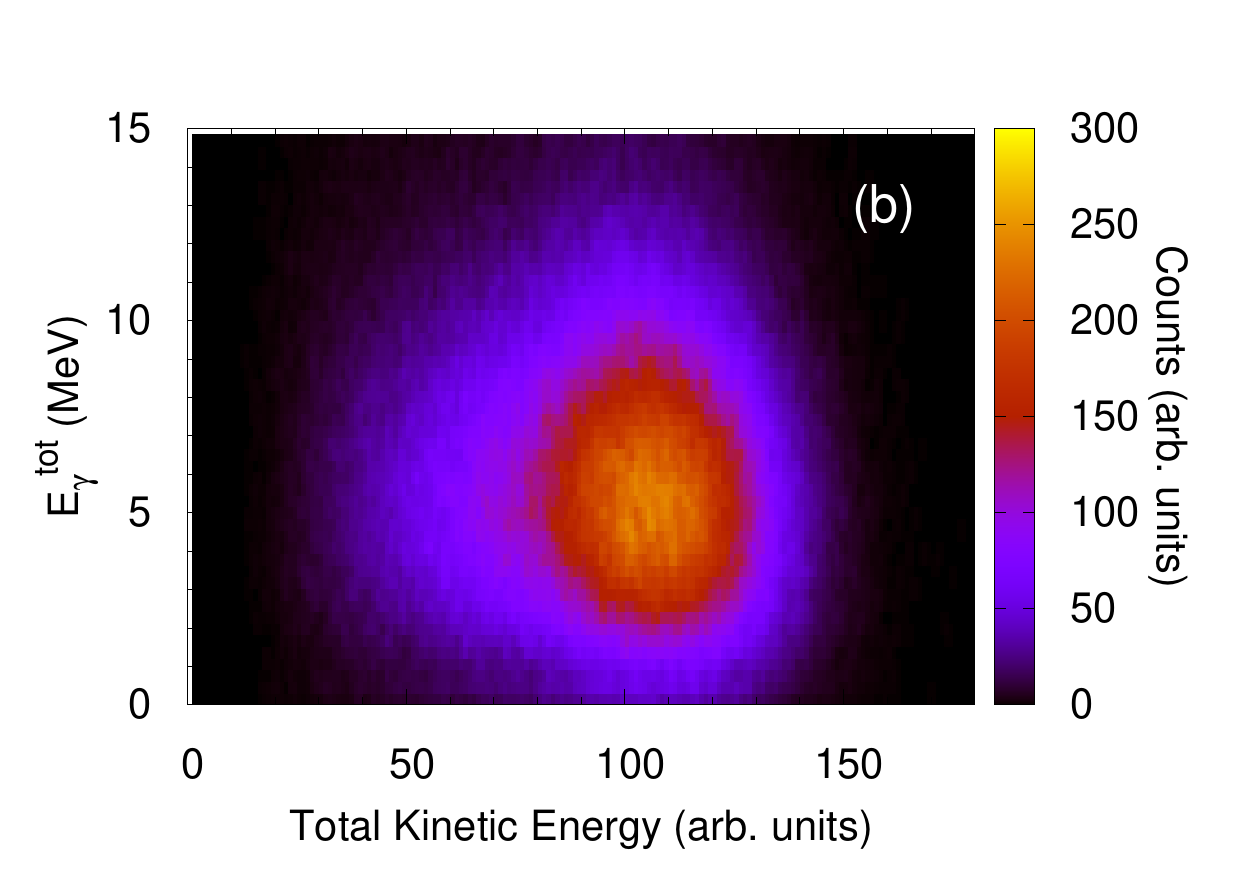}
\caption{(Color online) \CGMF-calculated (a) and DANCE experimental (b) data for the total prompt \gray~energy as a function of TKE in $^{252}$Cf(sf). Note the slightly different $E_\gamma^{\rm tot}$ limits on the $y$-axes and the arbitrary units for TKE on the DANCE data.}
\label{fig:EgtotTKE}
\end{figure}

Figure~\ref{fig:EgtotTKE} shows the predicted (a) and measured (b) $E_\gamma^{\rm tot}$ as a function of TKE. The experimental data were obtained with the DANCE array, in coincidence with two silicon detectors to measure the kinetic energy of the fragments~\cite{Rusev:2017}. While the very poor resolution of the fission fragment kinetic energies obtained with this preliminary setup prevents a fair comparison to the predicted results, a new experiment with improved energy resolution is planned. 
No detector response corrections have been applied to the experimental data in Fig.~\ref{fig:EgtotTKE}. 
The regularities in the calculation shown in the top panel of Fig.~\ref{fig:EgtotTKE} can be interpreted as follows for a single decaying fission fragment. If TKE decreases, TXE increases and the excitation energy available in the fragment for particle emission increases. For fragment excitation energies lower than the neutron emission threshold, all the excitation energy is available for \g~emission.  If the excitation energy is above the threshold for neutron emission, the probability for emitting a neutron is larger than the probability of \g~decay. In this case, the fragment $A$ emits one neutron and the residual $(A-1)$ fragment is now produced with a much smaller residual excitation energy, resulting a rather low total energy available for \g~emission. For still higher values of TXE, more energy is available to contribute to $E_\gamma^{\rm tot}$ until the TXE reaches the threshold for two-neutron emission. This pattern is repeated every time a new neutron threshold is reached, thereby explaining the somewhat regular pattern observed in Fig.~\ref{fig:EgtotTKE}.

\begin{figure}[ht]
\centering
\includegraphics[width=\columnwidth]{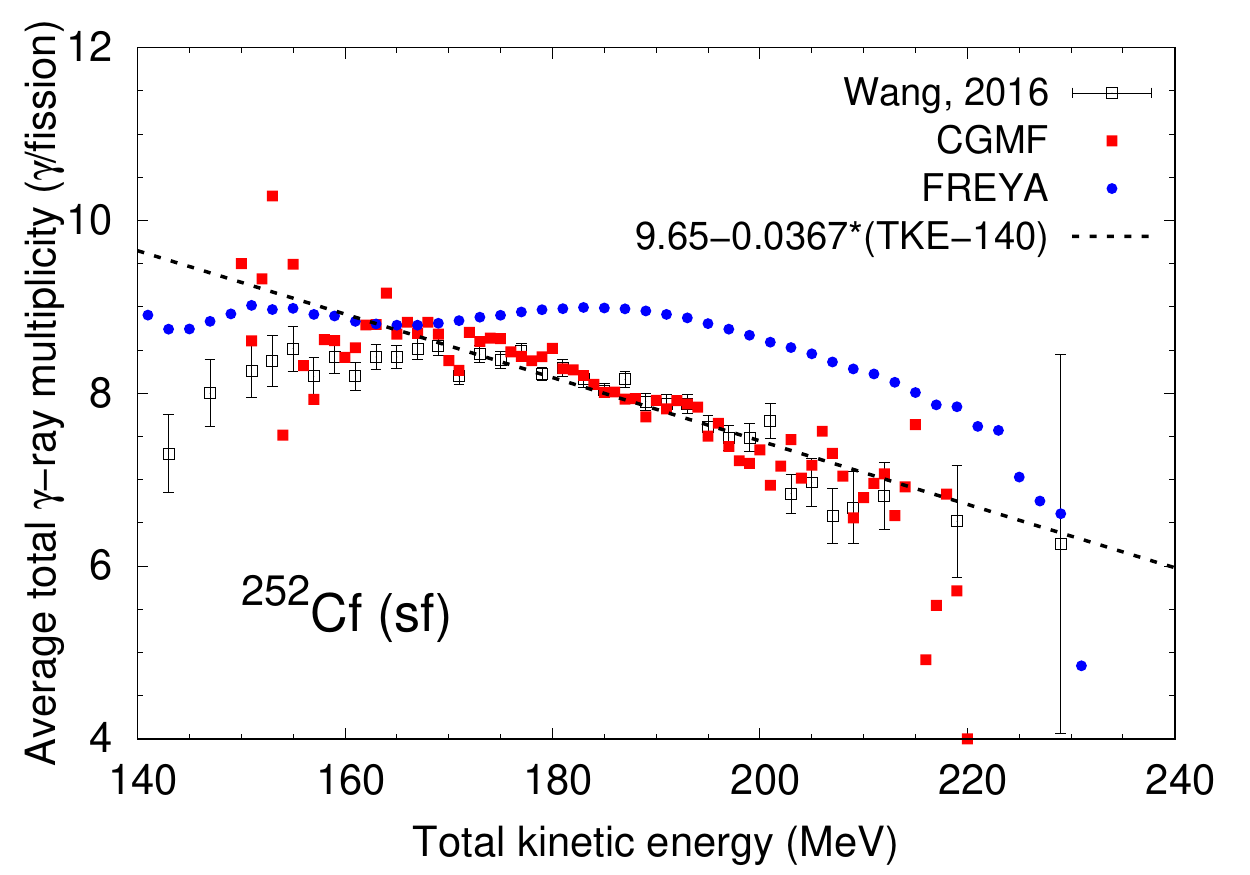}
\caption{\label{fig:MgTKE}(Color online) The average total \gray~multiplicity as a function of the total kinetic energy of the fragments, as measured by Wang \etal~\cite{Wang:2016}.}
\end{figure}

The average total \gray~multiplicity as a function of TKE has been measured by Wang \etal~\cite{Wang:2016}, and is shown in Fig.~\ref{fig:MgTKE} in comparison with \FREYA~and \CGMF~results. The agreement between \CGMF~and the experimental data is remarkable, while \FREYA~tends to overpredict the \g~multiplicity for most TKE values, as observed previously in Ref.~\cite{Wang:2016}.  The calculations use $g_{\rm min} = 0.05$~MeV and $t_{\rm max} = 5$~ns, as in Ref.~\cite{Wang:2016}.

\subsection{Multiplicity-dependent spectra}

\begin{figure}[ht]
\centering
\includegraphics[width=\columnwidth]{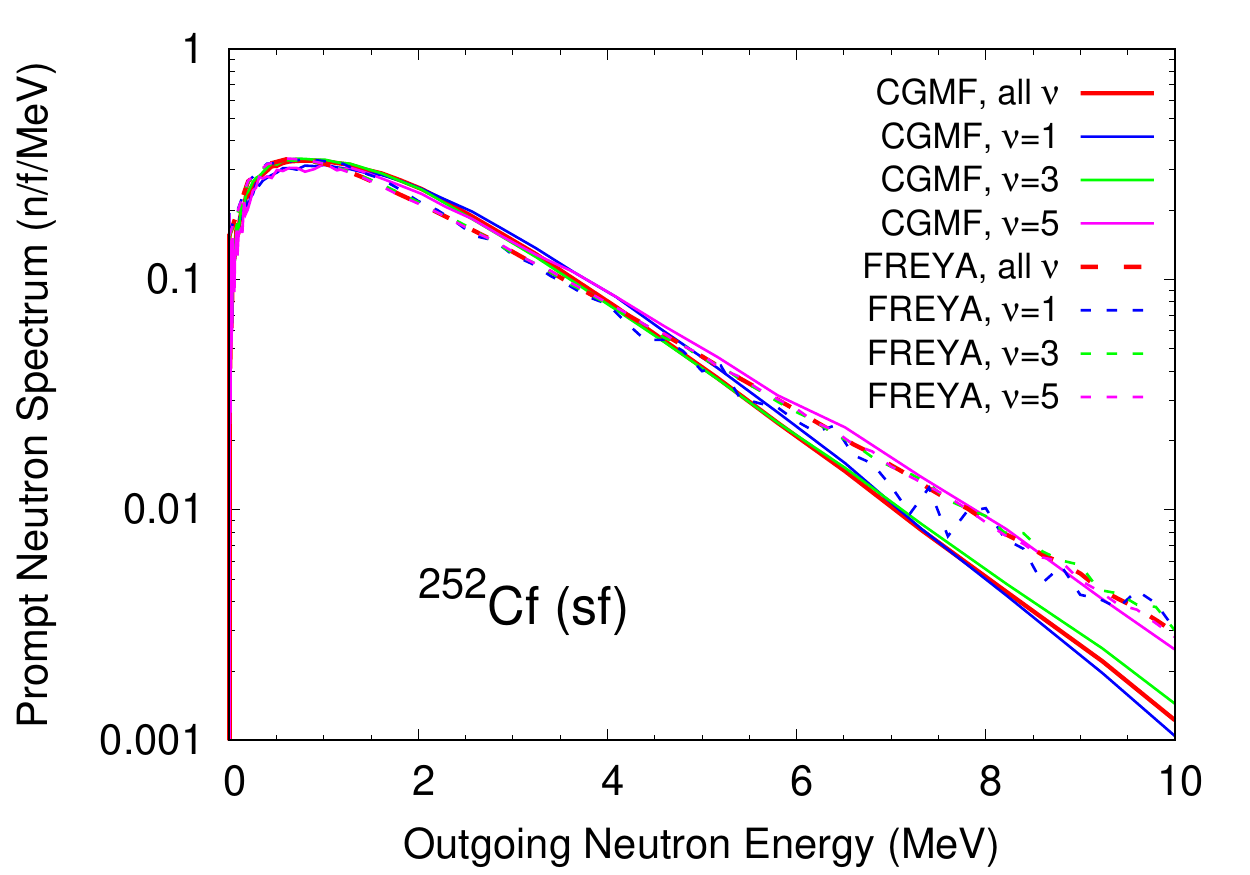}
\caption{\label{fig:nuPFNS}(Color online) Neutron multiplicity-dependent PFNS calculated for $^{252}$Cf(sf).}
\end{figure}

\begin{figure}[ht]
\centering
\includegraphics[width=\columnwidth]{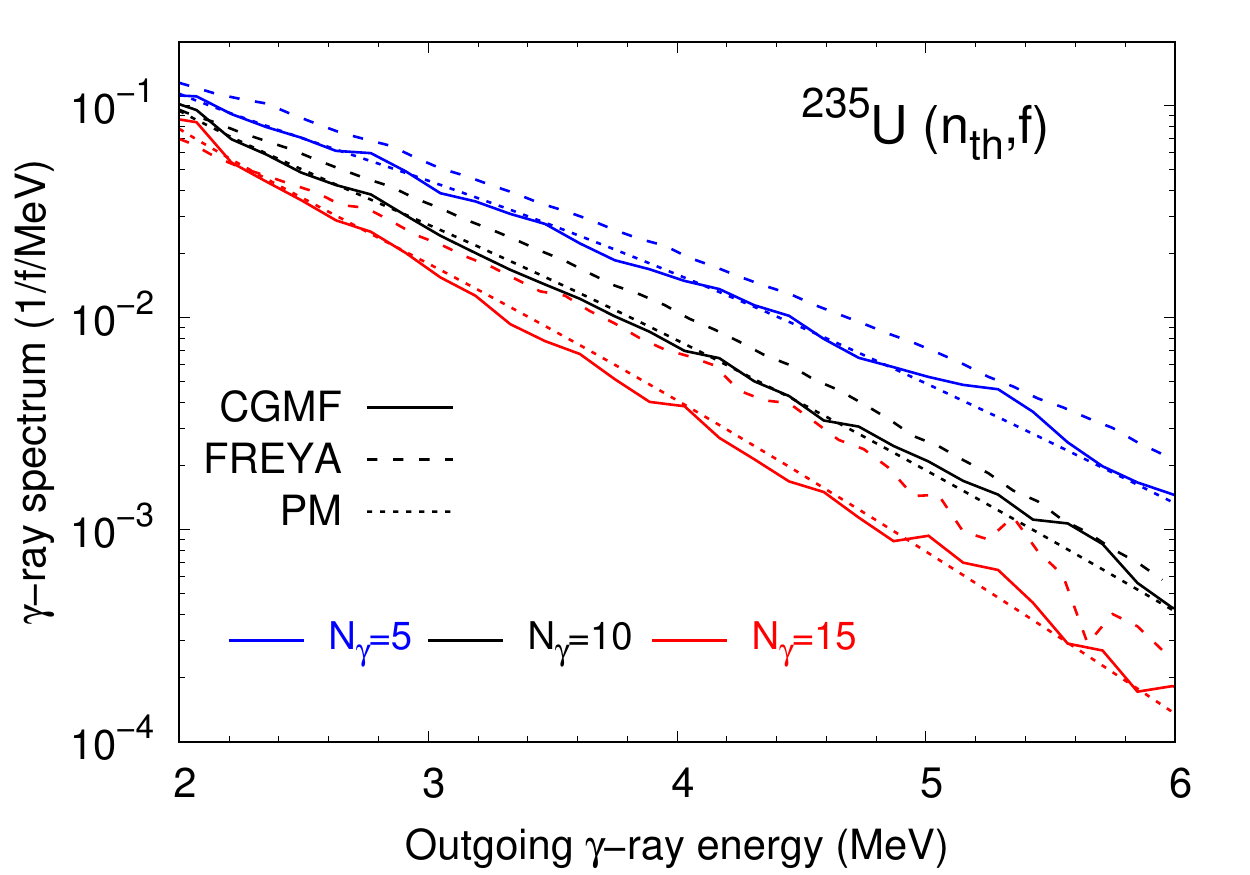}
\caption{\label{fig:nuPFGS}(Color online) Multiplicity-dependent prompt fission \gray~spectra calculated with \CGMF, \FREYA, and the parameterized model (PM) for $^{235}$U($n_{\rm th}$,f), for $N_\gamma =5$, 10, and 15. All the calculations are normalized to unity.}
\end{figure}

The average prompt fission neutron spectrum (PFNS) and the average prompt fission \gray~spectrum (PFGS) are not what would be called correlated data in the present context. Those quantities can generally, albeit not always, be found in the ENDF evaluated libraries, and are commonly used in transport simulations. What is not present, however, is the multiplicity-dependent PFNS and PFGS, as shown in Figs.~\ref{fig:nuPFNS} and~\ref{fig:nuPFGS}, respectively. The \CGMF-predicted results show a slight hardening of the neutron spectrum with increasing neutron multiplicity, and a much stronger softening of the \g~spectrum with increasing \g~multiplicity. It is important to note that \CGMF~predicts a much softer spectrum than the current standard evaluated result~\cite{Carlson:2009}, which is most likely linked to an incorrect optical potential for neutron-rich, deformed nuclei. The \FREYA~results for the PFNS show very little dependence on the neutron multiplicity for this nucleus.  The larger uncertainty for $\nu = 1$ is because of the lower probability for single neutron emission from $^{252}$Cf(sf) with its higher than average $\overline \nu$.

The dotted lines in Fig.~\ref{fig:nuPFGS} correspond to the results of a ``parameterized model" (PM) developed to interpret and fit the DANCE experimental results~\cite{Jandel:2014}. In this model, the total prompt fission \gray~multiplicity is taken as a sum of two multiplicities, $N_1 + N_2 = N_\gamma$, one from each fragment, which are sampled from two independent distributions,
\begin{equation}
P(N_i) = C_i (2N_i+1)\exp\left(-N_i(N_i+1)/c_i^2\right) \, ,
\end{equation}
where $i=1,2$. The terms $C_i=C_{1,2}$ are constants to ensure proper normalization of the probability distribution. Following approximate relations based on the statistical model of the compound nucleus, the probability to emit one  \g~ray with energy $\varepsilon_\gamma$ is given by
\begin{equation}
\chi_1(\varepsilon_\gamma)=D_1 \varepsilon_\gamma^2\exp\left(-(a_1+b_1N_\gamma)\varepsilon_\gamma\right) \, ,
\label{eq:chiM1}
\end{equation}
for multiplicity $N_1$ and
\begin{equation}
\chi_2(\varepsilon_\gamma)=D_2\varepsilon_\gamma^3\exp\left(-(a_2+b_2N_\gamma)\varepsilon_\gamma\right) \, ,
\label{eq:chiM2}
\end{equation}
for multiplicity $N_2$.  In Eqs. (\ref{eq:chiM1}) and (\ref{eq:chiM2}), $N_\gamma$ is the total \gray~multiplicity, while $D_{1,2}$ are normalization constants that ensure that
\begin{equation}
\int_0^\infty \chi_{1,2}(\varepsilon_\gamma)\:d\varepsilon_\gamma=1 \, .
\end{equation}

Therefore, for a given multiplicity $N_\gamma$, the multiplicity-dependent prompt-fission $\gamma$-ray spectrum (also unit normalized) is given by:
\begin{eqnarray}
\chi_{N_\gamma}(\varepsilon_\gamma) =& \frac{1}{N_\gamma}\sum_{N_1=0}^{N_\gamma}\left\{ N_1 P(N_1)\chi_1(\varepsilon_\gamma)+\right. \nonumber \\ 
& \left. N_2 P(N_\gamma-N_1)\chi_2(\varepsilon_\gamma) \right\} \, .
\label{eq:chiM}
\end{eqnarray}

The six model parameters \{$a_{1,2}$, $b_{1,2}$, $c_{1,2}$\} were fit to the DANCE data for $^{235}$U\nfth\ \cite{Jandel:2012}, $^{239}$Pu\nfth\ \cite{Ullmann:2013}, and $^{252}$Cf(sf)~\cite{Jandel:2014}. This model, while not used in \CGMF\ and \FREYA\ is useful to speed up calculations and correctly predict the tails of the multiplicity-dependent PFGS. However, predictions based on this simplified statistical model would fail to predict the low-energy part of the spectrum where non-statistical transitions between discrete excited states in fission fragments cause strong fluctuations in the average total \gray~spectrum as observed repeatedly for various fissioning systems~\cite{Oberstedt:2013}.

\begin{figure}[ht]
\centering
\includegraphics[width=\columnwidth]{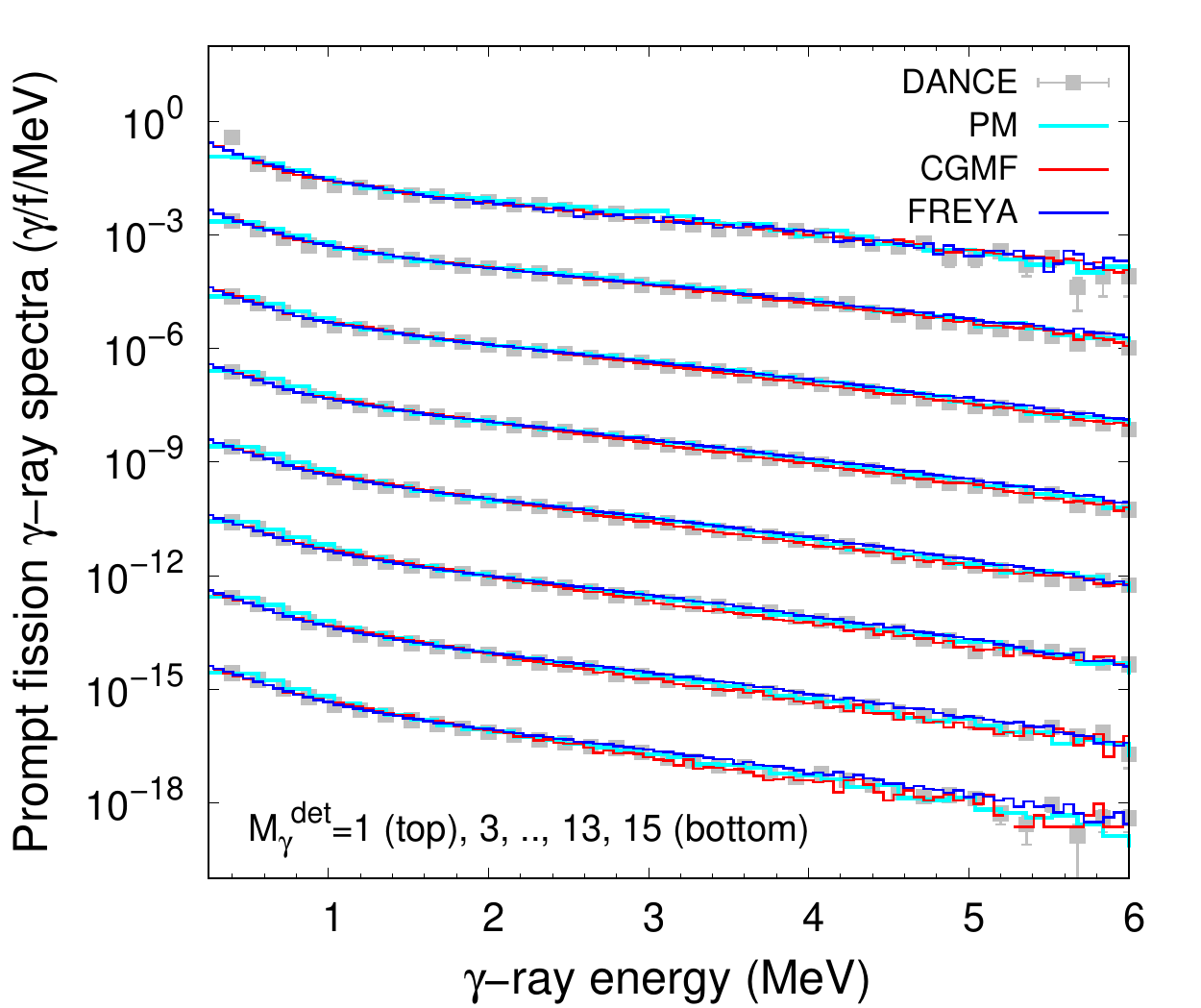}
\caption{\label{fig:nuPFGS-DANCE}
(Color online) The prompt fission \gray~spectra for \g~detector multiplicities, $1 \leq M_\gamma^{\rm det} \leq 15$, in steps of $M_\gamma^{\rm det} = 2$, observed~\cite{Jandel:2014} in the DANCE detector array in the spontaneous fission of $^{252}$Cf, are compared with the results of the PM (red lines), \CGMF~(cyan lines) and \FREYA~(blue lines). These spectra correspond to the raw data as measured by DANCE and have not been processed to unfold the complicated detector response. Instead, Monte Carlo output from the PM, \CGMF\ and \FREYA~have been processed through a $\mathtt{GEANT}4$ simulation~\cite{GEANT4} of DANCE response. All curves are unit normalized, but scaled down by a factor of 10$^{M_\gamma^{\rm det}-1}$. }
\end{figure}

The PM results are compared with \CGMF\ and \FREYA~calculations for $^{235}$U\nfth\ in Fig.~\ref{fig:nuPFGS}.  There is very good agreement between the PM and \CGMF\ above $\epsilon_\gamma > 2$~MeV.  The \FREYA\ calculations exhibit the same trend but with a somewhat harder slope at higher \gray\ energy.  A follow-up comparison between the model calculations propagated through the $\mathtt{GEANT4}$ model of DANCE and the experimental results for $^{252}$Cf(sf) is shown in Fig.~\ref{fig:nuPFGS-DANCE}. The \gray~spectra shown correspond to a 
\gray~detector multiplicity, $M_\gamma^{\rm det}$, indicative of how many DANCE detectors are fired in coincidence. The relation between $N_\gamma$ and $M_\gamma^{\rm det}$ is not trivial and has to be simulated through $\mathtt{GEANT}$ or \MCNP~simulations. The dependence of the DANCE \gray~spectra on $M_\gamma^{\rm det}$ is very well reproduced by the calculations.  The same trends observed in Fig.~\ref{fig:nuPFGS} for $^{235}$U($n_{\rm th}$,f) are seen here.  

\subsection{Multiplicity distributions $P(\nu)$ and $P(N_\gamma)$}

Neutron multiplicity and neutron coincidence counting methods are employed to assess the mass and multiplication of fissile materials in safeguards and international treaty verification. In these methods, the arrival times of neutrons in detectors are recorded and analyzed. Bursts of neutrons are indicative of the presence of fissile material and are used to statistically infer the fissile material properties. The first, second and third factorial moments of the prompt neutron multiplicity distribution $P(\nu)$, Eqs.~(\ref{eq:nu1})-(\ref{eq:nu3}) are important input data to these techniques. Higher moments (fourth and fifth) are even being considered~\cite{Santi:2015} as a way to extract more useful information about fissile materials. 

\begin{figure}[ht]
\centering
\includegraphics[width=\columnwidth]{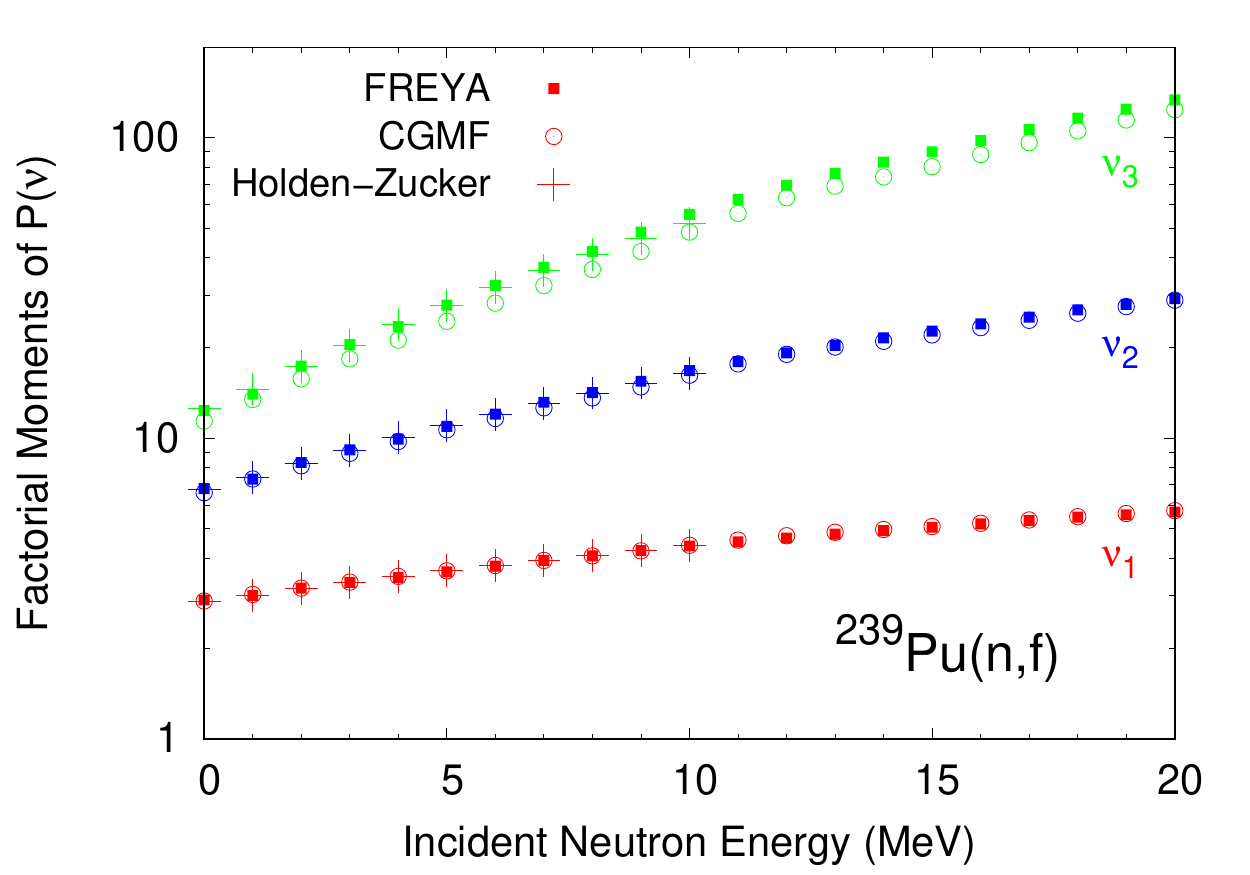}
\caption{\label{fig:Pnu-Einc}(Color online) The first three factorial moments $\nu_1$ (red), $\nu_2$ (blue) and $\nu_3$ (green) as defined in Eqs.~(\ref{eq:nu1})-(\ref{eq:nu3}), of the prompt neutron multiplicity distribution $P(\nu)$ for the $^{239}$Pu$(n,{\rm f)}$ reaction, as a function of incident neutron energy.}
\end{figure}

The neutron multiplicity distributions of most common spontaneously fissioning nuclei, see Fig.~\ref{fig:Pnu}(b), and important thermal-neutron-induced fission reactions are relatively well known.  These distributions have been revisited recently by Santi and Miller for many spontaneous fission cases~\cite{Santi:2008}.  Recall, however, that much less is known about the energy dependence of these distributions. The Zucker and Holden evaluation \cite{Zucker:1986} in Fig.~\ref{fig:Pnu-Einc} is based on the single data set of Soleilhac {\it et al.}~\cite{Soleilhac:1969} and Terrell's model~\cite{Terrell:1957}, see Eq.~(\ref{eq:Terrell}).  The \CGMF~calculations for neutron-induced fission reactions on $^{239}$Pu up to 20~MeV incident neutron energy are shown in Fig.~\ref{fig:Pnu-Einc}. \CGMF~model input parameters $R_T({\rm A})$ and $\alpha$ were adjusted to match the experimental $\overline{\nu}$ and $\overline N_\gamma$ values below 5~MeV only. The agreement with Terrell's model at higher energies and for the higher moments of the distribution is quite good. In this comparison, the \FREYA~parameters $c_S$ and $e_0$ were fixed by $^{252}$Cf(sf) data while $c_T$ was adjusted to the thermal point to improve agreement with the shape of $P(\nu)$ but was left energy independent.  The $d$TKE, adjusted to agree with \nubar\ for all energies, is the only energy dependent parameter in \FREYA.

\begin{figure}[ht]
\centering
\includegraphics[width=\columnwidth]{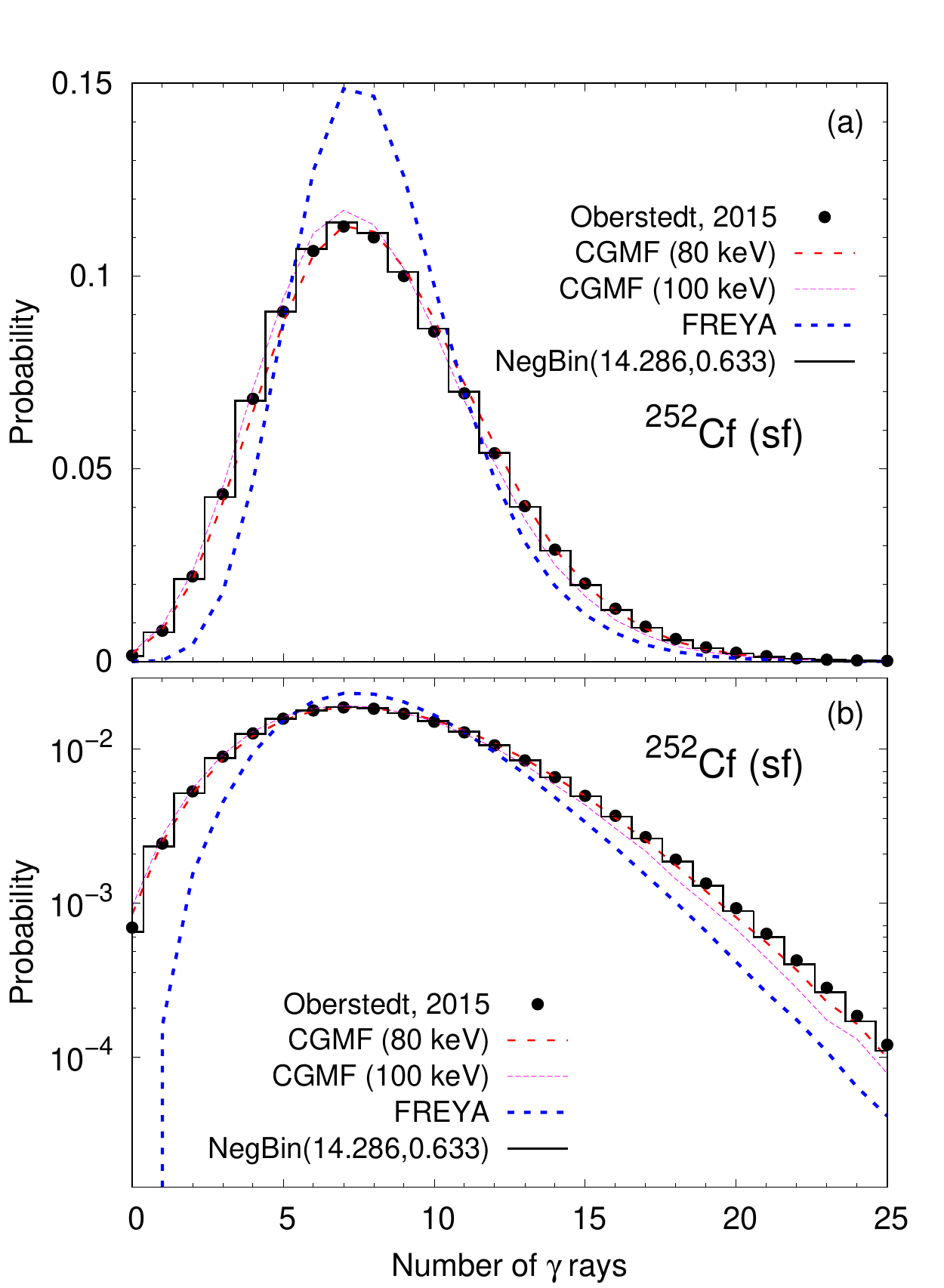}
\caption{\label{fig:Pnug}(Color online) The prompt \gray~multiplicity distribution $P(N_\gamma)$ for $^{252}$Cf(sf) measured by Oberstedt \etal~\cite{Oberstedt:2015b} is compared to \CGMF\ and \FREYA\ results on linear (a) and logarithmic (b) scales. The negative binomial function with parameters $r$=14.286 and $p$=0.633 represents a best fit to the experimental data.}
\end{figure}

The prompt \gray~multiplicity distribution $P(N_\gamma)$ measured~\cite{Oberstedt:2015b} for $^{252}$Cf(sf) is compared to \CGMF~and \FREYA~calculations in Fig.~\ref{fig:Pnug}. This distribution and in particular the average \gray~multiplicity $\overline{N}_\gamma$ is very sensitive to the threshold energy and time coincidence window since fission considered. However, the overall shape of the \CGMF-calculated distribution is very nicely
represented by a negative binomial distribution NB($r,p$) with $r$=14.286 and
$p$=0.633, in agreement with the distribution inferred by Oberstedt
\etal~\cite{Oberstedt:2015b},
albeit with a \g~energy threshold of 80~keV instead of the 100~keV reported by the authors.  The \FREYA\ results, although in rather good agreement with the average multiplicity, are narrower than the data.  This is still under investigation.

\subsection{\g-$n$ multiplicity correlations}

\begin{figure}[ht]
\centering
\includegraphics[width=\columnwidth]{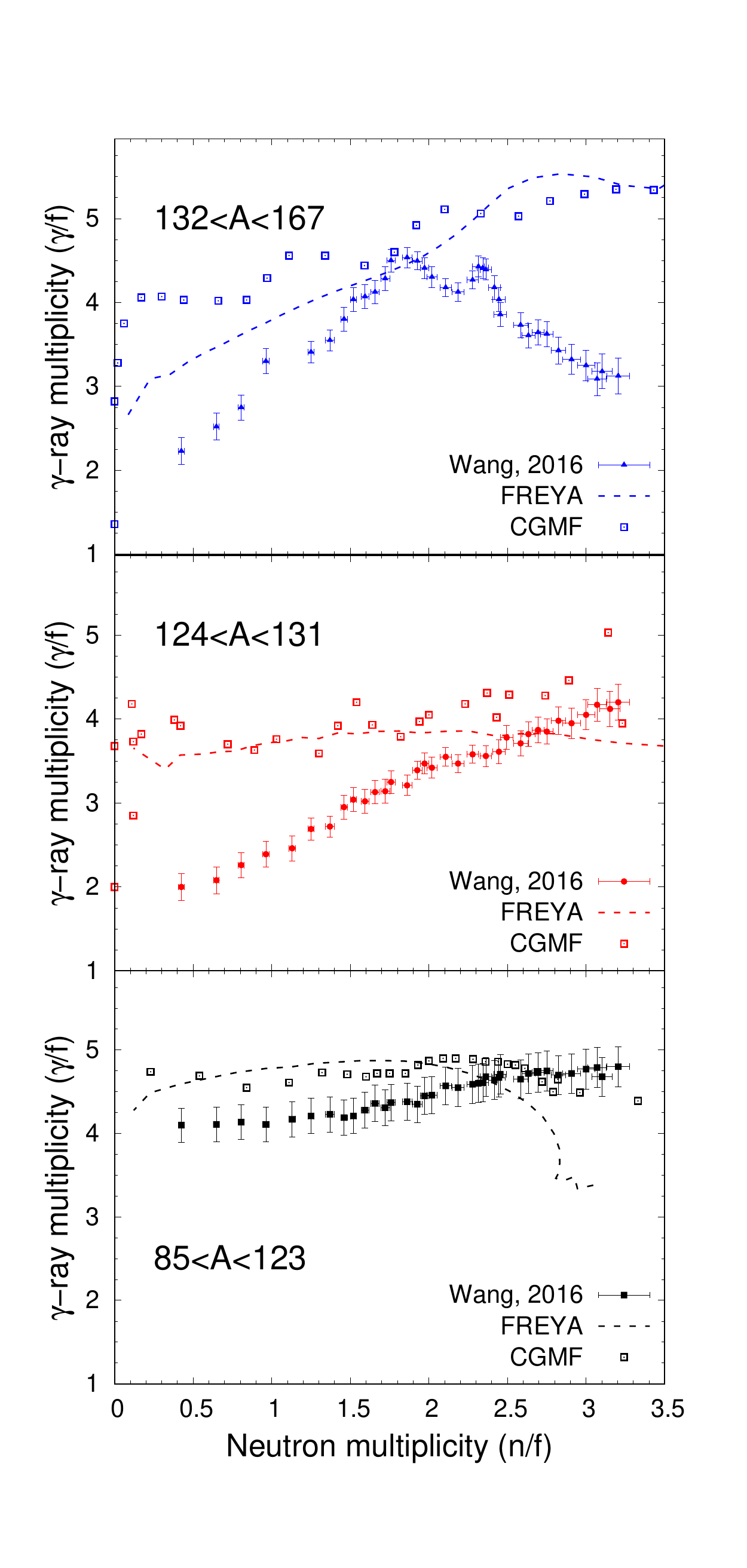}
\caption{(Color online) Prompt \gray~multiplicity as a function of the neutron multiplicity $\nu$, recently measured by Wang \etal~\cite{Wang:2016} for $^{252}$Cf(sf), and compared to \FREYA~and \CGMF~calculations. The mass selection is performed on the mass $A$ of the initial pre-neutron emission fission fragments.}
\label{fig:nu-nug-calc}
\end{figure}

Nifenecker \etal\cite{Nifenecker:1972} found that the average total \gray~energy from $^{252}$Cf(sf) was linearly proportional to the average neutron multiplicity, see Eq.~(\ref{eq:Nifenecker}) and Fig.~\ref{fig:Nifenecker}. The experiments were conducted with a $^{252}$Cf fission source and fragment detectors surrounded by a large spherical gadolinium-loaded liquid scintillator tank. Prompt neutron and \gray~detections were discriminated primarily through timing cuts. Recently, Wang \etal\cite{Wang:2016} expanded upon Nifenecker's work by measuring the \gray~multiplicity as a function of the neutron multiplicity for the light ($85<A<123$), symmetric ($124<A<131$), and heavy ($132<A<167$) fission fragments, shown in Fig.~\ref{fig:nu-nug}. For these experiments, two surface barrier detectors were used to estimate the fragment mass. A high purity germanium detector was used to count \grays~and a LS301 liquid scintillator was used to count neutrons.  Figure~\ref{fig:nu-nug-calc} shows a slightly increasing, a strongly increasing, and a non-monotonic trend in \gray~multiplicity as a function of neutron multiplicity, for those three mass regions respectively. Each data point is a 3~MeV-wide bin in TKE. While the predictions from \FREYA~also show some fragment mass dependence, the experimentally observed trends are not well reproduced.  (Note that the \FREYA~calculations differ from those in Ref.~\cite{Wang:2016}, performed before the RIPL-3 lines were included.)  The \CGMF\ results exhibit the same trends as the \FREYA\ calculations.

It is worth explaining the calculated trends.  The calculated $\nu$(TKE) decreases with increasing TKE in all mass regions, see Fig.~\ref{fig:nuTKE-selectA}.  However, for \FREYA, there is somewhat of a plateau for ${\rm TKE} < 175$~MeV in the low mass region.  Also, the dependence of the photon multiplicity with TKE is more complex and changes with mass region: it slightly increases with TKE until ${\rm TKE} \sim 185$~MeV and then remains relatively independent of TKE in the low mass region; it is independent of TKE for masses near symmetry; and it decreases with TKE in the high mass region, leading to the behavior shown in Fig.~\ref{fig:nu-nug-calc} when $\nu$(TKE) and $N_\gamma$(TKE) are plotted against each other.  The decrease in \g-ray multiplicity for $\nu > 2.5$ is due to the lower $N_\gamma$ at low TKE.  When averaged over all masses, the $N_\gamma$(TKE) for \FREYA\ is as observed in Fig.~\ref{fig:MgTKE}.  In general, for a positive for $N_\gamma$ relative to $\nu$, both must decrease with increasing TKE. 

\begin{figure}[ht]
\centering
\includegraphics[width=\columnwidth]{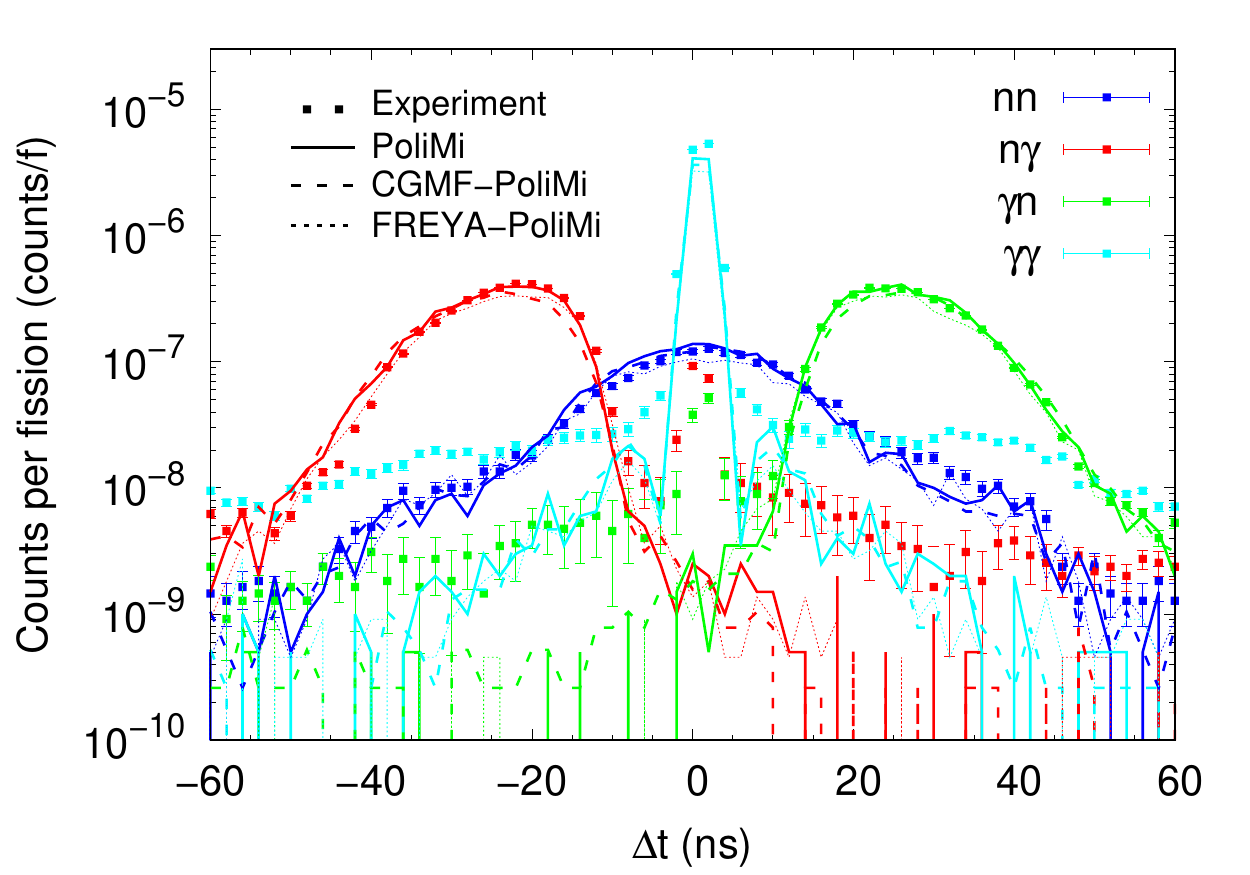} 
\caption{(Color online) $^{252}$Cf(sf) cross-correlated time distributions for neutron and \g~coincidences for measured (points) and simulated (lines) data from Ref.~\protect\cite{Marcath:2016}. The three simulated cases utilize \polimi~for particle transport and vary the fission models: \CGMF, \FREYA~and the built-in \polimi~model are employed.}
\label{fig:UMtime}
\end{figure}

A recent experiment performed at the University of Michigan~\cite{Marcath:2016} used a scintillator detector array (see Fig.~\ref{fig:UMexpsetup}) to measure correlations between prompt neutrons and \grays~from fission. The measured time cross-correlations are in relatively good agreement with the results simulated by \polimi, \FREYA, and \CGMF, as shown in Fig.~\ref{fig:UMtime} for the cross-correlation time distributions for $n$-$n$, $n$-\g, \g-$n$, and \g-\g~coincidences.
Note that the \g-\g and $n$-$n$ correlations are expected to be symmetric around $\Delta t = 0$ while $n$-\g and \g-$n$ correlations are reflected around $\Delta t = 0$ since the time axis represents the difference between the time of detection in detector 1 and detector 2.  Therefore an $n$-\g event (a photon detected in 1 and a neutron in 2) results in $\Delta t < 0$ while a \g-$n$ event results in $\Delta t > 0$. 
The three simulated cases utilize \polimi~for particle transport but vary the fission models to include \CGMF~and \FREYA~as well as the \polimi~model. Rather large discrepancies appear in the \g-\g~correlation points (cyan), most likely due to background contamination which could be removed with the use of a fission chamber.

Neutron energies were estimated by time-of-flight with a \gray~trigger. Tables~\ref{tab:polimi-gamma} and~\ref{tab:polimi-neutron} show the mean neutron energy as a function of coincident particle detections for the measurements as well as the simulated \polimi, \CGMF, and \FREYA~results. A small increase in calculated neutron energy is observed in the calculations with the number of coincident detections for both neutron and $\gamma$ coincidences, primarily between one and two emitted neutrons and \g-rays. 
More data are required to resolve any trend in the neutron energy.

\begin{center}
\begin{table}[ht]
\setlength{\tabcolsep}{0.25em}
\def\arraystretch{1.15}
\begin{tabular}{cccccc} 
\hline
\g & $\mathtt{PoliMi}$-1 & \CGMF-\polimi & \FREYA-\polimi & Meas. \\
coinc. & & & \\
\hline
1 & 2.344(2) & 2.320(6) & 2.397(6) & 2.475(2) \\
2 & 2.411(9) & 2.37(3) & 2.49(3) & 2.48(1) \\
3 & 2.45(4) & 2.4(1) & 2.4(2) & 2.49(6) \\
4 & 2.5(3) & - & - & 2.4(4) \\
\hline
\end{tabular}
\caption{\label{tab:polimi-gamma} (Color online) The average detected neutron energy (in MeV) by time-of-flight over the sensitive range of the detectors, $1.1-6.6$~MeV, as a function of the number of \gray~coincidences. Omitted entries had too few statistics. The parentheses record the uncertainty in the last significant figure.}
\end{table}
\end{center}

\begin{center}
\begin{table}[ht]
\setlength{\tabcolsep}{0.25em}
\def\arraystretch{1.15}
\begin{tabular}{cccccc} 
\hline
$n$ & \polimi-1 & \CGMF-\polimi & \FREYA-\polimi & Meas. \\
coinc. & & & \\
\hline
1 & 2.347(2) & 2.322(6) & 2.400(6) & 2.475(2) \\
2 & 2.359(9) & 2.34(3) & 2.44(3) & 2.52(1) \\
3 & 2.34(6) & 2.4(2) & 2.5(2) & 2.51(9) \\
4 & 2.3(5) & - & - & 2.4(9) \\
\hline
\end{tabular}
\caption{\label{tab:polimi-neutron}Average detected neutron energy (in MeV) by time-of-flight over the sensitive range of the detectors, $1.1-6.6$~MeV, as a function of the number of neutron coincidences. Omitted entries had too few statistics. The parentheses record the uncertainty in the last significant figure.}
\end{table}
\end{center}

\begin{figure}[ht]
\centering
\includegraphics[width=\columnwidth]{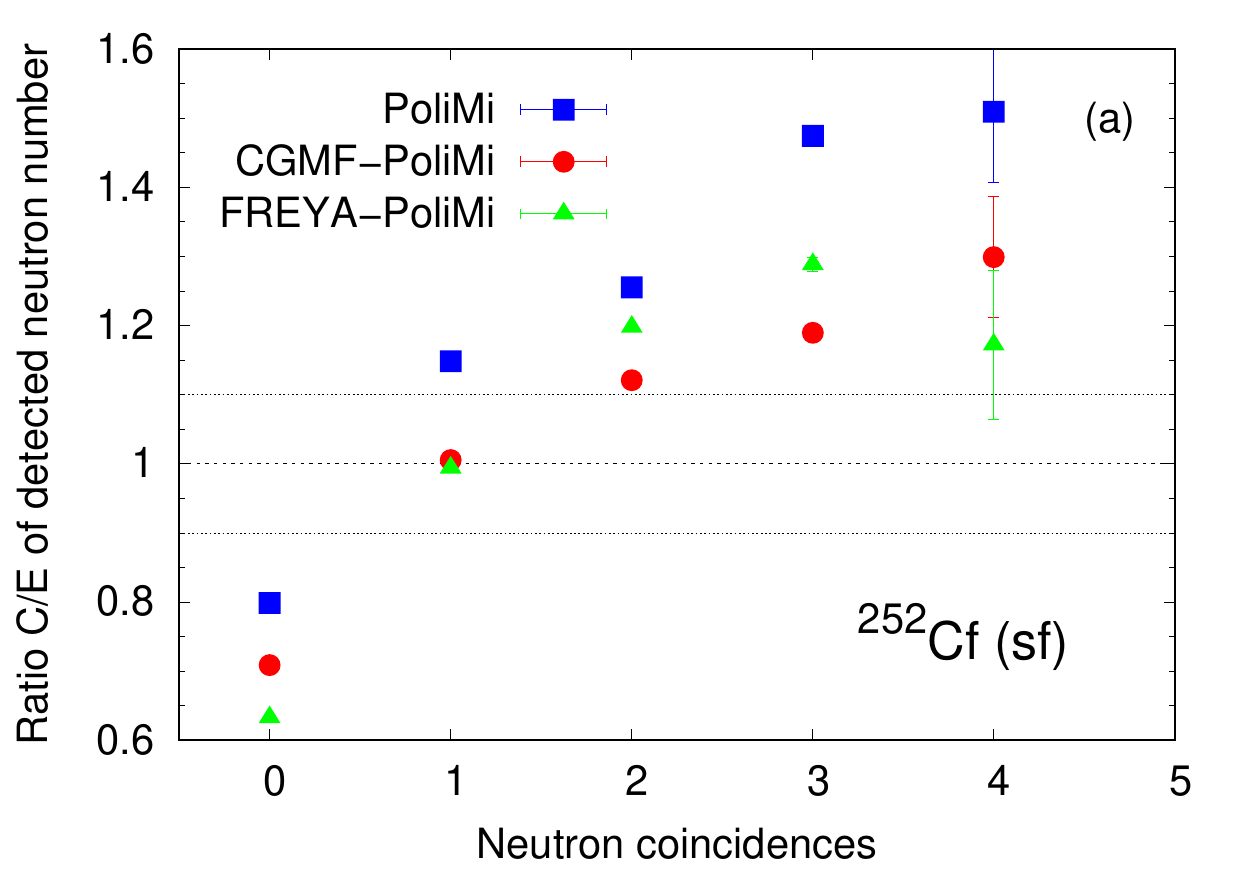} 
\includegraphics[width=\columnwidth]{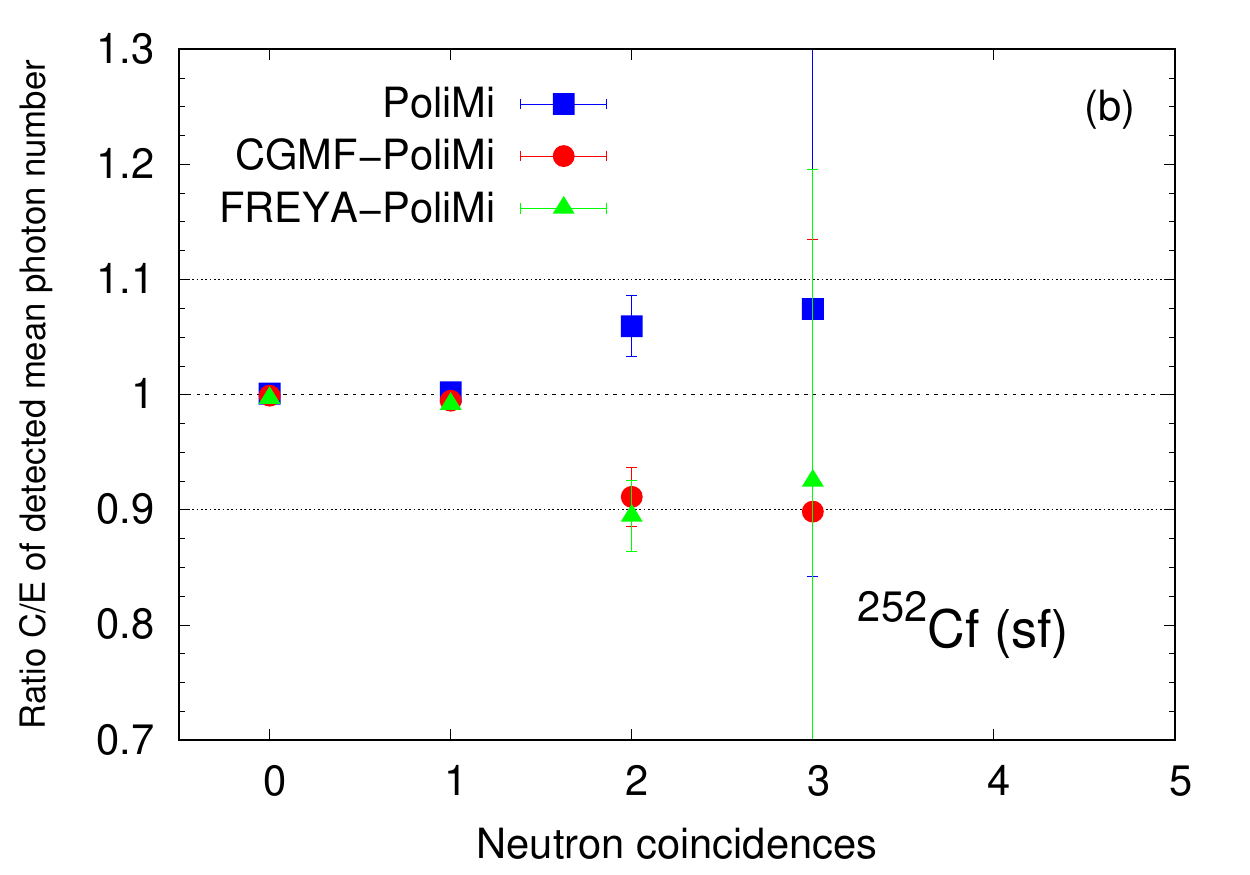} 
\caption{\label{fig:um-nn}(Color online) (a) Ratio of calculated to measured (C/E) neutron number as a function of neutron coincidences. (b) Ratio of calculated to measured (C/E) average number of \grays\ as a function of neutron coincidences.}
\end{figure}

The two-particle coincidence events were binned by the number of neutrons and \grays\ detected within the 80~ns time-coincidence window.  The measurements are shown in Fig.~\ref{fig:um-nn}(a) and (b) respectively as the ratio of calculated to experimental results, C/E. The simulations overpredict the observed counts for all neutron coincidences except zero. Despite a basic background subtraction, the number of coincidences is underpredicted at zero because background photon coincidences contribute disproportionately.  However, the observed discrepancies do not necessarily reflect a problem with the calculated $n$-\g~correlations and may instead be due to inaccuracies in the calculated PFNS.  The C/E for $\gamma$ multiplicity relative to neutron coincidences agrees well, ${\rm C/E}\, \sim 1$, for zero counts.  In this case \polimi\ alone overpredicts while \CGMF\ and \FREYA\ underpredict.  (Note that \polimi\ employs a Watt distribution for the PFNS except for $^{252}$Cf(sf) where the Mannhart evaluation is used directly.)  
Here, C/E is within 10\% of unity.

\subsection{Neutron-fragment angular correlations}

\begin{figure}[ht]
\centerline{\includegraphics[width=\columnwidth]{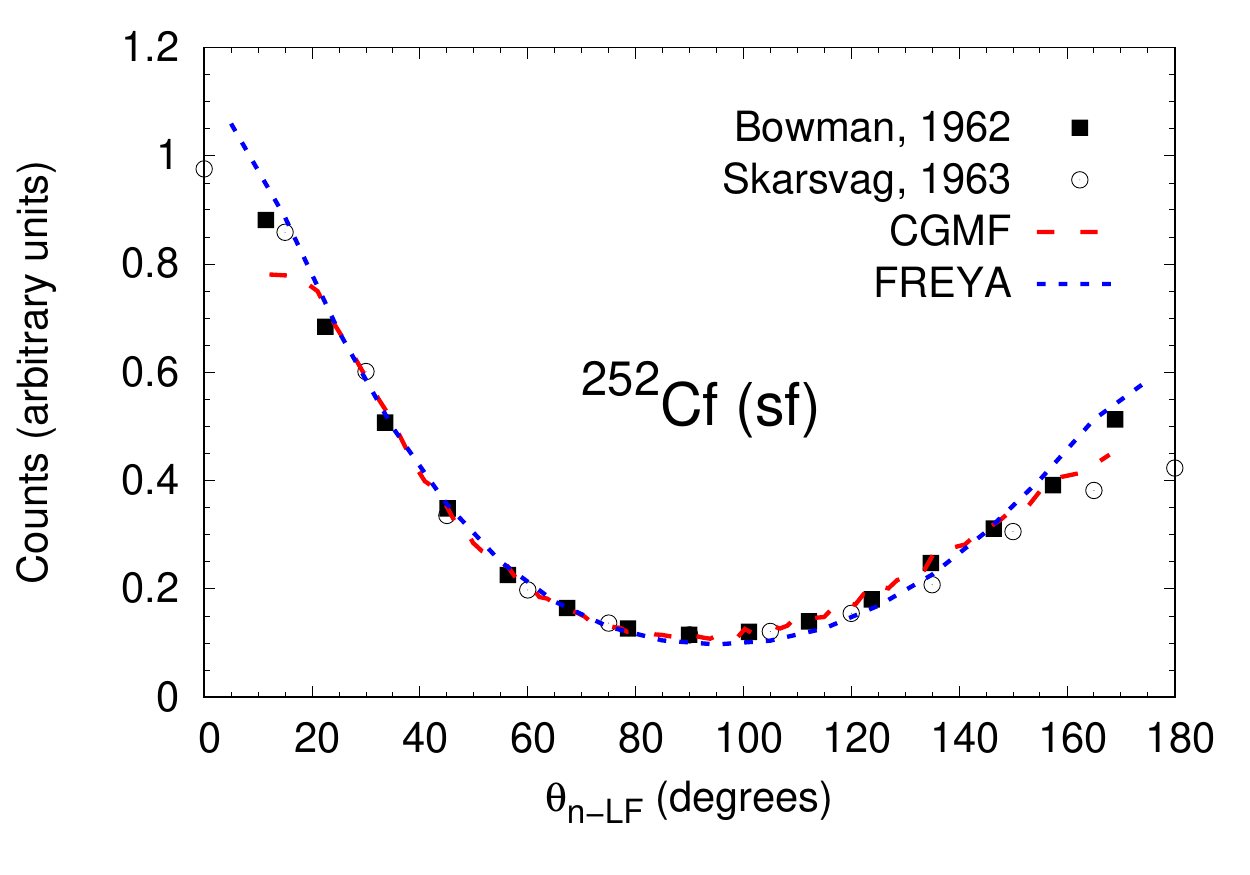}}
\caption{\label{fig:nLF}(Color online) The neutron-light fragment angular distribution for $^{252}$Cf(sf). The experimental data are from Bowman~\cite{Bowman:1962} and Skarsv\r{a}g~\cite{Skarsvag:1963}.  Only neutrons with kinetic energies above 0.5~MeV were analyzed.}
\end{figure}

Angular correlations between the fission fragments and the emitted neutrons emerge naturally from the kinematics of the reaction. Assuming that neutrons are emitted from fully accelerated fragments, the kinematic boost of the fragments from the center-of-mass frame to the laboratory frame induces a significant focusing of the neutrons in the direction of the fragments. Therefore, it is expected, and observed, that neutrons are emitted preferentially near zero and 180 degrees from the direction of the light fragment. Figure~\ref{fig:nLF} illustrates this feature in the case of $^{252}$Cf(sf). The ratio of neutrons emitted in the direction of the heavy fragment relative to the light fragment direction corresponds rather well to the average number of neutrons emitted from the light relative to the heavy fragment.  Both the \CGMF\ and \FREYA\ calculations compare rather well with the data.

\begin{figure}[ht]
\centering
\includegraphics[width=\columnwidth]{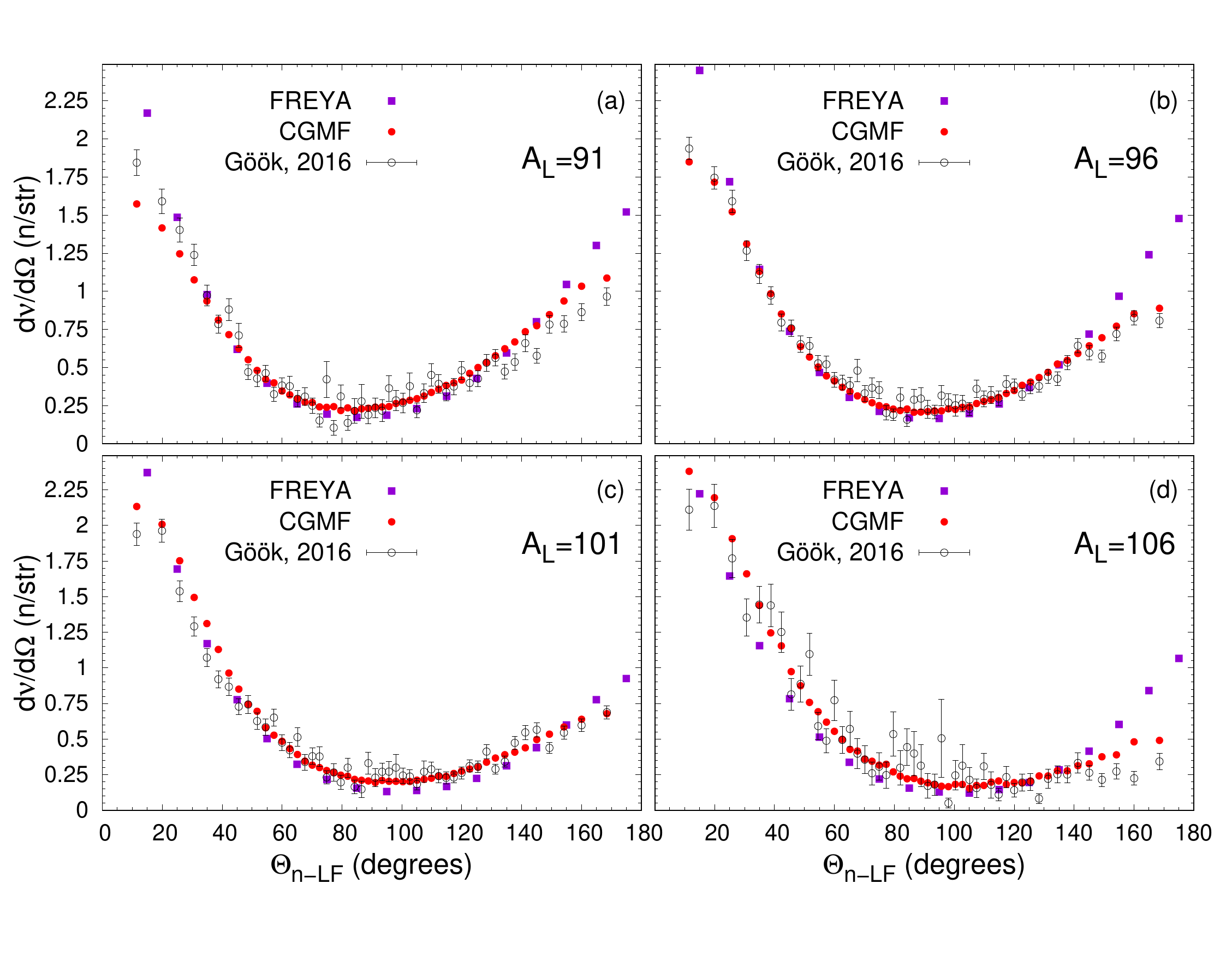} 
\caption{(Color online) The angular distributions of prompt fission neutrons in the thermal-neutron-induced fission reaction on $^{235}$U, for different fragment mass splits. The experimental data are by G\"o\"ok \etal~\cite{Gook:2016}. Only neutrons with kinetic energies above 0.5~MeV were analyzed.}
\label{fig:nLF-A}
\end{figure}

A more detailed view of this process can be obtained by isolating the contribution from each mass split. This was recently achieved experimentally by G\"o\"ok \etal\cite{Gook:2016} as shown in Fig.~\ref{fig:nLF-A} for $^{235}$U.  The data were taken with incident neutron energies $0.3 \,\, {\rm eV} \leq E_{\rm inc} \leq 60$~keV, with a mean energy of 1.6~keV.  The \CGMF\ and \FREYA\ results for thermal-neutron induced fission are also shown.  The agreement is very satisfactory between the experimental and calculated results, except in a few cases:  \CGMF~overestimates the experimental data near 180 degrees for the 106/130 mass split while the \FREYA\ calculation overestimates the data for $\theta_{n-{\rm LF}} > 140$ degrees for $A_L = 96$ and 106.

\subsection{$n$-$n$ and $n$-\g~correlations}

In the absence of a fission fragment detector, correlations between the prompt neutrons and \grays~are invaluable to distinguish a fission reaction from other neutron-induced reactions such as inelastic scattering. Because of the boost imparted to the prompt neutrons emitted from the excited fission fragments, $n$-$n$ angular correlations are expected to be peaked near 0 and 180 degrees. On the other hand, neutrons emitted in an $(n,2n)$ reaction, for example, would be emitted isotropically in the laboratory frame, except at higher incident neutron energies where the pre-equilibrium component would tend to focus the neutrons slightly more in the direction of the incident beam. Observing significant enhancements near 0 and 180 degrees is therefore a clear signature of a fission event. 

\begin{figure}[ht]
\centerline{\includegraphics[width=\columnwidth]{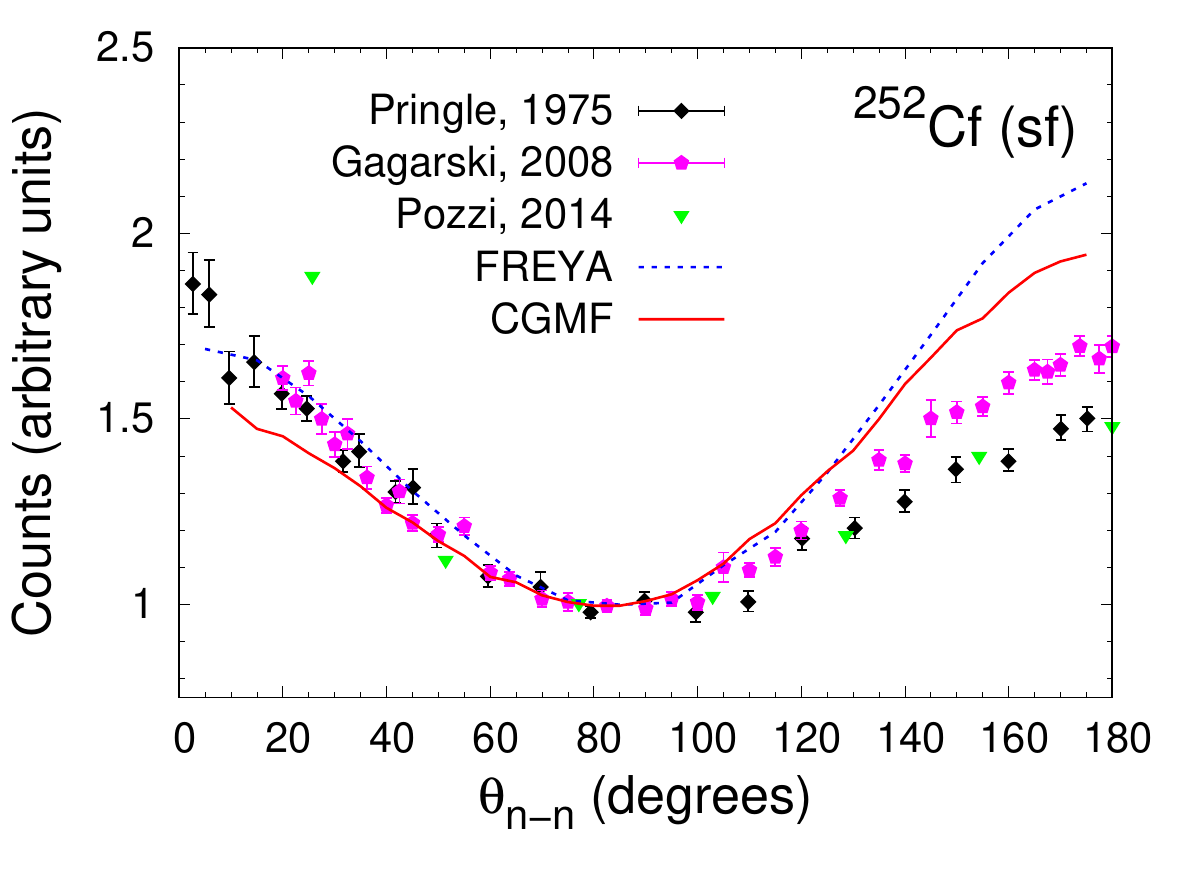}}
\caption{\label{fig:nn}(Color online) Neutron-neutron angular distribution for $^{252}$Cf(sf). Experimental data are from Pringle~\cite{Pringle:1975}, Gagarski \etal\cite{Gagarski:2008} and Pozzi \etal\cite{Pozzi:2014}.  No detector response was folded onto the calculations. The experimental neutron detection thresholds are 0.7~MeV for Pringle~\cite{Pringle:1975} and 0.425~MeV for Gagarski~\cite{Gagarski:2008} and Pozzi~\cite{Pozzi:2014} data.}
\end{figure}

Figure~\ref{fig:nn} shows neutron-neutron angular distributions in the spontaneous fission of $^{252}$Cf, as measured by several experimental groups~\cite{Pringle:1975,Gagarski:2008,Pozzi:2014}, compared to calculations using \CGMF~and \FREYA. The Pozzi~\cite{Pozzi:2014} and Pringle~\cite{Pringle:1975} data agree very well, except at the lowest angles where cross-talk corrections should be taken into account.  The effect of cross talk is evident in the Pozzi point at $\theta_{n-n} = 22$ degrees which has not been corrected for this effect.  The Gagarski~\cite{Gagarski:2008} data lie higher than the other two data sets at the largest angles. The experiments use somewhat different cutoffs:  Pringle~\cite{Pringle:1975} had a minimum neutron energy cutoff of 0.7~MeV, compared to the Gagarski measurement with a cutoff of 0.425~MeV \cite{Gagarski:2008}.  The Pozzi~\cite{Pozzi:2014} data used the same cutoff as the Gagarski result shown here.  The \FREYA\ calculations agree well for $\theta_{n-n} < 90$~degrees while \CGMF\ somewhat underestimates here.  At angles close to back-to-back neutron emission, the calculations both overestimate the data.  The dependence of the calculated correlation on $\theta_{n-n}$ for a fixed neutron energy threshold is determined by the excitation energy sharing, $x$ for \FREYA\ and $R_T(A)$ for \CGMF. 

\begin{figure}[ht]
\centerline{\includegraphics[width=\columnwidth]{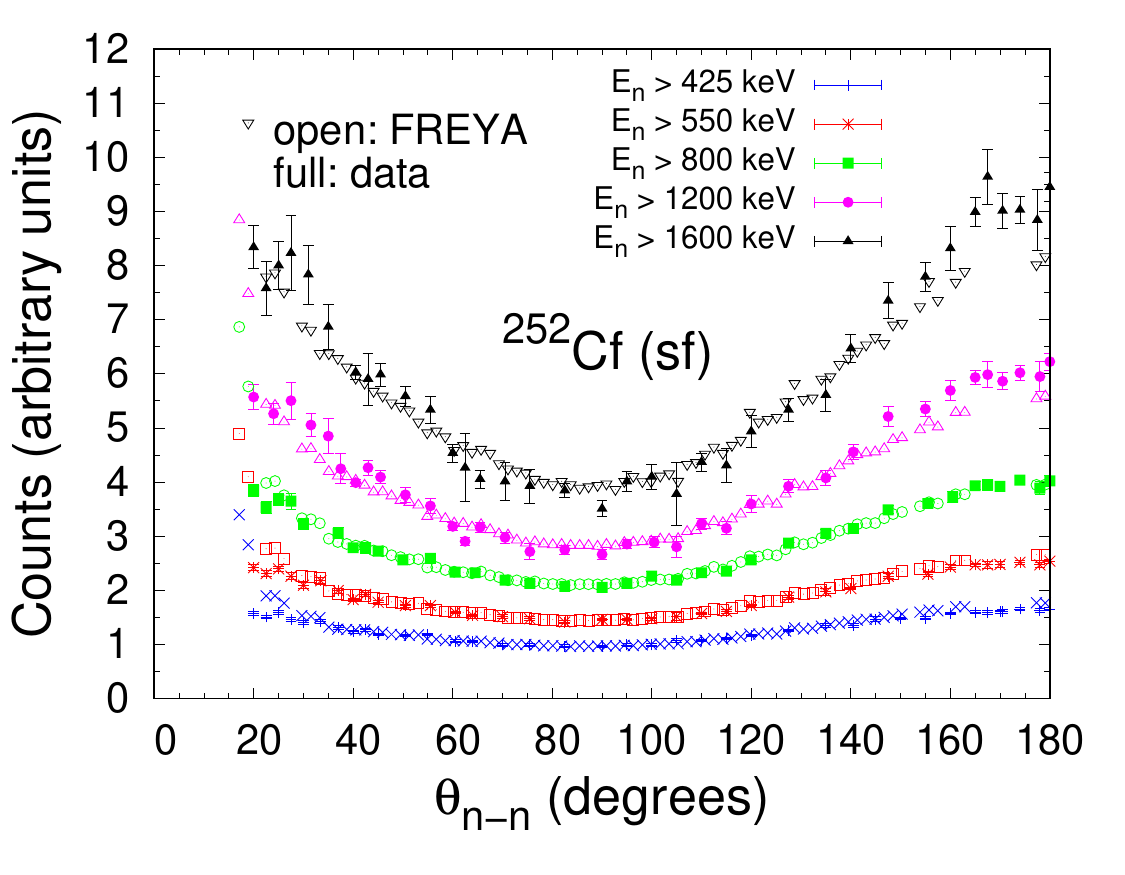}} 
\caption{\label{fig:nn-Eth} (Color online) Same as Fig.~\ref{fig:nn} but for different neutron energy detection thresholds. The data are from Gagarski \etal~\cite{Gagarski:2008}.}
\end{figure}

Also of interest is the evolution of these $n$-$n$ correlations as the neutron energy threshold increases, as shown in Fig.~\ref{fig:nn-Eth} with the Gagarski~\cite{Gagarski:2008} data.  The results obtained using $\mathtt{MCNPX}$ with \FREYA\ are overlaid in Fig.~\ref{fig:nn-Eth}.  For each energy-matched pair of data with simulation, the integral are matched in the range $40 < \theta_{n-n} < 140$~degrees.  The lower angular range is excluded to avoid the region at low $\theta_{n-n}$ where no cross-talk correction has been applied.
The correlation rises with threshold energy at 0 and 180 degrees.  One may expect a steeper rise since the higher energy neutrons are more likely to be emitted in the direction of the fragment before the boost to the laboratory frame.  Note that in Fig.~\ref{fig:nn-Eth} the \FREYA\ calculations are run through a full detector simulation while the calculations in Fig.~\ref{fig:nn} were not.

\begin{figure}[ht]
\centerline{\includegraphics[width=\columnwidth]{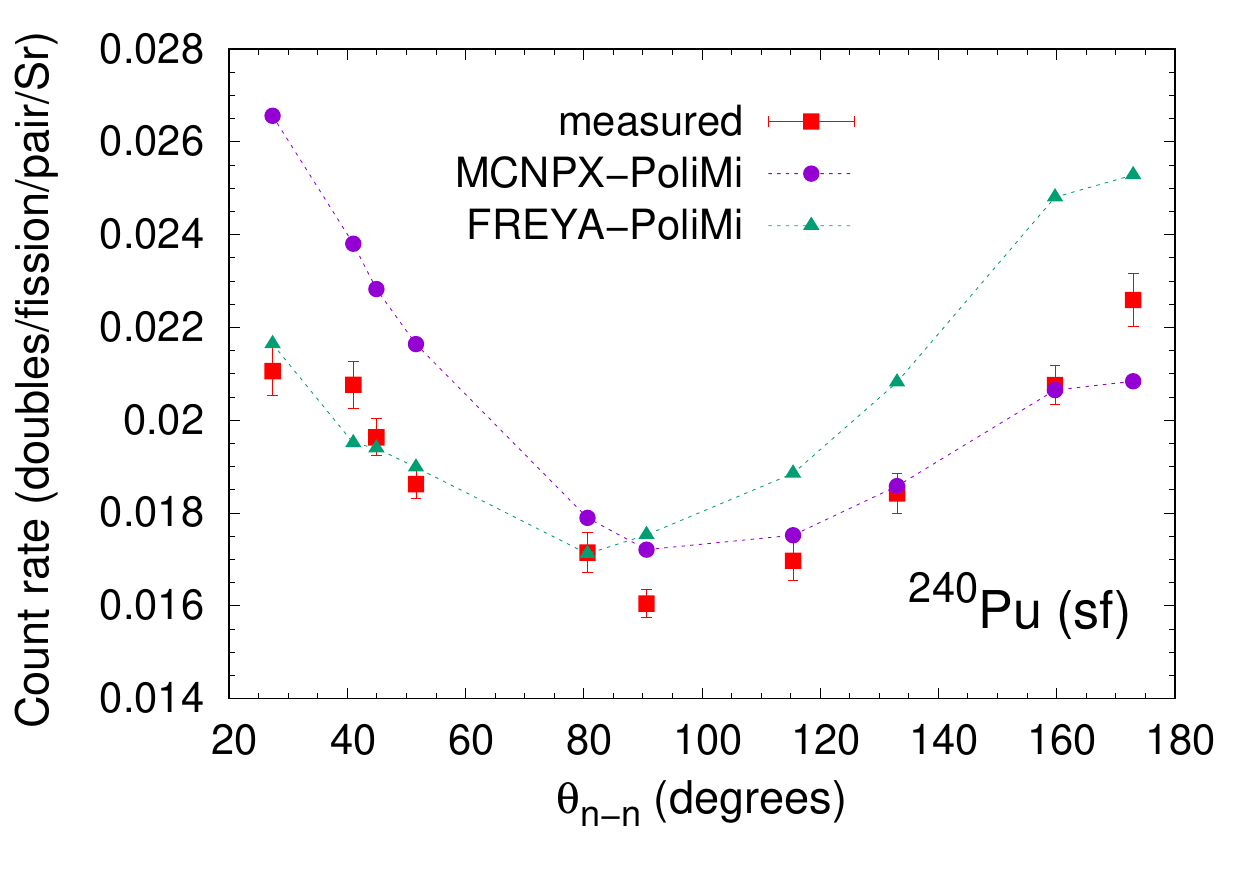}}
\caption{\label{fig:nn-Pu240sf}
(Color online) The neutron-neutron angular distribution for $^{240}$Pu(sf) was measured recently~\protect\cite{Marcath:2016}  (70 keVee distribution in Fig. 10 of Ref.~\protect\cite{Marcath:2016}) and compared to \POLIMI built-in fission model and \FREYA\ simulations with $x = 1.2$. The neutron detection threshold for all results is 0.65~MeV, equivalent to 70~keVee in light output.}
\end{figure}

Neutron-neutron angular correlation measurements of $^{240}$Pu(sf) were performed recently~\cite{Marcath:2016} at the European Commission Joint Research Center, in Ispra, Italy. Figure~\ref{fig:nn-Pu240sf} shows the experimental points (red) in comparison with \POLIMI~and \FREYA~simulations, with cross-talk events removed.  The cross-talk contribution was estimated at each detector angle pair using \POLIMI\ simulations and removed from the measured doubles \cite{Marcath:2016}. 
The \POLIMI~results (purple) overpredict the measured values below 100 degrees and agree above, whereas the \FREYA~results overpredict the data above 80 degrees but agrees below.  Note that the \FREYA~calculation in Fig.~\ref{fig:nn-Pu240sf} is using a value of the $x$ parameter discussed in Sec.~\ref{sec:FREYA} that has not been adjusted to $\nu(A)$ data since none are available.  The default value of $x = 1.2$ from Ref.~\cite{FREYA} was used here, giving a stronger correlation for $\theta_{n-n} > 90$ degrees.

Results on $n$-$n$ angular correlations in spontaneous fission were also obtained at LLNL with the detector setup shown in Fig.~\ref{fig:birthdayCake}, with a $^{240}$Pu source located at the center. The measurements were carried out using a 4.5~mg sample of $^{240}$Pu (98\% pure) of intensity 4,590 neutrons/s. The contribution of fission neutrons originating from other plutonium isotopes present in the sample are negligible. The data were obtained over a 23-hour period.  Only fission neutrons detected within a 40~ns time window were considered correlated.

\begin{figure}[ht]
\centering
\includegraphics[width=\columnwidth]{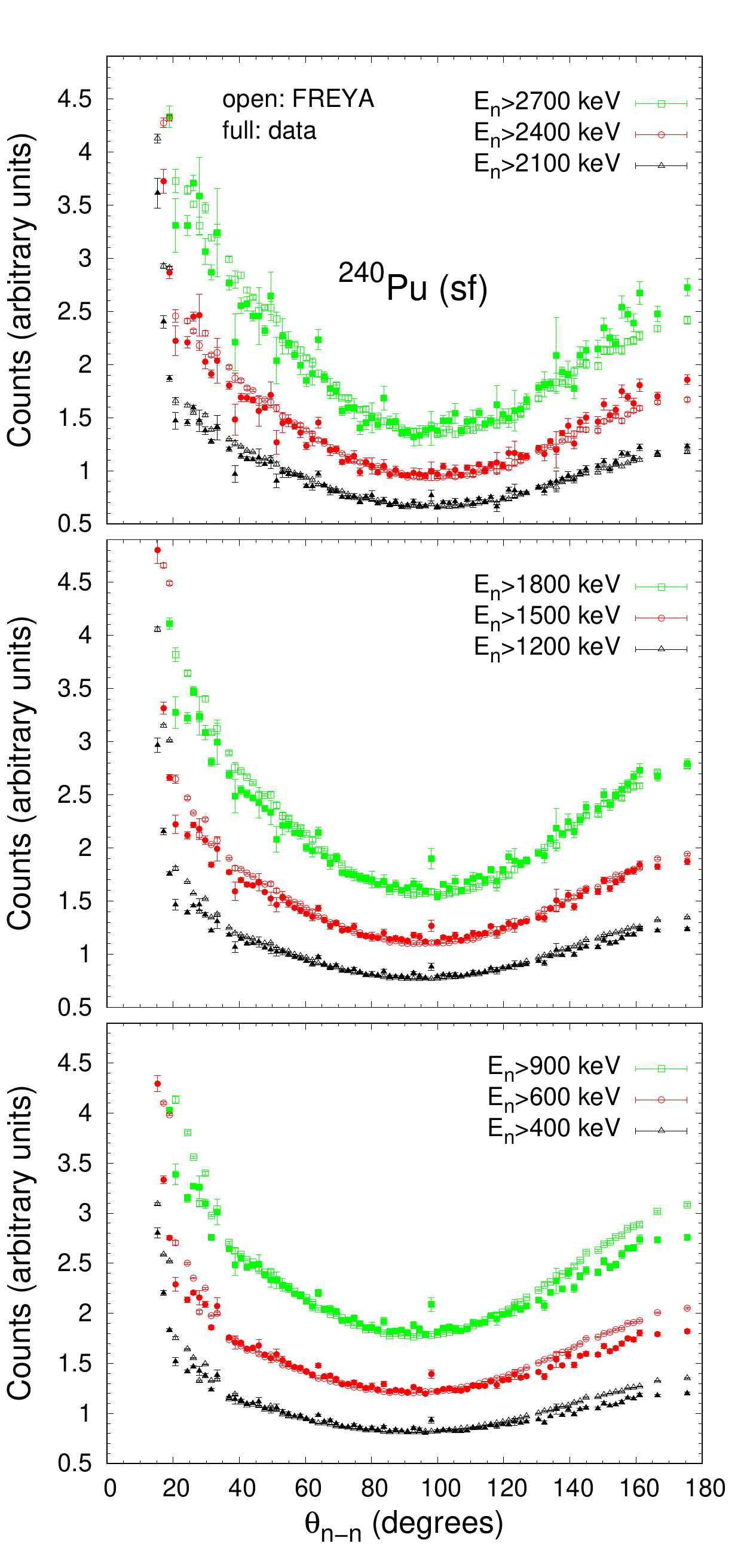}
\caption{(Color online) The neutron-neutron angular correlations for $^{240}$Pu(sf) as a function of a neutron kinetic energy threshold, determined from the neutron time-of-flight using a spontaneous fission photon trigger. The data have been corrected for cross-talk, and compared with \FREYA~simulations using $x=1.3$ \protect\cite{Verbeke:2017}.}
\label{fig:nn-Pu240sf-LLNL}
\end{figure}

Figure~\ref{fig:nn-Pu240sf-LLNL} shows the measured cross-talk corrected correlations for several neutron kinetic energy cutoffs with full symbols.  The angle-dependent cross-talk corrections were estimated by Monte-Carlo simulation.  The corrections were calculated by comparing the number of neutron-neutron coincidences obtained from simulating time-correlated neutrons from spontaneous fission to those obtained from simulating single neutron emission from the same spontaneous fission neutron spectrum.  More details on the experimental setup, cross-talk correction, and analysis can be found in Ref.~\cite{Verbeke:2017}. An \MCNPX~model of the detector setup was developed and \FREYA~simulations were folded in to obtain the results shown with open symbols. In this case, the angular correlations were used to adjust the value of $x$ used, to $x = 1.3$, since these correlations are sensitive to the excitation energy sharing \cite{Verbeke:2017}.   The number of fissions simulated with \FREYA~was equivalent to the 23 hours of data taking in the experiment. The agreement between experiment and simulations is very good for most energies and angles. A default \MCNPX~calculation without \FREYA~would have resulted in flat distributions, except for a peak at 0 degrees due to neutron cross talk before subtraction of cross-talk. 

\subsection{Late-time emission of prompt \grays}

Prompt \gray~emissions can be delayed due to the presence of long-lived isomers in the fission products~\cite{Talou:2016}. Depending on the specific half-lives of those isomers, the observed prompt fission \gray~spectrum and multiplicity can change significantly with the time coincidence window. This is particularly true if one singles out a specific fragment, through, for example, \gray~tagging.

\begin{figure}
\centering
\includegraphics[width=\columnwidth]{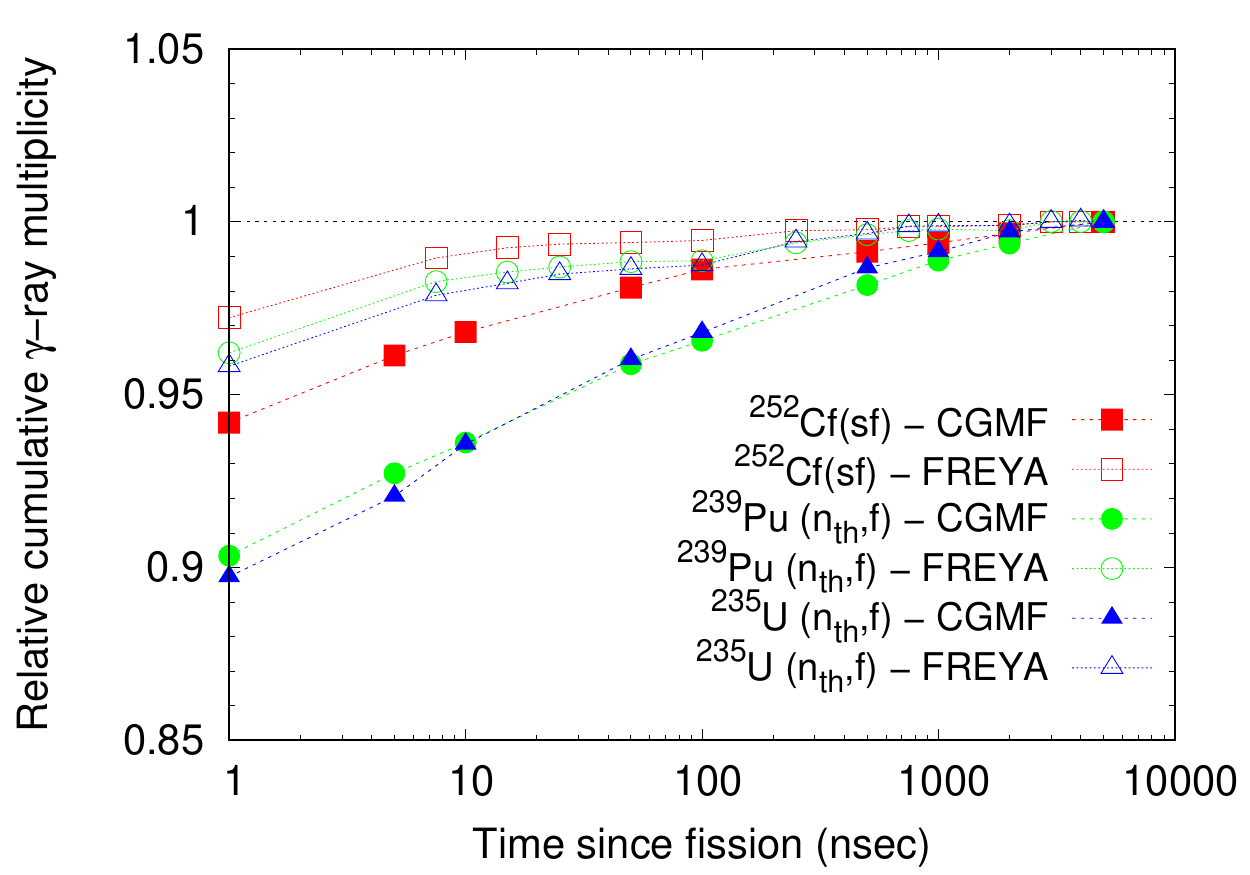}
\caption{(Color online) The relative cumulative prompt fission \gray~multiplicity calculated with \FREYA\ and \CGMF~as a function of time since fission. As can be seen in this figure, 3$-$7\% of the \grays~are emitted after 10 ns following fission. This delay is due to the presence of isomers in the fission fragments that are populated during the evaporation cascade. }
\label{fig:Ngt}
\end{figure}

Figure~\ref{fig:Ngt} shows the relative cumulative \gray~multiplicity as a function of time since fission, normalized to 1.0 at 5 $\mu$s, before any $\beta$-delayed contributions. A typical time coincidence window used for identifying prompt fission neutrons, in coincidence with a fission event, is a few nanoseconds. Up to 8\% of the prompt \grays~are emitted after 5 ns since fission. The determination of the detected \gray~multiplicity can therefore be biased by as much as 8\% for $^{239}$Pu and $^{235}$U thermal neutron-induced fission reactions. In addition, if one uses prompt fission \grays~to estimate a neutron detector efficiency, as discussed in Refs.~\cite{Granier:2015,Chatillon:2014}, the effect of these late prompt \grays~is to artificially bias the efficiency curve for the most energetic neutrons.  Some differences in the calculations may be expected since \CGMF\ and \FREYA\ treat gaps in the RIPL-3 tables in different ways.

Although \MCNP6.2 does not include this time dependent \gray~information, plans are to include it in a future version of the code.

\subsection{\label{sec:timeCorrelations}Time chain correlations} 

Methods based on time-correlated signals have been developed over many decades to characterize fissile materials~\cite{Feynman:1956,Cifarelli:1986,Prasad:2012}. Starting in the 1940s, Neutron Multiplicity Counting (NMC) techniques have enabled quantitative evaluation of masses and multiplications of fissile materials. In NMC, sequences of thermal neutron captures are recorded in $^3$He tubes. The $^3$He$(n,p)$ reaction produces two ions that generate charges within the gas.  These charges are collected by a voltage-biased wire running through the tube. To determine the dominant constituents of the measured objects, the sequences were split into time windows and the numbers of neutrons arriving in each window were recorded to build statistical count distributions.

Some materials such as $^{252}$Cf simply emit several neutrons effectively simultaneously without multiplication (independent fission), whereas others like uranium and plutonium multiply the number of neutrons by subsequent, time-correlated fissions, to form bursts of neutrons. This multiplication translates into unmistakable count distribution signatures. To determine the type of materials measured, one can reconstruct measured count distributions with theoretical ones generated by a fission chain model. When the neutron background is negligible, the theoretical count distributions can be completely characterized by a few parameters: the mass of the object; the multiplication $M$; the $\alpha$-ratio (the ratio of the rate of $(\alpha,n)$ source neutrons to the rate of spontaneous fission neutrons); and the neutron detection efficiency $\epsilon$. For such reconstruction to be successful, the precise knowledge of the multiplicity distributions of the isotopes is important. Indeed, to determine parameters of the object to be characterized by neutron multiplicity counting measurements, it is necessary to solve equations involving factorial moments of the multiplicity distribution, given in Eqs.~(\ref{eq:nu1})-(\ref{eq:nu3}).   Any error in this distribution will thus lead to errors in the parameters of the reconstructed object.
 
The neutron capture cross section in $^3$He is only large enough to record fission neutrons after they have been thermalized in a moderating material. Therefore the time windows must be at least tens to hundreds of microseconds long to collect sufficient counting statistics to pick up counts from the same spontaneous fission or fission chain. In the case of a strong neutron source such as plutonium, many fission chains will thus be generated within individual time windows and therefore overlap within a window. While the neutron time correlations of interest are generated by individual fission chains, the signal received by the $^3$He tubes is a convolution of multiple fission chains. To disentangle the contributions from separate fission chains, counting with $^3$He requires high statistics and thus long measurement times.

Scintillators, on the other hand, can detect fission neutrons directly without the need for a moderator.  Scintillators detect neutrons through inelastic scattering, primarily on hydrogen, emitting a recoil proton and producing prompt scintillation light. Consequently, counting happens on a nanosecond time scale. To detect a correlation signal, microsecond time windows are not required, as with $^3$He, but only of order $\sim 100$~ns. These shorter time windows enormously reduce the number of overlapping chains within a window, so that windows encompass neutrons from far fewer fission chains.

\begin{figure}[!hbt]
\centering
\includegraphics[width=\columnwidth]{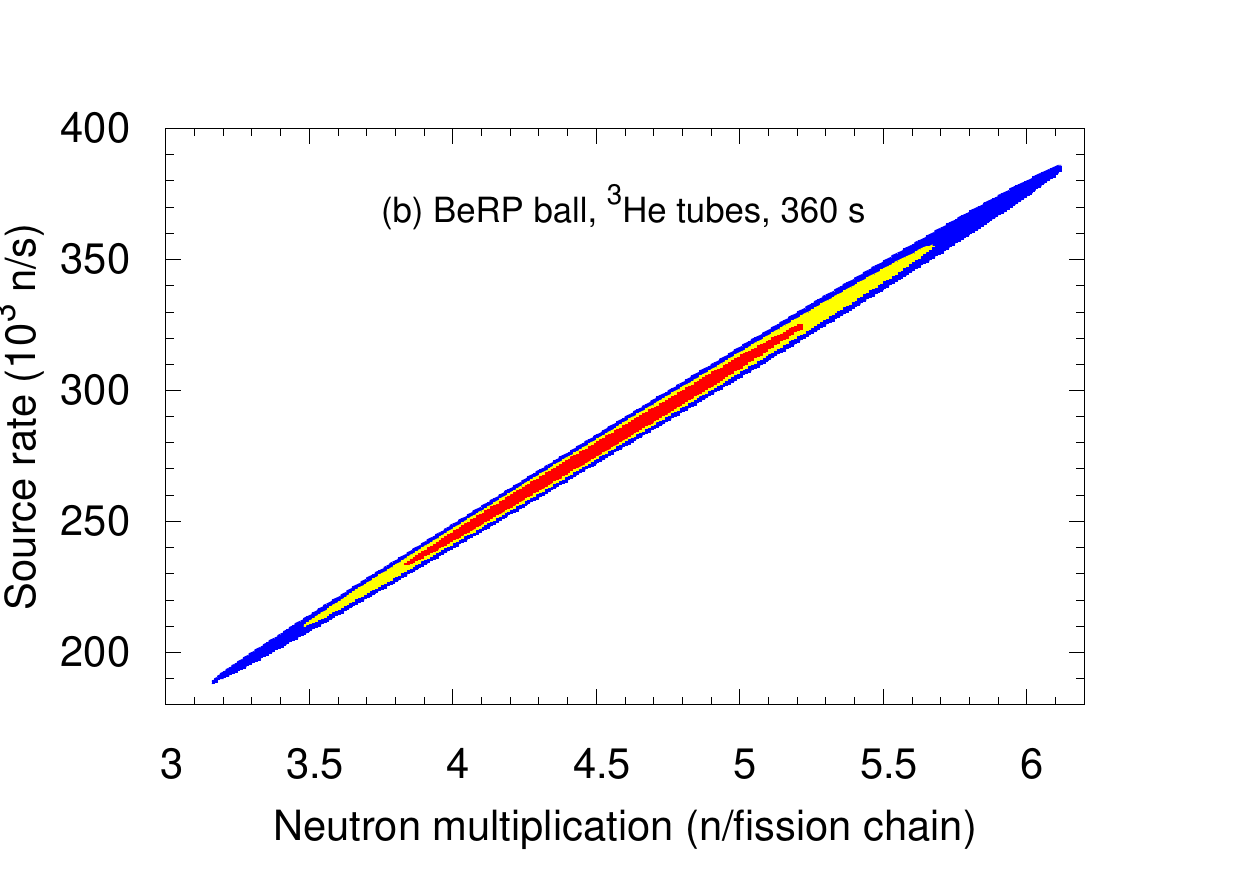}
\includegraphics[width=\columnwidth]{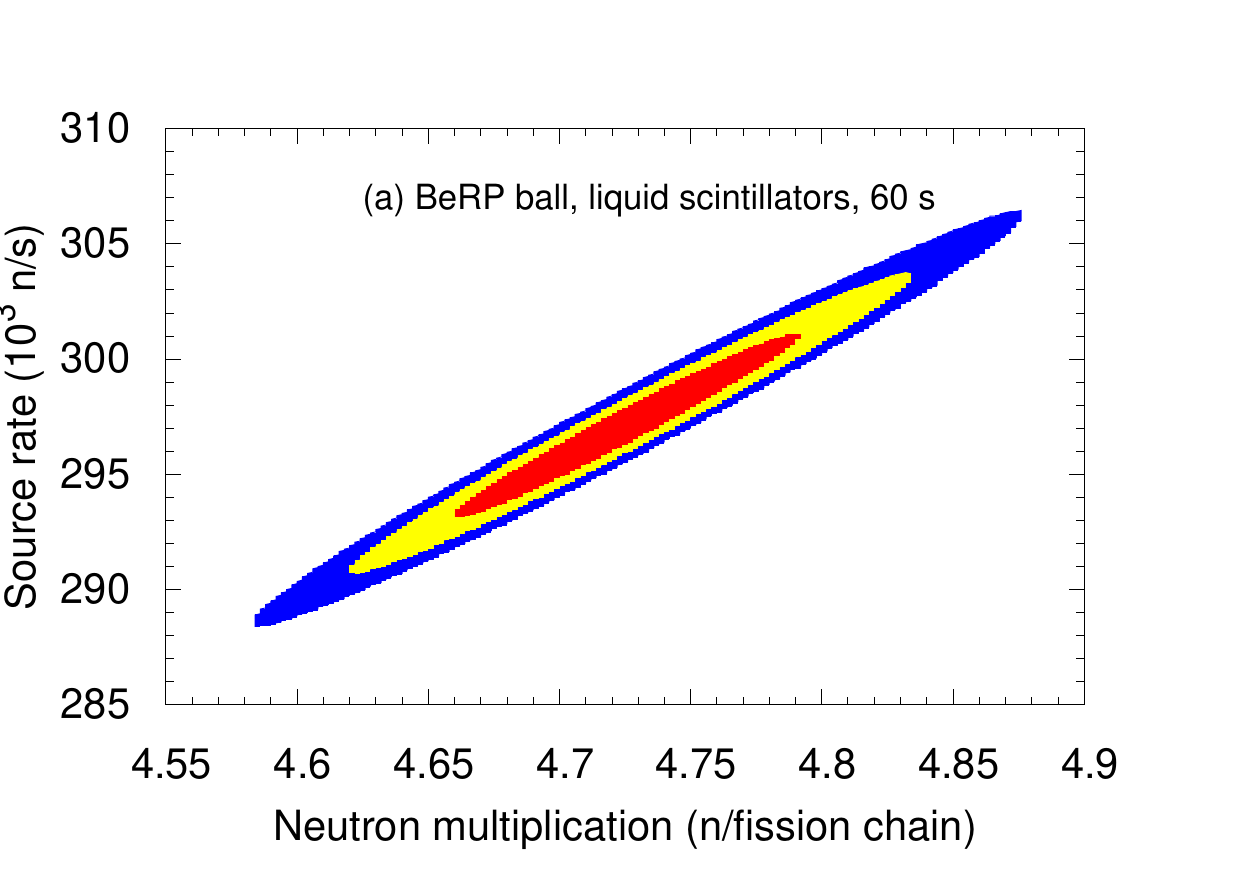}
\caption{(Color online) Reconstruction of source neutron rate, the spontaneous fission neutron yield of the source,  
  as a function of neutron multiplication for BeRP ball simulations~\cite{Verbeke:2016}. The colors outline the regions where the reconstructed  solution lies within confidence levels of 68.27\% (red), 95.45\% (yellow) and 99.73\% (blue). Note the difference in scale between the two plots. (a)  A 360~s measurement with $^3$He tubes.  (b) A 60~s measurement with a liquid scintillator array.}
\label{fig:BeRP}
\end{figure}

In terms of fissile material detection and authentication, scintillators prove to be more efficient than $^3$He tubes. Indeed, smaller time gates allow a larger number of samples to be studied in a given measurement time, leading to reduced uncertainties. Figure~\ref{fig:BeRP} shows the uncertainties on the reconstruction of the mass and multiplication of a simulated Beryllium-Reflected Plutonium (BeRP) ball, often used in criticality-safety measurements~\cite{Verbeke:2016}. When one reconstructs these quantities, the uncertainties on mass and multiplication are much smaller with the scintillators (a) than with the $^3$He tubes (b), even for six times longer $^3$He measurements.  (Note the difference in the axes for the two types of detectors.)

It has also been shown that these new fast counting signatures can help distinguish fast multiplication from thermal multiplication in a system where neutrons restart chains after they have been thermalized by a
moderator~\cite{Chapline:2011}.

Discrimination between neutrons and \grays~opens a new door to coincidence counting applications. Whereas one has historically focused on neutron multiplicity counting, correlations between neutrons and \grays~can now be measured. Feynman's original point model theory, which uses the factorial moments of $P(\nu)$ given in Eqs.~(\ref{eq:nu1})-(\ref{eq:nu3}) to determine the composition of an unknown object, was originally developed for neutron detectors based on $^3$He. The neutron capture cross section of this isotope is such that these detectors are blind to fast neutrons and most sensitive to neutrons that have undergone thermalization, a process which requires fission neutrons to down-scatter for tens of microseconds. To model the fission neutron detection, the original point model theory assumed that neutrons were detected on a time scale much longer than the fission chain evolution time scale: all neutrons in a chain were assumed to be emitted at time zero and then slowly diffused to the detectors over a time scale of tens of microseconds. While this is a good approximation for $^3$He detectors, scintillators can detect $\gamma$ rays and fast neutrons within nanoseconds of their production, long before the fission chain has ended. Therefore, Feynman€'s original assumption of instantaneous fission chain evolution and slow neutron diffusion to the detectors no longer holds on a nanosecond time scale. Feynman's theory has recently been extended to the detection of \grays~\cite{Enqvist:2006,Enqvist:2009,Pazsit:2008} and to fast counting~\cite{Kim:2015}.

\begin{figure}[!hbt]
\begin{center}
\includegraphics[width=0.95\columnwidth]{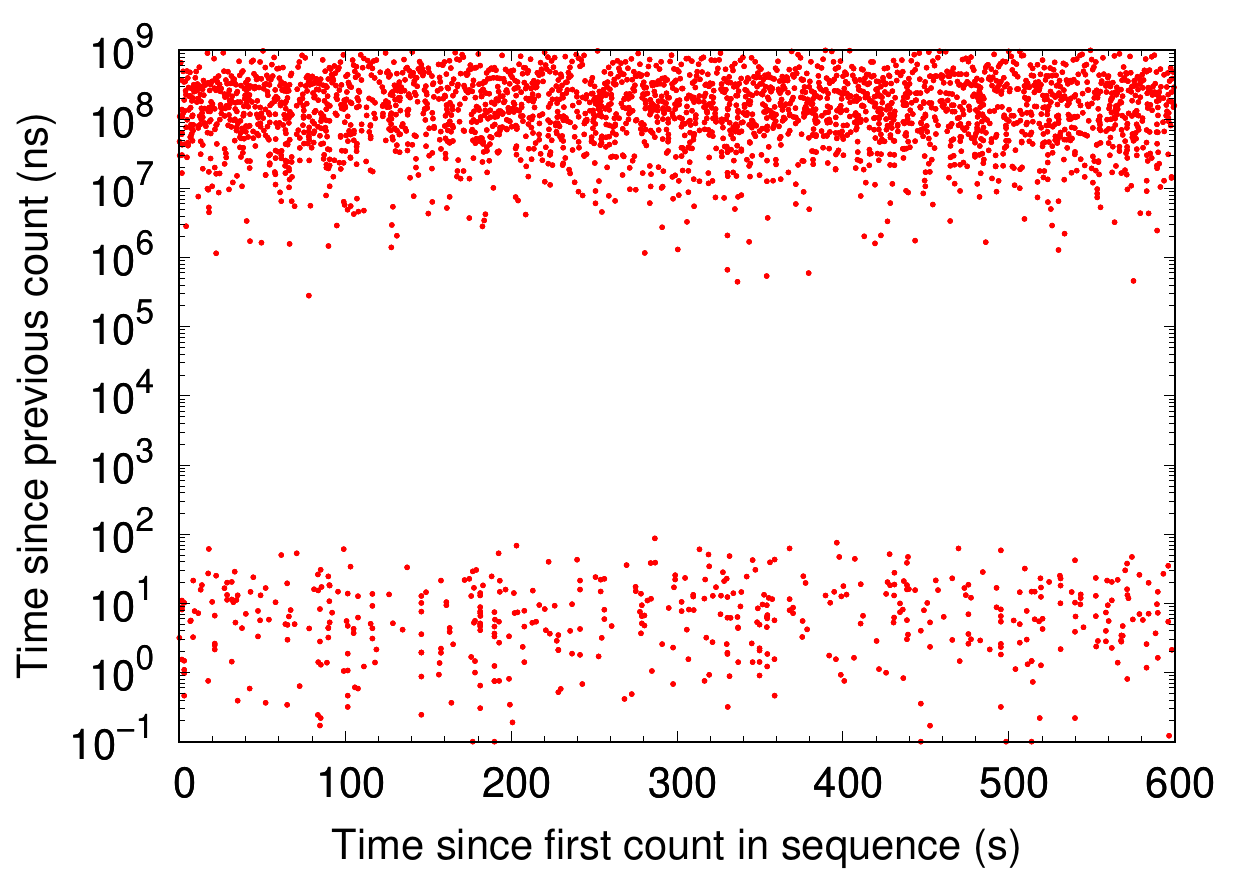} 
\end{center}
\caption{(Color online) The time interval between fast neutrons in a lead pile.}
\label{fig:timeIntervalsLead}
\end{figure}

In terms of data visualization, the shorter time scale over which scintillators detect neutrons enables the study of fission chains on the time scale over which they evolve. By plotting time intervals between fast neutron detections as a function of time, one can easily observe fission chains when fissile isotopes are present. Indeed, when no fissile materials are present, two bands are observed in Fig.~\ref{fig:timeIntervalsLead}: one around 10~ns, which is mainly due to individual cosmic-ray-induced fast neutrons registering multiple counts in adjacent detectors (neutron cross-talk) and a second around 0.1~s, which represents the average time interval between cosmic-ray secondaries interacting with the lead pile. The region between the two time bands is empty.

\begin{figure}[!hbt]
\begin{center}
\includegraphics[width=0.95\columnwidth]{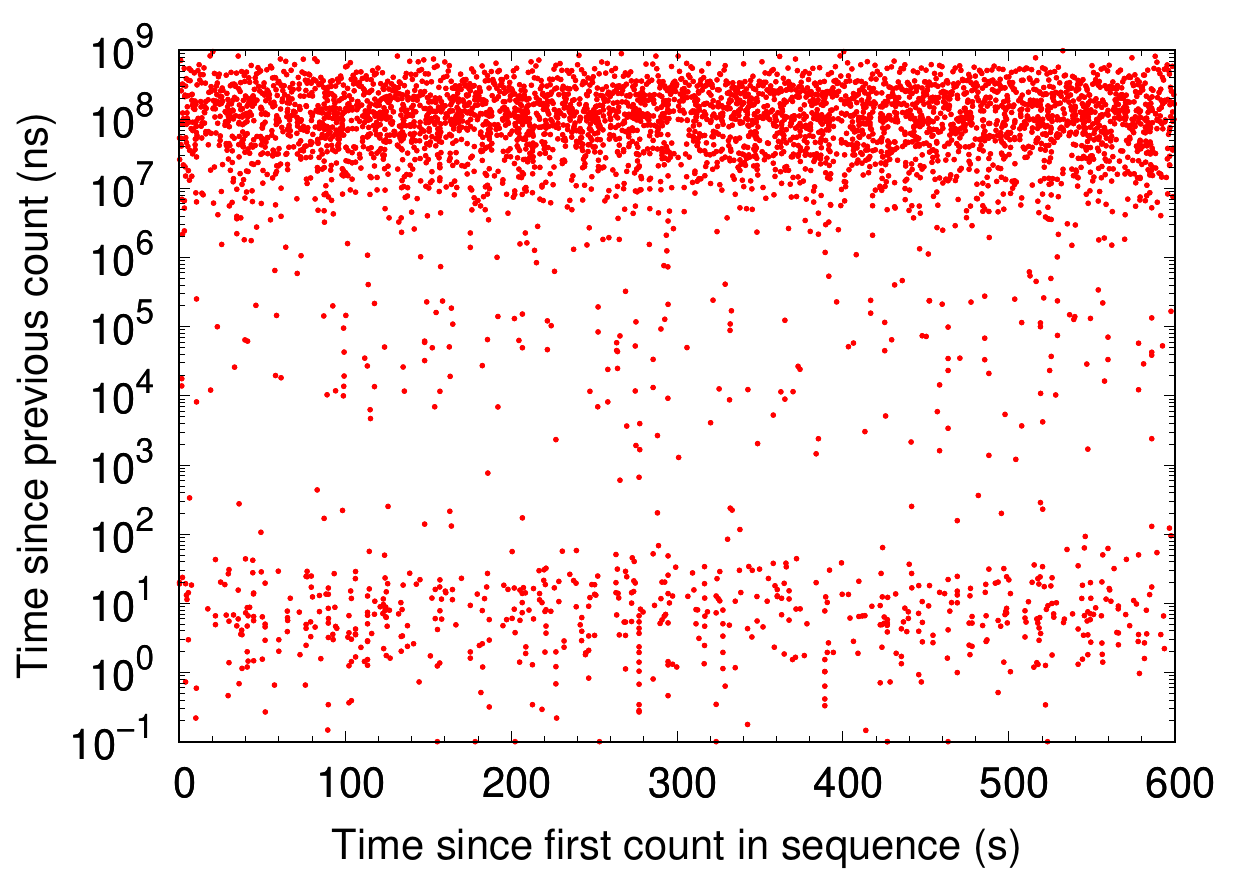} 
\end{center}
\caption{(Color online) The time interval between fast neutrons for a ball of highly-enriched uranium in a lead pile.}
\label{fig:timeIntervalsHEU}
\end{figure}

When fissile materials are present, the gap between the two bands in Fig.~\ref{fig:timeIntervalsLead} fills in, as shown in Fig.~\ref{fig:timeIntervalsHEU}.  The band around 10~ns is now a mix of fast neutrons coming from individual fission reactions, fast neutrons coming from fast fission chains, and neutron cross-talk.  The second band around 0.1~s represents the average time interval between spontaneous fissions as well as cosmic-ray secondaries interacting with the lead pile. The empty region has now filled in. The vertical streaks? filling the gap between the two bands point to the presence of neutron bursts due to fission chains that evolve over time scales of multiple microseconds. Such fission chains can only be due to neutrons thermalizing and restarting new fission chains after a thermal neutron induced fission of $^{235}$U.

\begin{figure}[!hbt]
\begin{center}
\includegraphics[width=0.95\columnwidth]{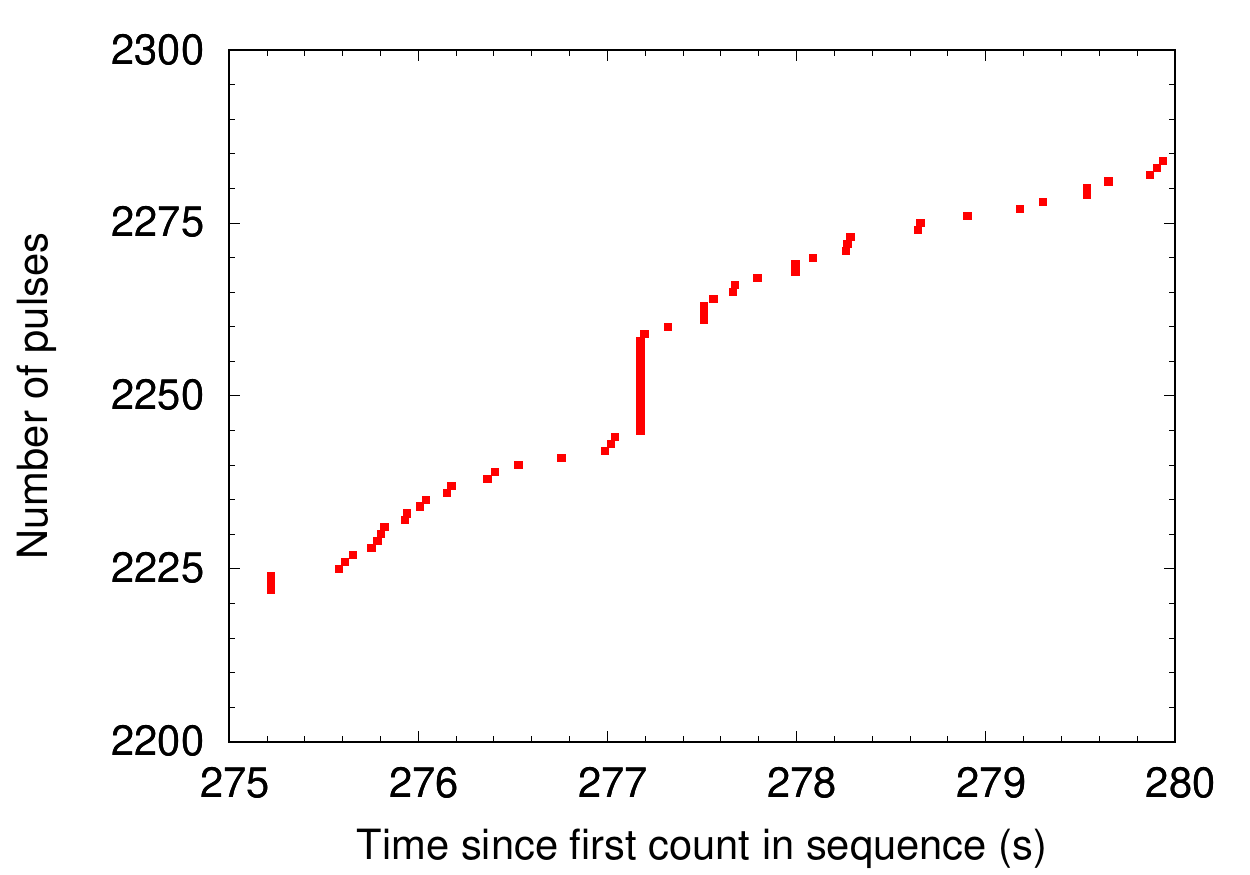} 
\end{center}
\caption{(Color online) The accumulation of fast neutrons from a single fission chain for a ball of highly-enriched uranium in a lead pile.}
\label{fig:accumulation}
\end{figure}

Figure~\ref{fig:accumulation} focuses on a single fission chain observed in Fig.~\ref{fig:timeIntervalsHEU} at time 277~s. It shows the accumulation of fast neutrons as a function of time. The step of 14 neutrons emitted over less than 20~ns is indicative of a fast-neutron fission chain.

In contrast to $^3$He tubes, which detect only thermal neutrons, liquid scintillators detect neutrons above a higher energy threshold with minimum kinetic energies of 500~keV to 1~MeV. There is a threshold because the recoil protons do not produce a sufficiently unique scintillation light pulse to confidently distinguish them from the light pulse produced by \gray~Compton interactions. Reducing this threshold is an area of very prolific research, and the discrimination between neutrons and \grays~has improved over the years through material research and advanced signal processing.

A further advantage of scintillators is their ability to determine the neutron energies. The total light collected from the fast proton recoil in the scintillator is statistically proportional to the incident neutron energy. This enables a statistical energy spectrum to be determined which, for example, can be used to distinguish plutonium metal from plutonium oxide~\cite{Verbeke:2015}.

\section{Correlated fission data: status and challenges} \label{sec:status}

This section presents the current status of \CGMF\ and \FREYA\ and discusses some of the future challenges that have to be met.  The first part deals with a brief discussion of the sensitivity of the results to input data and parameters.  It then goes on to assess the current relevance of these codes for important applications.  The next part highlights some of the modeling issues these codes need to address and describes some ways that current theory improvements may help address these issues.  Since \CGMF\ and \FREYA\ depend upon data for model constraints, some of the basic experimental needs for modeling are touched upon.  Finally, the current capabilities of the codes are briefly summarized.

\subsection{Accuracy and sensitivity of fission event generators}

While modern codes such as the ones presented in this paper can predict many prompt fission data of interest, it is equally important to judge how accurate those predictions are and how sensitive they are to model assumptions and input parameters. Here an initial and somewhat limited attempt to answer this question is reported.

The sensitivity of the calculated observables to a particular model parameter
can be readily determined by simply varying that parameter.
However, because the model parameters are not necessarily mutually independent, 
it is important to consider their correlations.
Furthermore, by sampling the parameter values from probability distributions
obtained by analyzing the experimental errors on the input data 
and performing a statistical analysis of the observables generated by the codes
it is possible to identify the most critical measurements needed for improving
the predictions.

An important ingredient of these calculations is the pre-neutron emission fission fragment yields in mass, charge and kinetic energy. In particular, the kinetic energy carried away by the fission fragments determines, to a large degree, the intrinsic excitation energy left in the fragments and thus the number of prompt neutrons that they will evaporate. Any uncertainty in \TKEbar~will therefore produce an uncertainty in \nubar.  Some preliminary studies of the sensitivity of the calculations to the input yields have been made in the case of $^{252}$Cf(sf), for which a covariance matrix associated with the yields $Y(A,Z,{\rm TKE})$ was estimated based on experimental data. Yields sampled from this covariance matrix were then used as input to \FREYA~and results on prompt neutrons and \grays~analyzed. Preliminary results were reported in~\cite{Randrup:2016} confirming the strong anti-correlation between \TKEbar~and \nubar. In the case of $^{252}$Cf(sf), \nubar~is a ``standard''~\cite{Carlson:2009} and is known with a reported uncertainty of $\sim 0.13$\%. This very small uncertainty places stringent constraints on the values of \TKEbar, much stronger than the reported experimental uncertainty of 1.5~MeV~\cite{Wagemans:1991}.

Although \nubar~is very sensitive to \TKEbar, the average prompt fission neutron spectrum, $n$-$n$ angular correlations, and all \gray~data are only weakly impacted by any change in \TKEbar. Such sensitivity studies are very useful to assess what can be considered as {\it robust} predictions by the event generators, regardless of the precise knowledge of the input parameters, as opposed to quantities that show great sensitivity to those same parameters. 

Accurate simulations of critical assemblies are very sensitive to any change in the underlying nuclear data. Typical examples include the GODIVA and JEZEBEL critical assemblies, made almost entirely of $^{235}$U and $^{239}$Pu respectively, which are considered as the most accurate criticality-safety benchmarks and play a somewhat disproportionate role in validating evaluated nuclear data libraries. Their simulations are very sensitive to the PFNS and \nubar. At this moment, calculations of those two quantities by \FREYA~and \CGMF~cannot reproduce the quality of results present in evaluated libraries such as ENDF/B-VII.1. Therefore the use of those two codes for very accurate criticality-safety applications should be avoided.

A complete uncertainty quantification study would require more than a simple estimate based on parameter sampling. Assumptions and limits of the phenomenological physics models used in the description of prompt neutron and \gray~emission carry some systematic and correlated biases that are more difficult to assess.

Another important example of target accuracy for fission event generators can be found in the assay of nuclear materials. To assay a spontaneous fission source like $^{252}$Cf surrounded by unknown materials, a system of two equations needs to be solved for the source strength $F_{\text{sf}}$ and the neutron detection efficiency $\epsilon$:
\begin{eqnarray}
    R_1 & = & \epsilon \overline{\nu} F_{\text{sf}}  \label{Eq:R_1} \\ 
    R_{2F} & = & \epsilon \nu_2/\overline{\nu} \label{Eq:R2F}
\end{eqnarray}
where $R_1$ is the measured neutron count rate in the detectors. (Note that $\overline \nu F_{\rm sf} = R_1/\epsilon$ is the source rate on the $y$-axis of Fig.~\ref{fig:BeRP}.)  $R_{2F}$ is the measured number of correlated neutron pairs relative to the number of counts, sometimes referred to as the Feynman correlated moment.  The spontaneous fission multiplicity distribution produced by \FREYA~is very close to the distribution from the Santi evaluation~\cite{Santi:2008}. However, when solving Eqs.~(\ref{Eq:R_1}) and (\ref{Eq:R2F}) for $\epsilon$ and $F_{\text{sf}}$, the small differences between the two results lead to differences of 0.13\% in the neutron detection efficiency $\epsilon$ and 0.8\% in the $^{252}$Cf source strength.

\subsection{Challenges for theory and modeling}

Fission event generators such as \FREYA~and \CGMF, coupled with a transport code such as \MCNP, provide a very powerful simulation tool for fission reactions, producing a wealth of correlated data on prompt fission neutrons and \grays~in multiplicity, energy and angle for every single fragmentation in mass, charge, kinetic energy, excitation energy and spin of the initial fragments. Given this daunting task, it is remarkable that such codes provide as good results as they do for a large quantity of data.

However, fundamental theoretical questions about the fission process and the de-excitation of the fission fragments remain. Here a few are mentioned that should be explored in the future to improve the quality and predictive power of these unique tools. 

Accurate pre-neutron emission fission fragment distributions are a crucial input for those codes to produce reliable results. Depending on the type of output data one is interested in, the accuracy with which those yields need to be known varies. The total kinetic energy distribution $Y$(TKE) is important to accurately predict the prompt neutron multiplicity distribution $P(\nu)$. Several promising theoretical efforts~\cite{Moller:2017,Sierk:2017} are underway to generate fragment yields in mass, charge and kinetic energy and provide, for the first time, a rather predictive capability for this quantity. Using the results of such model predictions in fission event generators is still at the early stage but results are very promising. 

For a particular pair of fission fragments, for which the ground-state masses are relatively well known, and for a given TKE, the energy balance of the reaction gives the total excitation energy that will eventually be released through prompt neutron and \gray~emission. How energy is shared between the two fragments at scission has been the object of various theoretical studies~\cite{Schmidt:2010,Morariu:2012}, with reasonable success. The importance of the extra-deformation energy induced by the collective deformation of the fragments near scission compared to their ground-state shapes, and the role of the level density in the fragments remain to be quantified more precisely. 

There is rather conclusive evidence that the average angular momentum of the fission fragments following scission is much higher ($\sim 8 \hbar$) than the value observed in low-energy compound nuclear reactions ($\sim 4 \hbar$). Various mechanisms for producing such high angular momentum have been discussed~\cite{Bonneau:2007,Goennenwein:2007,Kadmensky:2009}. A quantifiable and predictive theory of the angular momentum distribution for a given fragmentation in mass, charge and kinetic energy remains to be developed. Again, microscopic calculations may be best equipped to address this point.  The Hartree-Fock+BCS theoretical framework using a Skyrme nucleon-nucleon interaction was used~\cite{Bonneau:2007} in the discussion of angular momentum production in the quantum pumping process. 

All those considerations, which point to the initial conditions of the fission fragments immediately following scission can only be addressed theoretically through an appropriate treatment of the dynamics of the fission process from the saddle to the scission point. Semi-classical macroscopic-microscopic approaches to this problem, which rely on a description of the fissioning nucleus via a set of shape parameters and which treat the dynamics with random walk or Langevin-type equations, have enjoyed some recent successes in describing the initial fragment yields~\cite{Sierk:2017,Randrup:2011,Aritomo:2014}. Microscopic approaches such as the Time-Dependent Hartree-Fock~\cite{Staszczak:2009}, the Time-Dependent Generalized Coordinate  Method~\cite{Regnier:2016} or even the Time-Dependent Superfluid Local Density Approximation~\cite{Bulgac:2016} are promising methods for describing fission dynamics in a more fundamental, microscopic approach. 

Fission dynamics right around the scission point is important for at least one other reason. So-called ``scission neutrons", emitted right at the time of the rupture of the neck have been postulated for many years, and have been used at various times to explain discrepancies between observed and calculated PFNS and angular distributions of neutrons. However, such interpretations are highly model-dependent and lead to a large spread of predictions for this extra source of neutrons. That is not to say that such a neutron source cannot exist. However, its existence should be addressed through a more consistent approach, leaving less room for fitting.  Neither scission neutrons nor neutrons emitted during fragment acceleration are included in \CGMF\ and \FREYA.

A high value for the initial angular momentum of the fission fragments could lead to anisotropic neutron emission in their center-of-mass frame. Such an assumption led Terrell~\cite{Terrell:1959} to derive an analytical formula that has been commonly used to infer a softer PFNS, in better agreement with selected experimental data. The anisotropy parameter used in this formula is adjusted to reproduce the PFNS below a few hundred keV, regardless of the neutron angular distributions, which are often unknown.  Model predictions of the angular momentum distribution should be able to constrain this parameter more effectively.

\FREYA~and \CGMF~treat the de-excitation of the fission fragments in the framework of the statistical Weisskopf-Ewing and Hauser-Feshbach nuclear reaction theories respectively.  They rely on extensive databases and systematics of nuclear reaction model parameters, such as those of the RIPL-3 ``Reference Input Parameter Library''~\cite{RIPL3}. Many of these models are phenomenological in nature and their parameters have been tuned to reproduce experimental data available in subsets of the nuclear chart, most extensively near the valley of stability.  Some of the difficulties noticed in reproducing the PFNS in well-known fission reactions, {\it e.g.}, $^{252}$Cf(sf) with the present calculations, are certainly due in part to the inaccuracy of such models. Improvements in a global deformed optical potential applicable across a large suite of deformed, neutron-rich nuclei, are necessary.

In those models, the $\gamma$ decay probabilities are estimated using the strength function formalism. Significant efforts have been devoted to this topic in recent years, leading in particular to the discovery 
of the importance of the ``scissors mode'' to the $(n,\gamma)$ cross section calculations~\cite{Ullmann:2014}. Systematics for the description of this mode throughout the nuclear chart are being developed.  It will be interesting to study the impact of these systematics on the predictions of fission event generators.

The nuclear level densities used to represent the continuum of states above the known energy region of resolved excitations are an important input to statistical nuclear reactions. They are commonly described using a constant temperature formula at low energy with a Fermi gas formula at higher energies. For many nuclei away from the valley of stability the parameters entering in these formulae are obtained from systematics not necessarily accurate in these regions. In the present context, the level densities can also be used to share the excitation energy between the fragments. 

\subsection{Experimental needs and status}

A tremendous amount of work has been accomplished recently in measurements of the fission fragment yields for various fission reactions and at a number of excitation energies. The SOFIA experimental program~\cite{Martin:2014,Pellereau:2017} at GSI, Darmstadt, has been measuring fission product yields with unprecedented accuracy through Coulomb excitation in reverse kinematics. While such data are not monoenergetic, they still provide very valuable benchmarks for theoretical developments. The SPIDER project at LANSCE is a $2E$-$2v$ experiment measuring fission fragment yields for incident neutron energies from thermal up to 200~MeV. So far, $n+^{235}$U and $n+^{238}$U data have been released~\cite{Duke:2015} with a two-arm spectrometer configuration. The incident neutron energy dependence of fission product yields was also measured at TUNL~\cite{Bhatia:2014} for several important actinides using monoenergetic neutrons through $(d,p)$ and $(d,t)$ reactions. Other efforts such as the VERDI~\cite{Fregeau:2016} and EXILL~\cite{Materna:2015} experiments are also trying to provide new information on fission fragment yields and their dependence on excitation energy.

Measurements of \TKEbar~as a function of incident neutron energy in the fast region and above were performed recently at LANSCE, providing invaluable data for $^{235,238}$U\nf~\cite{Duke:2015,Duke:2016} and $^{239}$Pu\nf~\cite{Meierbachtol:2016} for $E_{\rm inc}$ up to $\sim 200$~MeV.  These data have been used to constrain \CGMF~calculations below the second-chance fission threshold in order to accurately reproduce the energy dependence of the average neutron multiplicity. The initial rise of \TKEbar~at low incident energies observed in $^{235,238}$U remains somewhat of a mystery and constitutes a test of current theoretical approaches to the fission dynamics. Other measurements of similar systems would be useful.
Note that high-energy resolution measurements of $\langle TKE \rangle$ in the
resonance regions have been performed~\cite{Hambsch:1989,Hambsch:2011} to study
the correlation between TKE and $\overline{\nu}$ fluctuations.

Experiments aimed at measuring correlations between prompt neutrons and \grays~as a function of fragment characteristics are obviously ideal to benchmark the type of studies presented here. The average neutron multiplicity as a function of the fragment mass, \nubar$(A)$, has been measured for only a handful of spontaneous and low-energy fission reactions. Only two experimental results have been reported for higher-energy neutrons: $^{237}$Np\nf\ and $^{235}$U\nf~\cite{Mueller:1984,Naqvi:1986}, although additional information can be somewhat inferred from proton-induced fission reactions~\cite{Burnett:1971}. The lack of good experimental data on \nubar($A,E^*$) as a function of increasing excitation energy, as well as on the average \gray~multiplicity as a function of fragment mass and excitation energy, \nubarg($A,E^*$), is preventing the emergence of a clear theoretical model of the energy sorting and angular momentum production mechanisms at scission. 

Recent results~\cite{Gook:2014} on the PFNS and the neutron-light fragment angular distributions $\theta_{n, {\rm LF}}$ as a function of the fragment mass are very useful for testing assumptions made about the energy sharing mechanisms while at the same time providing constraints on the magnitude of any anisotropy parameter, often introduced rather arbitrarily in PFNS evaluations. Such measurements should be extended to neutron-induced reactions up to at least $E_{\rm inc} = 20$~MeV.

Measurements of the average neutron multiplicity \nubar~as a function of TKE have been reported for various low-energy and spontaneous fission reactions. All fission event generators such as \FREYA~and \CGMF~have been able to reproduce the observed trend for $^{252}$Cf(sf) but have failed to reproduce the ones reported for $^{235}$U\nfth\ and $^{239}$Pu\nfth. Recently, Hambsch argued~\cite{Hambsch:2015} that all previously reported experimental trends have been biased due to poor mass and energy resolution and that more recent and more accurate results would tend to agree better with theoretical calculations.

\subsection{Current capabilities}

At this time, the \FREYA~and \CGMF~fission event generators now integrated into \MCNP6.2 can compute correlated fission data for a limited set of fission reactions and isotopes. Those are listed in Table~\ref{tab:applicability}. Only spontaneous fission and neutron-induced fission have been considered so far. Photofission reactions play a special role. A simple hack of both code input files, with yields and TKE based on these quantities in neutron-induced fission, 
can be used to calculate photofission reactions, but only in an approximate way, which could lead to incorrect results if used improperly.

\begin{table}[htb]
\centering
\begin{tabular}{c|c|c|c}
\hline
\hline
Isotope & $E_{\rm inc}$ (MeV) & \FREYA & \CGMF \\
\hline
\hline
$n+^{233}$U & 0$-$20 & \checkmark & - \\
$n+^{235,238}$U & 0$-$20 & \checkmark & \checkmark \\
$^{238}$U & sf & \checkmark & \checkmark \\
$n+^{239,241}$Pu & 0$-$20 & \checkmark & \checkmark \\
$^{238}$Pu & sf & \checkmark & - \\
$^{240,242}$Pu & sf & \checkmark & \checkmark \\
$^{244}$Cm & sf & \checkmark & - \\
$^{252}$Cf & sf & \checkmark & \checkmark \\
\hline
\hline
\end{tabular}
\caption{\label{tab:applicability}List of isotopes and fission reactions that the current versions of \FREYA~and \CGMF~can handle. This list will eventually be extended to all isotopes and fission reactions, in particular photofission, in the near future.}
\end{table}

It is very important to note that, at this point, the use of those fission event generators should be limited to the simulations of correlated data only. As discussed earlier, the average prompt fission neutron spectra predicted by those codes are not necessarily in agreement with the evaluated PFNS present in libraries such as ENDF/B-VII.1. Two main reasons contribute to this situation.  First, evaluated PFNS are often obtained by combining model calculations and experimental data~\cite{Capote:2016}.  In some cases, only experimental PFNS data are used, as in the case of the $^{252}$Cf(sf) standard.  Next, the \FREYA~and \CGMF~model parameters have not (yet) been tuned to obtain a better agreement with the evaluated PFNS.  One should note that the evaluated PFNS are not necessarily correct.  Because they are part of a coherent ensemble of evaluated data, which has been validated against various integral benchmarks, it is, however, difficult to modify them without negatively impacting the benchmark results {\it e.g.} by modifying other quantites such as the neutron multiplicity distribution. 
At this stage, the correlated fission option in \MCNP6.2 cannot be used for criticality safety calculations. In the future, however, the hope is to be able to reconcile the event generator calculations with benchmark simulations.

\section{Conclusions and future work} \label{sec:conclusion}

Fission event generators such as \FREYA~and \CGMF~that can follow the de-excitation of fission fragments through neutron and \gray~emissions in detail provide a unique and powerful view into the post-scission physics of a nuclear fission reaction. They can be used to push for a more fundamental understanding of the fission process as well as for developing new powerful applications that implicitly or explicitly make use of the natural correlations between particles emitted from the fragments.

Fundamental questions about nuclear fission remain unanswered or only partially answered to this day, such as: What are the configurations of the nascent fragments near the scission point? What is the possible contribution of scission or pre-scission neutrons on the average total prompt fission neutron spectrum? Are the prompt neutrons emitted from fully accelerated fragments only? Are prompt \g~rays emitted in competition with prompt neutrons? Fission event generators are based on the statistical nuclear theory of compound nuclei, which should apply quite well in the case of excited fission fragments. Significant differences exist between \FREYA~and \CGMF~in the way they compute the probabilities of neutron and \gray~emissions at a given stage in the de-excitation cascade. Those differences, instead of being a drawback, can help reveal interesting and overlooked features of the fission process. However, both codes rely on a suite of phenomenological model parameters whose values remain somewhat uncertain. In most cases, those parameters have been tuned to nuclear reactions close to the valley of stability, as opposed to the neutron-rich region where fission fragments are produced. Work to better understand and quantify the uncertainties associated with the model parameters, as well as provide reasonable bounds on the applicability of the physics models used in those cases, is in progress.

From a more applied point of view, fission event generators can already be efficiently used to extract much more information from a set of neutron and \gray~data than by simply relying on average quantities. For instance, neutron chain reactions cannot be simulated properly if only an average neutron multiplicity is used in the transport calculations. Experimental data and systematics on $P(\nu)$ have been used for a long time but, in most cases, assuming that the neutron energy spectrum does not depend on how many neutrons are emitted. The more realistic descriptions used in these Monte Carlo codes provide a way to go beyond such approximations. Also, most $P(\nu)$ systematics concern spontaneous fission or thermal neutron-induced fission reactions only. Both \FREYA~and \CGMF~can now provide results up to 20~MeV incident neutron energies. Angular correlations between emitted neutrons are also accessible since the successive emissions are obtained from any fission event. Such data can be invaluable for unambiguously detecting a fission signature. Prompt \grays~are simulated either using an appropriate statistical \gray~strength function model or from specific transitions between known discrete nuclear levels. Those discrete lines show up mostly in the low-energy part of the \gray~spectrum, and the calculations reproduce some of the most recent measurements on a few fission reactions quite well. As some of those discrete states are isomers whose half-lives range from a few nanoseconds to a few microseconds, the prompt \gray~spectrum changes in time, and can be used to tag specific fission fragments.

As has been seen throughout this work, \FREYA~and \CGMF~can reproduce many fission data reasonably well, although the validity and accuracy of the predictions depends significantly on the specific quantity of interest. It also depends to a large extent on many accurate experimental data that are already known for a particular fission reaction. For instance, a wealth of experimental information has been accumulated on the spontaneous fission of $^{252}$Cf and further predictions of prompt neutron and \gray~observables from \FREYA~and \CGMF~are therefore more trustworthy for this nucleus than in making similar predictions for lesser known actinides, such as Cm. The incident neutron energy dependence of the results is also more questionable as multi-chance fission and pre-equilibrium components become important. Some predictions such as neutron-neutron angular correlations are also fairly robust against variations in the model input parameters while others are very sensitive to particular inputs, {\it e.g.}, $P(\nu)$ varies strongly with the distribution in total kinetic energy of the pre-neutron emission fission fragments.

The past few years have seen a resurgence in the accurate measurement and modeling of fission fragment yields in mass, charge and kinetic energy, $Y(Z,A,{\rm TKE})$. An accurate representation of those yields is critical to providing reliable predictions for many prompt fission neutron and \gray~observables. Several experimental efforts are already underway to address some of the most important limitations of past experiments, such as poor mass resolution and low statistics and thus offer accurate measurements of the fission fragment yields of many isotopes as a function of excitation energy for the first time. Nuclear theories have also made great strides toward developing rather predictive tools, with several approaches 
in healthy competition. Evaluated nuclear data files related to fission yields should see large improvements in the next few years, thereby significantly improving our understanding of prompt fission data as well.

Finally, experiments aimed at capturing correlated fission data are also being devised, like DANCE+NEUANCE at LANL. While it is possibly foolish or even misguided to search for an all-encompassing experimental fission setup that would measure everything, it is also quite interesting and important to study correlated data as they place stringent constraints on theoretical models of fission, which in turn help devise more predictive tools.

In the near future, the plan is to significantly extend the number of fission reactions, energies and isotopes that \FREYA~and \CGMF~can handle. Uncertainty quantification and the optimization of model input parameters to reproduce most of the experimental data is also underway. It is an important step towards ensuring good agreement between simulations and very accurate criticality benchmarks while at the same time predicting correlations beyond anything than can be reasonably stored in a tabulated file. The power of incorporating such fission event generators in \MCNP~will become more and more evident as time goes by and the simulations get faster and more reliable.

\section*{Acknowledgements}

We are grateful to A. J. Sierk, M. B. Chadwick and J. P.~Lestone for discussions.  This work was supported by the Office of Defense Nuclear Nonproliferation Research \& Development (DNN R\&D), National Nuclear Security Administration, US Department of Energy. The work of P. T., M. E. R., M. T. A., P. J., M. J., T. K., K. M., G. R., A. S., I. S. and C. W. was performed under the auspices of the National Nuclear Security Administration of the U.S.\ Department of Energy at Los Alamos National Laboratory under Contract DE-AC52-06NA25396. The work of R. V., L. N., and J. V. was performed under the auspices of the U.S. Department of Energy by Lawrence Livermore National Laboratory under Contract DE-AC52-07NA27344.  The work of J. R.\ was performed under the auspices of the U.S.\ Department of Energy by Lawrence Berkeley National Laboratory  under Contract DE-AC02-05CH11231.



\bibliography{references}
\bibliographystyle{epj}

\end{document}